\documentclass[twocolumn]{aastex63}

\newcommand{\askap}{ASKAP J173608.2$-$321635}
\newcommand{\change}[1]{{#1}}

\received{XXX}
\revised{YYY}
\accepted{ZZZ}
\submitjournal{ApJ}

\shorttitle{Discovery of \askap}
\shortauthors{Wang et al.}

\begin{document}

\title{Discovery of \askap\ as a Highly-Polarized Transient Point Source with the Australian SKA Pathfinder}

\correspondingauthor{Ziteng Wang}
\email{zwan4817@uni.sydney.edu.au, tara.murphy@sydney.edu.au}

\author[0000-0002-2066-9823]{Ziteng Wang}
\affiliation{Sydney Institute for Astronomy, School of Physics, University of Sydney, Sydney, New South Wales 2006, Australia.}
\affiliation{ATNF, CSIRO Space and Astronomy, PO Box 76, Epping, New South Wales 1710, Australia}
\affiliation{ARC Centre of Excellence for Gravitational Wave Discovery (OzGrav), Hawthorn, Victoria, Australia}

\author[0000-0001-6295-2881]{David~L.~Kaplan}
\affiliation{Center for Gravitation, Cosmology, and Astrophysics, Department of Physics, University of Wisconsin-Milwaukee, P.O. Box 413, Milwaukee, WI 53201, USA}

\author[0000-0002-2686-438X]{Tara Murphy}
\affiliation{Sydney Institute for Astronomy, School of Physics, University of Sydney, Sydney, New South Wales 2006, Australia.}
\affiliation{ARC Centre of Excellence for Gravitational Wave Discovery (OzGrav), Hawthorn, Victoria, Australia}

\author[0000-0002-9994-1593]{Emil Lenc}
\affiliation{ATNF, CSIRO Space and Astronomy, PO Box 76, Epping, New South Wales 1710, Australia}

\author[0000-0002-9618-2499]{Shi Dai}
\affiliation{School of Science, Western Sydney University, Locked Bag 1797, Penrith South DC, NSW 2751, Australia}

\author[0000-0001-8715-9628]{Ewan Barr}
\affiliation{Max-Planck-Institut f\"{u}r Radioastronomie, Auf dem H\"{u}gel 69, D-53121 Bonn, Germany}

\author[0000-0003-0699-7019]{Dougal Dobie}
\affiliation{ARC Centre of Excellence for Gravitational Wave Discovery (OzGrav), Hawthorn, Victoria, Australia}
\affiliation{Centre for Astrophysics and Supercomputing, Swinburne University of Technology, Hawthorn, Victoria, Australia}

\author[0000-0002-3382-9558]{B. M. Gaensler}
\affiliation{Dunlap Institute for Astronomy and Astrophysics, University of Toronto, 50 St. George St., Toronto, ON M5S 3H4, Canada}
\affiliation{David A. Dunlap Department of Astronomy and Astrophysics, University of Toronto, 50 St. George St., Toronto, ON M5S 3H4, Canada}

\author[0000-0002-2155-6054]{George Heald}
\affiliation{CSIRO Space and Astronomy, PO Box 1130, Bentley WA 6102, Australia}

\author[0000-0002-9415-3766]{James K. Leung}
\affiliation{Sydney Institute for Astronomy, School of Physics, University of Sydney, Sydney, New South Wales 2006, Australia.}
\affiliation{ATNF, CSIRO Space and Astronomy, PO Box 76, Epping, New South Wales 1710, Australia}
\affiliation{ARC Centre of Excellence for Gravitational Wave Discovery (OzGrav), Hawthorn, Victoria, Australia}

\author[0000-0003-4609-2791]{Andrew O'Brien}
\affiliation{Center for Gravitation, Cosmology, and Astrophysics, Department of Physics, University of Wisconsin-Milwaukee, P.O. Box 413, Milwaukee, WI 53201, USA}

\author[0000-0003-3860-5825]{Sergio Pintaldi}
\affiliation{Sydney Informatics Hub, The University of Sydney, NSW 2008, Australia}

\author[0000-0003-1575-5249]{Joshua Pritchard}
\affiliation{Sydney Institute for Astronomy, School of Physics, University of Sydney, Sydney, New South Wales 2006, Australia.}
\affiliation{ATNF, CSIRO Space and Astronomy, PO Box 76, Epping, New South Wales 1710, Australia}
\affiliation{ARC Centre of Excellence for Gravitational Wave Discovery (OzGrav), Hawthorn, Victoria, Australia}

\author[0000-0003-2177-6388]{Nanda Rea}
\affiliation{Institute of Space Sciences (ICE, CSIC), Campus UAB, Carrer de Can Magrans s/n, 08193, Barcelona, Spain}
\affiliation{Institut d'Estudis Espacials de Catalunya (IEEC), Carrer Gran Capit\`a 2–4, 08034 Barcelona, Spain}

\author[0000-0001-6682-916X]{Gregory R. Sivakoff}
\affiliation{Department of Physics, University of Alberta, CCIS 4-181, Edmonton, AB T6G 2E1, Canada}

\author[0000-0001-9242-7041]{B. W. Stappers}
\affiliation{Jodrell Bank Centre for Astrophysics, Department of Physics and Astronomy, The University of Manchester, Manchester M13 9PL, UK} 

\author[0000-0001-8026-5903]{Adam Stewart}
\affiliation{Sydney Institute for Astronomy, School of Physics, University of Sydney, Sydney, New South Wales 2006, Australia.}

\author[0000-0002-4039-6703]{E. Tremou}
\affiliation{LESIA, Observatoire de Paris, CNRS, PSL Research University, Sorbonne Université, Université de Paris, Meudon, France}

\author[0000-0003-0203-1196]{Yuanming Wang}
\affiliation{Sydney Institute for Astronomy, School of Physics, University of Sydney, Sydney, New South Wales 2006, Australia.}
\affiliation{ATNF, CSIRO Space and Astronomy, PO Box 76, Epping, New South Wales 1710, Australia}
\affiliation{ARC Centre of Excellence for Gravitational Wave Discovery (OzGrav), Hawthorn, Victoria, Australia}

\author[0000-0002-6896-1655]{Patrick A. Woudt}
\affiliation{Inter-University Institute for Data-Intensive Astronomy, Department of Astronomy, University of Cape Town, Private Bag X3, Rondebosch 7701, South Africa}

\author[0000-0002-9583-2947]{Andrew Zic}
\affiliation{Department of Physics and Astronomy, and Research Centre in Astronomy, Astrophysics and Astrophotonics, Macquarie University, NSW 2109, Australia}
\affiliation{ATNF, CSIRO Space and Astronomy, PO Box 76, Epping, New South Wales 1710, Australia}

\begin{abstract}

We report the discovery of a highly-polarized, highly-variable, steep-spectrum radio source, \askap, located $\sim$4$\degr$ from the Galactic center in the Galactic plane.
The source was detected six times between 2020~January and 2020~September as part of the Australian Square Kilometre Array Pathfinder Variables and Slow Transients (ASKAP VAST) survey at 888\,MHz.   It exhibited a high degree ($\sim 25$\%) of circular polarization when it was visible.
We monitored the source with the MeerKAT telescope from 2020~November to 2021~February on a 2--4 week cadence. The source was not detected with MeerKAT before 2021~February~07 when it appeared and reached a peak flux density of 5.6\,mJy. The source was still highly circularly polarized, but also showed up to 80\% linear polarization, and then faded rapidly  with a timescale of one day. 
The  rotation measure of the source varied significantly, from $-11.8\pm0.8$\,rad\,m$^{-2}$ to $-64.0\pm1.5$\,rad\,m$^{-2}$, over three days.
No X-ray counterpart was found in follow-up \textit{Swift} or \textit{Chandra} observations about a week after the first MeerKAT detection, with  upper limits of $\sim 5.0\times10^{31}$\,erg\,s$^{-1}$ (0.3--8\,keV, assuming a distance $\sim10$~kpc). No counterpart is seen in new or archival near-infrared observations down to $J=20.8$\,mag.
We discuss possible identifications for \askap\ including  a low-mass star/substellar object with extremely low infrared luminosity, a pulsar with scatter-broadened pulses, a transient magnetar, or a Galactic Center Radio Transient: none of these fully explains the observations, which suggests that \askap\ may represent part of a new class of objects being discovered through radio imaging surveys.

\end{abstract}


\section{Introduction} \label{sec:intro}
Many types of Galactic sources are known to be variable at radio wavelengths, including pulsars, stars and magnetars. 
For example, \citet{1968Sci...162.1481S} detected giant radio pulses from the Crab pulsar, \citet{2007ApJ...663L..25H} found periodic radio bursts from the M9 dwarf \object[TVLM 513-46546]{TVLM~513--46546}, and \citet{2006Natur.442..892C} detected transient pulsed radio emission from the magnetar \object[XTE J1810-197]{XTE~J1810--197}.
Exploring the radio variability can help us better understand extreme astrophysical phenomena and probably find unexpected sources \citep{2015aska.confE..51F}.

The development of large field-of-view radio interferometers, such as the Australian Square Kilometre Array Pathfinder \citep[ASKAP;][]{2021PASA...38....9H}, enables us to investigate  variable and transient phenomena more systematically over a wider parameter space. 
The ASKAP survey for Variables and Slow Transients\footnote{\url{https://vast-survey.org/}} \citep[VAST;][]{2013PASA...30....6M}, is designed to search for such sources. 
The VAST Phase I Pilot Survey \citep[VAST-P1;][]{2021arXiv210806039M} was conducted between 2019 August and 2020 August. The footprint of VAST-P1 consists of six regions including a $\sim$250\,deg$^2$ region covering the Galactic Center (with $\sim356\degr<l<10\degr$, $\vert b\vert < \sim 5\degr$). We used the VAST Transient detection pipeline \citep{2021arXiv210105898P, 2021arXiv210806039M} to search for highly variable radio sources.

Given its high stellar density and ongoing star formation, the Galactic Center (GC) is a promising region for finding variable and transient radio sources \citep[e.g.,][]{2006JPhCS..54..110L}. Aside from transients of known origin like X-ray binaries \citep[e.g.,][]{2005ApJ...633..218B,2020ApJ...905..173Z},
\object[1A 1742-28]{1A~1742--28} \citep{1976Natur.261..476D} and the Galactic Center Transient \citep[\protect{\object[Name GCT]{GCT}};][]{1992Sci...255.1538Z} were the first two radio transients detected, and are only $\sim$arcmin away from the GC. 
Three Galactic Center Radio Transients (GCRTs) were discovered in the 2000s at lower frequencies: \object[GCRT J1746-2757]{GCRT~J1746--2757} \citep{2002AJ....123.1497H}, \object[GCRT J1745-3009]{GCRT~J1745--3009} \citep{2005Natur.434...50H}, and \object[GCRT J1742-3001]{GCRT~J1742--3001} \citep{2009ApJ...696..280H}. 
Unlike A1742--28 and GCT, the GCRTs are about a degree away from the GC, but they are all at low Galactic latitudes ($|b| < 0.6\degr$). Though the radio properties for these three GCRTs are not identical to each other, the spectra of all three GCRTs are very steep and none of them has a clear counterpart at other wavelengths. 
The most well-studied of the three, GCRT~J1745--3009, was detected in at least two different states: it emitted $\sim$1\,Jy bursts every 77 minutes in 2002, and gave off weaker ($\sim$50\,mJy) single bursts in 2003 and 2004.
\citet{2007ApJ...660L.121H} suggest that GCRT~J1745--3009 likely belongs to a new class of coherent emitters, while most radio transients are incoherent synchrotron sources.  And there are yet further candidates in need of confirmation and follow-up \citep[e.g.,][]{2016ApJ...833...11C}.

In this paper we report the discovery of a highly polarized, variable source near the Galactic Center, \askap, detected at 888\,MHz in VAST-P1 observations with  ASKAP, and redetected at 1.29\,GHz with MeerKAT \citep{2016mks..confE...1J,2018ApJ...856..180C}. 
We present the observations, including radio imaging, pulsar searching, X-ray searches, and near-infrared imaging in Section~\ref{sec:obs}, and discuss the possible nature of the source in Section~\ref{sec:discuss}.

\section{Observations, Data Reduction and Results} \label{sec:obs}

\begin{figure*}[hbt!]
\plotone{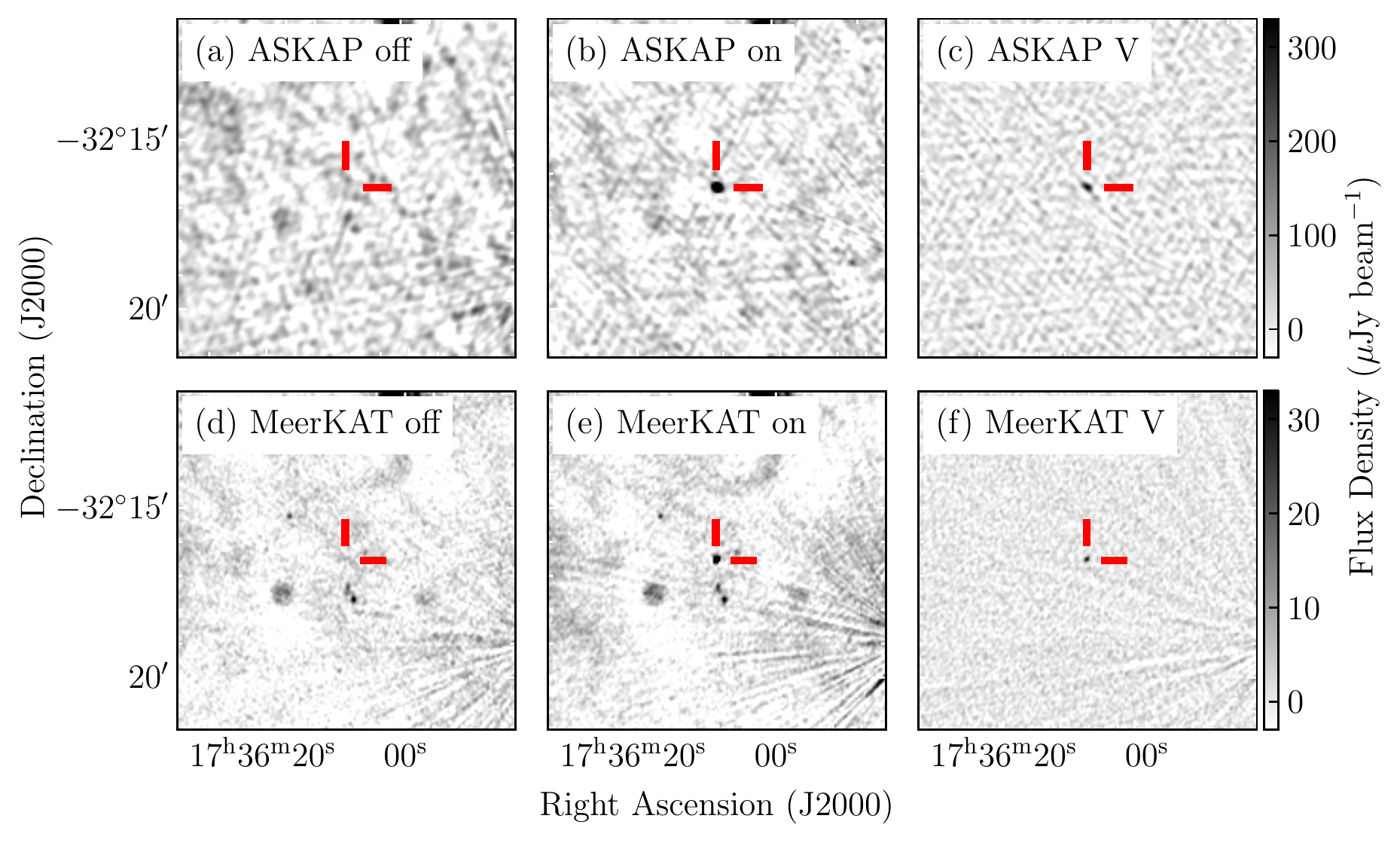}
    \caption{Upper panels: ASKAP images of \askap\ (centered at 888\,MHz). Each image is $10\arcmin$ on a side, with north up and east to the left. We show the ``off'' image observed on 2019~April~28 in panel~(a), the ``on'' image observed on 2020~January~11 in panel~(b) and Stokes V image from 2020~January~11 in panel~(c). The color scales are the same for all of these images. 
    Lower panels: MeerKAT L-band images of \askap. Each image is $10\arcmin$ on a side, with north up and east to the left. We show the ``off'' image observed on 2021~January~19 in panel~(d), the ``on'' image observed on 2021~February~07 in panel~(e) and Stokes V image from 2021~February~07 in panel~(f). The color scales are the same for all of these images.}
    \label{fig:J1736-3216_radio}
\end{figure*}

\subsection{ASKAP Observations}
\askap\ was first discovered as a compact radio source in a transients search of VAST-P1 data (Project Code\dataset[AS107]{http://hdl.handle.net/102.100.100/340959?index=1}) using the VAST transient detection pipeline (Figure~\ref{fig:J1736-3216_radio}). 
It was detected in the adjacent fields 1724$-$31A and 1752$-$31A, observed \change{13} times between 2019~April~28 and 2020~August~29.
The VAST-P1 survey incorporates the Rapid ASKAP Continuum Survey \citep[RACS, Project Code\protect{\dataset[AS110]{http://hdl.handle.net/102.100.100/374842?index=1}};][]{2020PASA...37...48M} as its first epoch.
Both RACS and VAST-P1 were conducted at a central frequency of 888\,MHz with a bandwidth of 288\,MHz and they shared the same tiling footprints. The integration time for RACS was 15\,mins while that for VAST-P1 was 12\,mins, achieving an rms noise of 0.36\,mJy\,beam$^{-1}$ and 0.40\,mJy\,beam$^{-1}$ for regions near the GC, respectively. Details of these survey observations and data reduction are given by  \citet{2020PASA...37...48M} and \citet{2021arXiv210806039M}.

\begin{figure*}[hbt!]
\plotone{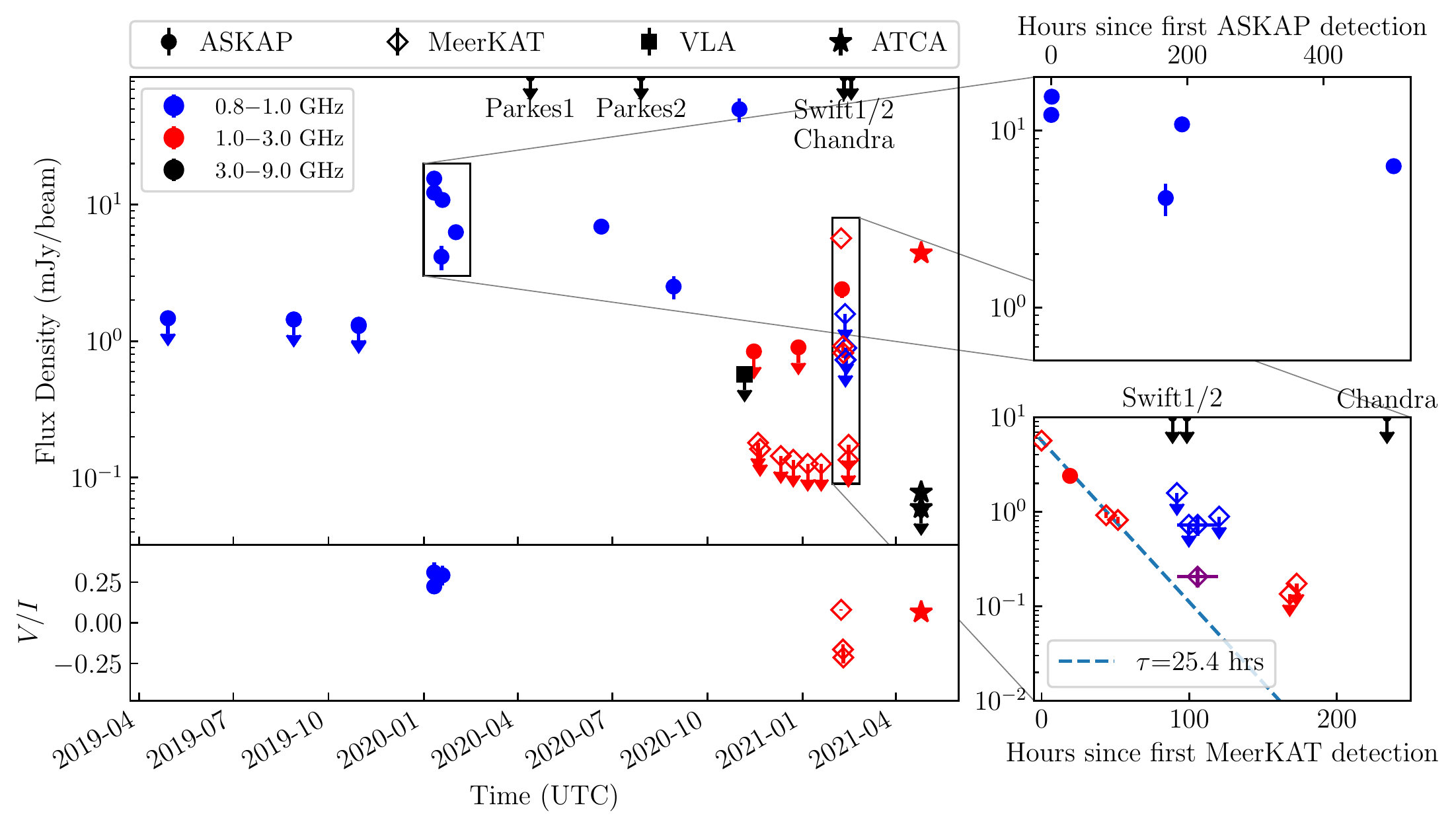}
    \caption{Full radio lightcurve for \askap, including non-detections (times of X-ray observations are also indicated).  The circular polarization fraction $V/I$ is shown in the bottom panel for the detections. 
    In the upper right panel, we show the detections with ASKAP from 2020~January.
    In the lower right panel, we show the observations close to the MeerKAT detections from 2021~February. We fit an exponential decay of the form $S\propto e^{-t/\tau}$ for the four 1.3\,GHz detections (blue dashes line) and find the timescale of decay to be $\sim$26 hours. We scale the UHF-band (800\,MHz) detection to L-band (1.3\,GHz) with the spectral index $\alpha\sim-2.7$ and show the scaled flux density as the \change{purple} diamond. }
    \label{fig:J1736-3216_lc}
\end{figure*}

Figure~\ref{fig:J1736-3216_lc} shows the full radio lightcurve of \askap, as well as the fractional circular polarization. 
Other than variability, \askap\ was highly circularly polarized with a fractional polarization ranging from 20\% to 30\% in VAST-P1 bright detections (see Figure~\ref{fig:J1736-3216_lc}, lower left panel).

There were four additional ASKAP observations that cover our source (Table~\ref{tab:radio_sum}). 
These observations were calibrated using 
\object[PKS B1934-638]{PKS~B1934--638} for both the flux density scale and the instrumental bandpass. 
All observations were processed using standard procedures in the {\sc ASKAPsoft} package \citep{2019ascl.soft12003G}.
We note that there was a $\sim$50\,mJy detection in a 10-hour observation at 943\,MHz on 2020~November~01. However, the systematic error is high due to the source being located near the edge of the beam. 

\begin{figure}[hbt!]
\plotone{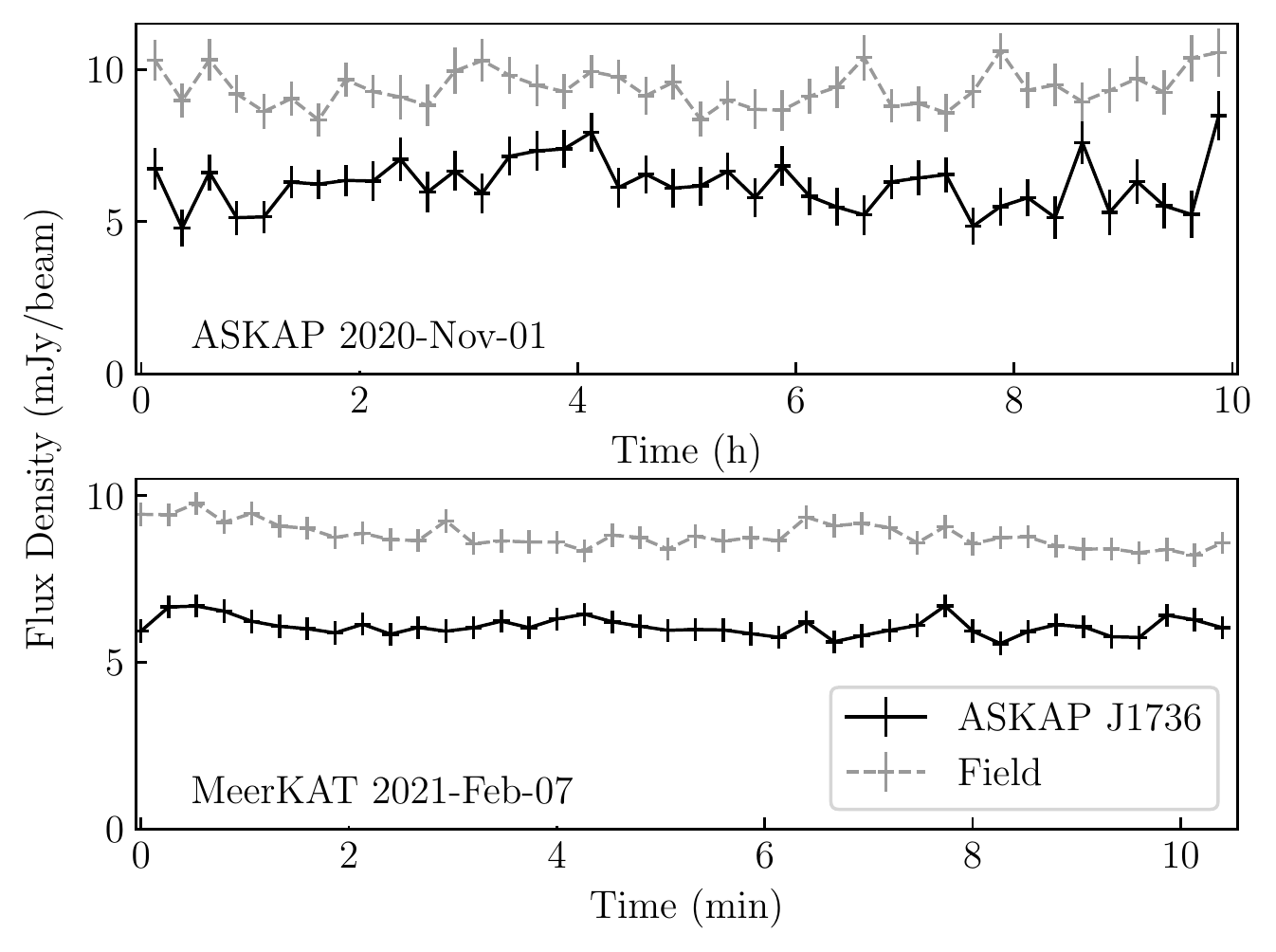}
    \caption{Short timescale lightcurves for \askap. Top: The observation on 2020~November~01 using ASKAP. The integration time for each point is 15\,min. Bottom: The observation on 2021~February~07 using MeerKAT. The integration time for each point is 16 seconds. The black solid lines are lightcurve for \askap\ and the gray dashed lines are lightcurve for a field source \change{(J173548.2$-$310811)} as a comparison.}
    \label{fig:J1736-3216_shortlc}
\end{figure}

To check for any shorter timescale variability we imaged the source using data from the 2020~November~01 ASKAP observation with an integration time of 15\,min (resulting in 40 images in total). This lightcurve showed a relatively low modulation index (standard deviation divided by the mean) of $\sim$13\%, and had a reduced $\chi^2$ \change{ relative to a constant model (a measure of the significance of the variability, see e.g., \citealt{2015A&C....11...25S})} of 1.6 for 39 degrees-of-freedom \change{(40 observations minus one parameter for the mean)}.  Overall  we did not see any evidence for hour-scale variability (Figure~\ref{fig:J1736-3216_shortlc}).

\subsection{Parkes Observations}
Motivated by the possibility that \askap\ is a pulsar, 
we conducted follow-up observations with the 64-m Parkes telescope of \askap\ on 2020~April~20 and 2020~July~29 using the pulsar searching mode with the Ultra-Wideband Low (UWL) receiver \citep{2020PASA...37...12H}, which provides simultaneous frequency coverage from 704 to 4032\,MHz.
Each observation was 30\,mins with 32\,$\mu$s time-resolution and high frequency resolution (1024 channels per 128\,MHz subband). 
We used {\sc Presto} \citep{2001PhDT.......123R} to perform a standard pulsar search.  We found no candidates in a search of dispersion measures (DMs) spanning 0--3000\,pc\,cm$^{-3}$, corresponding to 25\,kpc based on the YMW16 electron-density model \citep[][hereafter YMW16]{2017ApJ...835...29Y} or about two times the highest DM for pulsars discovered to-date \citep[e.g.,][]{2013MNRAS.435L..29S}, period $< \change{25}$\,s and accelerations up to $\sim$20\,m\,s$^{-2}$ (assuming a pulsation period of 1\,ms). 
We also found no single pulse above a SNR  of 8 using the single pulse search procedure for {\sc Presto}. 
However, the lack of simultaneous imaging meant we cannot determine whether the source was radio-loud during these observations. These non-detections (with an upper limit of $\sim$0.05\,mJy, assuming the duty cycle of the pulsar (W/P) to be 10\%) therefore do not rule out the presence of a pulsar.

\subsection{MeerKAT Observations}
To simultaneously search for pulsed and continuum emission from \askap, 
we observed it using the MeerKAT radio telescope with a central frequency of 1.28\,GHz and a two-week cadence starting from 2020~November~19 (project code DDT-20201005-DK-01). Each observation had 12 minutes on the target, achieving an rms noise of 40\,$\mu$Jy\,beam$^{-1}$. Imaging and pulsar searching were performed simultaneously in all MeerKAT observations. 
We used \object[PKS J1830-3602]{PKS~J1830--3602} for bandpass, flux density scale and phase calibration. We reduced the image data using {\sc Oxkat}\footnote{\url{https://github.com/IanHeywood/oxkat}}\citep[v1.0;][]{2020ascl.soft09003H}, where the Common Astronomy Software Applications \citep[CASA;][]{2007ASPC..376..127M} package and {\sc Tricolour}\footnote{\url{https://github.com/ska-sa/tricolour}} were used for measurement sets splitting, cross calibration, self-calibration, flagging and {\sc Wsclean} \citep{2014MNRAS.444..606O} was used for continuum imaging. 

We did not detect any source to a $5\sigma$ limit of 0.04\,mJy in the first five epochs. However, we detected a source in our observation on 2021~February~07 at a flux density of $5.67\pm 0.04$\,mJy but did not detect any pulsations.
The best-fit position of the source is: (J2000) RA $17^{\rm h}36^{\rm m}08\fs19\pm0\fs03$, Dec $-32\degr16\arcmin35\farcs0\pm0\farcs3$ with Galactic coordinates $l,b = (356.08\degr, -0.04\degr)$ based on the MeerKAT detection, where the uncertainties are based on a comparison of the positions of field sources to their RACS matches.
We imaged the source with an integration time of 16\,seconds (resulting in 40 images in total). The lightcurve showed a relatively low modulation index of $\sim$4\% and had a reduced $\chi^2$ of 0.8 for 39 degrees-of-freedom, with no evidence for minute-scale variablity (Figure~\ref{fig:J1736-3216_shortlc}).

\begin{figure}[hbt!]
\plotone{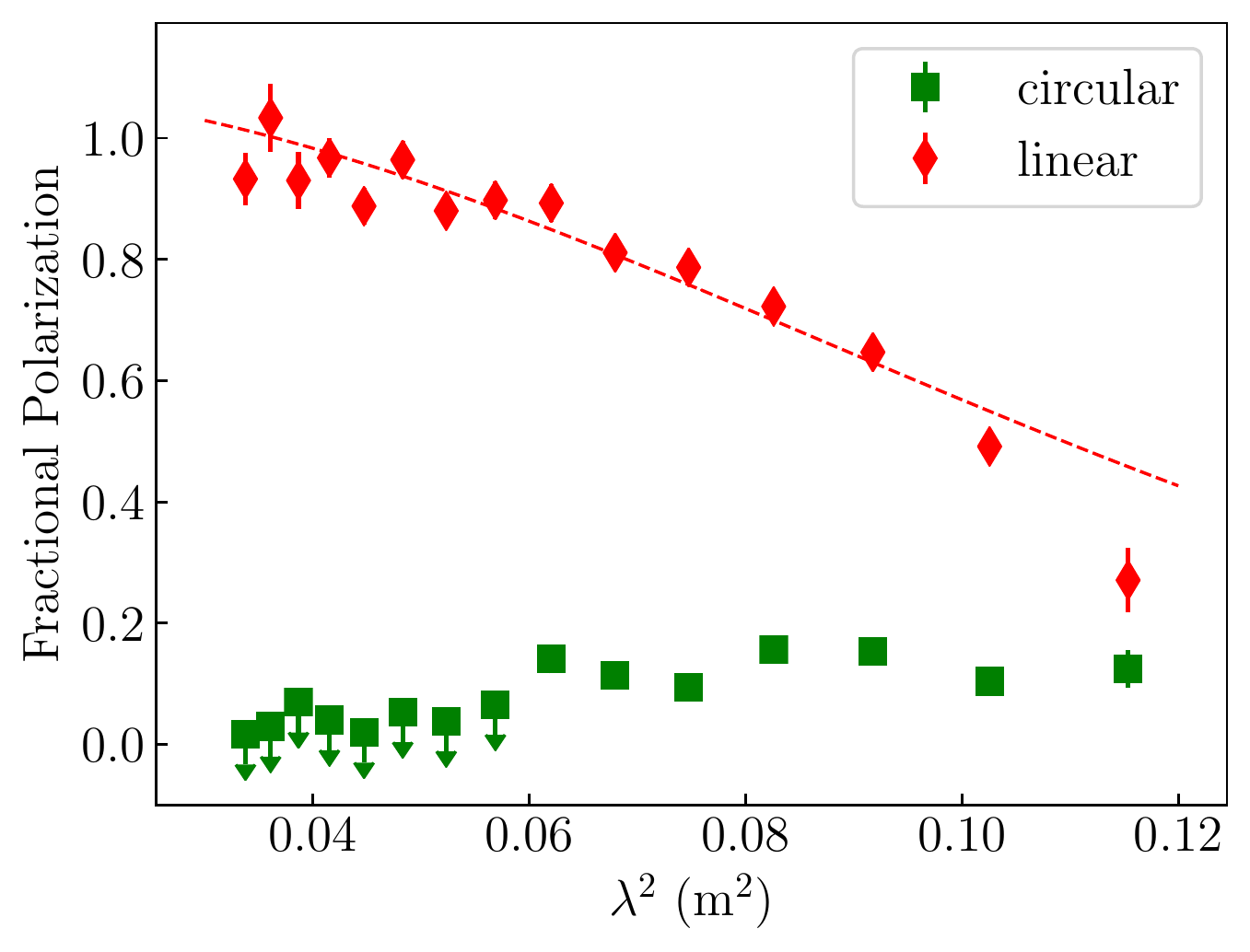}
    \caption{Fractional  polarization as a function of $\lambda^2$ in the MeerKAT L-band observation taken on 2021~February~07. We show the circular polarization as green squares, and linear polarization as red diamonds. We fit a simple depolarization equation $\Pi = \Pi_0\exp(-2\sigma^2\lambda^4)$ to the linear polarization data, which is shown as the red dashed line, where $\sigma=5.7\,{\rm m}^{-2}$ is the RM dispersion of the Faraday screen \citep{2014ApJS..212...15F}. 
    }
    \label{fig:J1736-3216_depol}
\end{figure}

The source was moderately circularly polarized ($V/I$=+8\%) and had a steep radio spectrum within the bandpass ($\alpha=-2.7\pm0.1$\footnote{Subband calibration has not been properly evaluated, and hence we are aware that our estimates may include \change{$\sim 10$\%} calibration error.}, where $S_\nu\propto \nu^{\alpha}$). 
We also found the source to be highly linearly polarized ($|L|/I\sim 80\%$) with a moderately low Faraday rotation measure (RM) of $-11.2\pm0.8$\,rad\,m$^{-2}$. 
The source also exhibited depolarization behavior towards lower frequencies: the fractional total polarization is nearly 100\% at 1.6\,GHz but only $\sim$20\% at 0.9\,GHz (Figure~\ref{fig:J1736-3216_depol}).  We performed further tests to verify the polarization and RM variability, as discussed in Appendix~\ref{sec:pol}.

Further radio observations showed a very rapid decline with an exponential timescale of $\sim 26\,$hrs (Figure~\ref{fig:J1736-3216_lc} inset). Our ASKAP observation 20 hours after the first MeerKAT detection gave a flux density of $2.4\pm0.3$\,mJy at $1.3\,{\rm GHz}$.
Two further MeerKAT observations over the following days demonstrated that the source continued to fade exponentially, while the spectral shape remained similar ($\alpha=-3.4\pm0.3$). We found the source was still highly linearly polarized in these observations, although the RM changed significantly, from $-11.2\pm0.8$\,rad m$^{-2}$ on 2021~February~07 to $-63.3\pm1.5$\,rad m$^{-2}$ on 2021~February~09. 
The ionosphere usually contributes to Faraday rotation of order $\sim 1\,{\rm rad}\,{\rm m}^{-2}$ \citep{2013A&A...552A..58S}, which can potentially cause RM variations between epochs. We used \textsc{IonFR}\footnote{\url{https://github.com/csobey/ionFR}} to model the ionospheric Faraday depth at the dates of the observations.  The ionospheric Faraday rotation is $+0.65\pm0.05\,{\rm rad}\,{\rm m}^{-2}$ and $+0.75\pm0.06\,{\rm rad}\,{\rm m}^{-2}$ on 2021~February~07 and 2021~February~09 respectively. The corrected RM of the source is therefore $-11.8\pm0.8$\,rad m$^{-2}$ and $-64.0\pm1.5$\,rad m$^{-2}$ on these days, after ionospheric RM corrections. 
The intrinsic polarization angle was consistent between the epochs (see justifications in Appendix \ref{sec:pol}).

We also obtained three 12-min observations in the Ultra high frequency band (UHF; 544--1088\,MHz) with MeerKAT, about one hundred hours after the first MeerKAT detection. There was no detection in these single observations, but there was a $\sim$5$\sigma$ detection when all three were summed coherently (see blue diamonds in Figure~\ref{fig:J1736-3216_lc}). This UHF-band detection is a factor of two higher than what we expected from the exponential decay (we corrected the UHF-band detection to 1.3\,GHz assuming a spectral index of $\alpha=-2.7$), 
suggesting that the spectrum may have steepened to $\alpha<-4$ or that the decay slowed.

During imaging observations with MeerKAT, the FBFUSE \citep[Filterbanking Beamformer User Supplied Equipment;][]{barr_2017} instrument was used to produce high-time-resolution Stokes-I beams to enable pulsar and fast-transient searching. At both L- and UHF-band, FBFUSE was configured to produce a tiling pattern of 7 coherent beams with the central beam positioned at (J2000) RA $17^{\rm h}36^{\rm m}08\fs20$, Dec $-32\degr16\arcmin33\farcs0$. The beams were arranged in a close-packed hexagonal grid with an overlap at their 70\% power points (see W.~Chen et al., submitted, for detail of FBFUSE beam tiling). At L-band FBFUSE produced 4096-channel data covering the 856\,MHz band with a time resolution of 76.56\,$\mu$s. At UHF-band the instrument produced 4096-channel data covering the 544\,MHz band with a time resolution of 120.47\,$\mu$s.

Data streams from FBFUSE were recorded to disk on the APSUSE \citep[Accelerated Pulsar Search User Supplied Equipment;][]{barr_2017} cluster. The data were dedispersed to dispersion measures in the range 0--\change{2000}\,pc\,cm$^{-3}$ at UHF-band and 0--3000\,pc\,cm$^{-3}$ at L-band, \change{with the different maximum DMs chosen to have roughly constant scattering timescales between the two bands}. The resultant trials were searched for periodicities up to 10\,s using the GPU-accelerated \textsc{Peasoup}\footnote{\url{https://github.com/ewanbarr/peasoup}} software with the resultant candidates folded modulo the detected periodicities using \textsc{PulsarX}\footnote{\url{https://github.com/ypmen/PulsarX.git}}. To retain sensitivity to binary systems, the data were time-domain resampled \citep{1991ApJ...368..504J} to constant acceleration values between $-150$ and 150\,m~s$^{-2}$ before searching.
Folded candidate signals were inspected by eye. No significant pulsed emission was detected above a signal-to-noise threshold of 9.

The MeerTRAP real-time single-pulse pipeline running on the TUSE instrument (Transients User Supplied Equipment; Stappers et al.\ in prep) was run in parallel with all of the MeerKAT observations. It operated on the same central beam that the pulsar search described above with a time resolution of $306.24\,\mu \textrm{s}$ for the L-band observations and $361.4\,\mu \textrm{s}$ for the UHF observations. Single pulses that are greater than a S/N limit of 8, were searched for over dispersion measures from 23--5000\,pc\,cm$^{-3}$ in L-band and 23--1500\,pc\,cm$^{-3}$ in the UHF over a range of widths from the time resolution up to 196\,ms and 231\,ms for the two frequencies respectively. No astrophysical pulses were detected above the S/N threshold.

\subsection{ATCA observations}

After our ASKAP and MeerKAT monitoring observations ended, we observed \askap\ with the Australia Telescope Compact Array (ATCA) in three bands (centered at 2.1\,GHz, 5.5\,GHz, 9.0\,GHz) for 80 mins each on 2021 April 25 (project code: C3431). 
The observation was calibrated using \object[PKS B1934-638]{PKS~B1934$-$638} for the flux density scale and the instrumental bandpass. \object[PMN J1733-3722]{PMN~J1733$-$3722} was used for phase calibration. 

\begin{figure}[hbt!]
\plotone{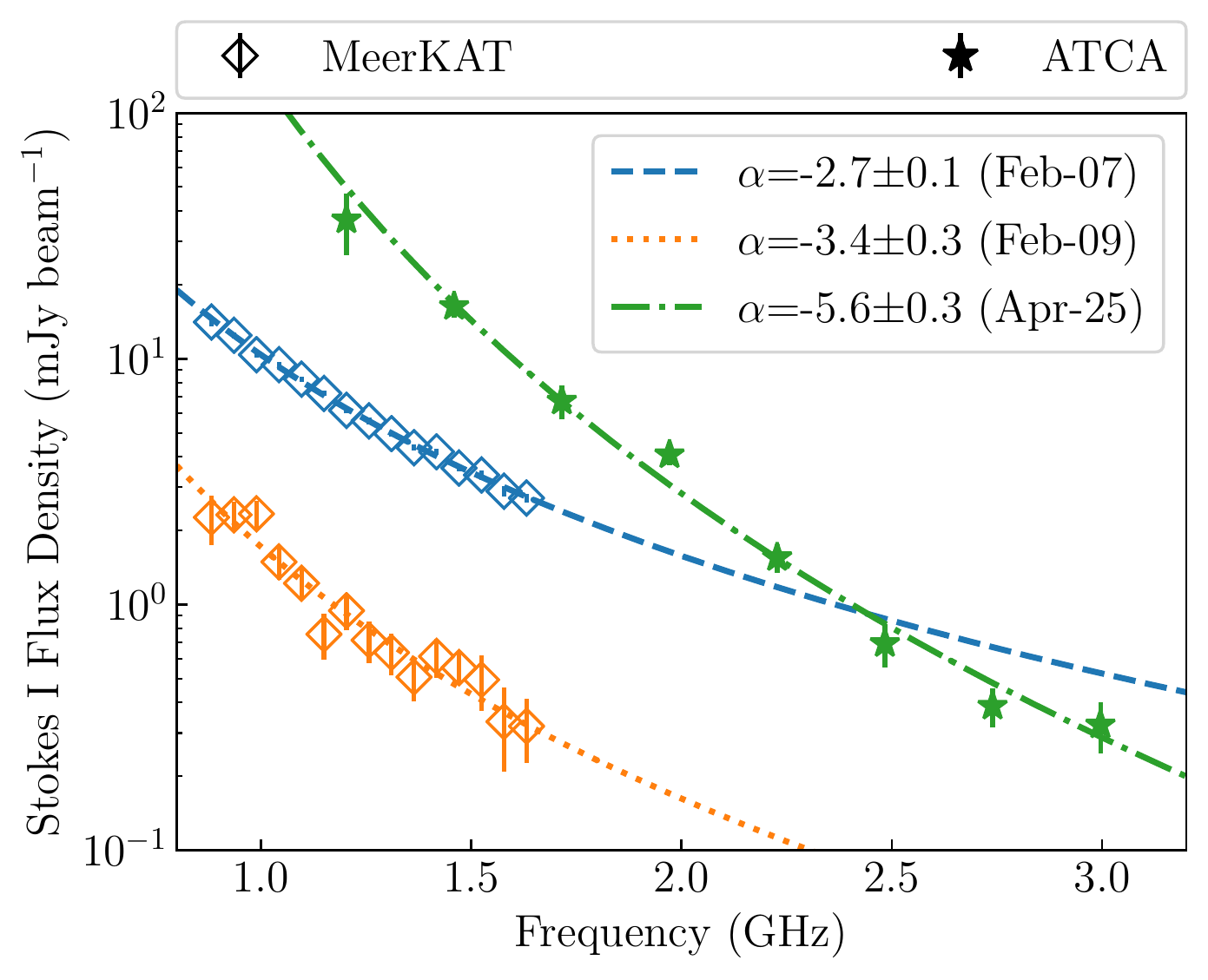}
    \caption{Stokes I spectral energy distributions for \askap\ from two observations with MeerKAT (0.9--1.6\,GHz; diamonds) and one with ATCA (1.4--3.0\,GHz; stars). We fit a power law relation for each observation and show the spectral index and the date of the observation in the legend.}
    \label{fig:J1736-3216_SPEIDX}
\end{figure}

We used {\sc Miriad} \citep{1995ASPC...77..433S} to perform the data calibration and {\sc Casa} to perform the continuum imaging. 
We detected a source with a flux density of $4.41 \pm 0.14$\,mJy  at 2.1\,GHz. 
We did not find any detection at 5.5\,GHz or 9.0\,GHz, which places 3$\sigma$ upper limits of 78\,$\mu{\rm Jy\,beam}^{-1}$ and 60\,$\mu{\rm Jy\,beam}^{-1}$ at 5.5\,GHz and 9.0\,GHz respectively. 
The non-detection at higher frequency (5.5\,GHz) constrains the spectral index to be $\alpha < -4.2$.
We measured the spectral index to be $\alpha = -5.6 \pm 0.1$ across the L-band (2.1\,GHz) bandpass (Figure~\ref{fig:J1736-3216_SPEIDX}), which is consistent with the constraints from the non-detection at 5.5\,GHz.
The source was moderately circularly polarized, with $V/I \sim +6\%$, which is consistent with the MeerKAT observation that fractional circular polarization is lower at higher frequencies (Figure~\ref{fig:J1736-3216_depol}). 

\subsection{X-ray Observations and Analysis}

We identified archival observations covering \askap\ with the \textit{Neil Gehrels Swift Observatory} (\textit{Swift}; \citealt{2004ApJ...611.1005G}), restricting observations to those using the X-ray Telescope (XRT; \citealt{2005SSRv..120..165B}) in the photon counting mode.  We used 4 observations between 2012~February~01 and 2012~September~09, with a summed exposure time of 2.3\,ks.  There was no source within $15\arcsec$ of \askap, and we determine an 95\% count-rate upper limit of $8.4\times 10^{-4}\,{\rm s}^{-1}$ (over the default energy range of 0.2--10\,keV).

Following the MeerKAT detections of \askap, we were awarded Director's Discretionary Time observations with \textit{Swift} (observation IDs \dataset[00014071001]{https://heasarc.gsfc.nasa.gov/FTP/swift/data/obs/2021_02/00014071001/} and \dataset[00014071002]{https://heasarc.gsfc.nasa.gov/FTP/swift/data/obs/2021_02/00014071002/}).  We obtained 1.7\,ks on 2021~February~10.95 and another 0.8\,ks on 2021~February~11.35. 
There was 1 count within $15\arcsec$ of \askap, but this is consistent with the background (mean expectation with $15\arcsec$ of 0.3\,counts).  So we set an upper limit of $1.2\times 10^{-3}\,{\rm s}^{-1}$ (0.2--10\,keV).
We estimated the upper limit of \ion{H}{1} column density for the position of our source based on the \ion{H}{1} 4$\pi$ survey \citep{2016A&A...594A.116H} using the HEASARC web-based PIMMS to be $1.59\times10^{22}$\,cm$^{-2}$ (through the entire Galaxy).
Assuming a power-law photon index of $\Gamma=2.0$ \citep{hyman2021extreme}, the non-detection in \textit{Swift} observations yields an upper limit on the unabsorbed flux (0.3--8\,keV) of $2.0\times10^{-13}$\,erg\,cm$^{-2}$\,s$^{-1}$. 
The upper limit for the X-ray luminosity at a distance of $d$ is $\sim2.4\times10^{33} (d/10\,{\rm kpc})^2$\,erg\,s$^{-2}$.

Finally, we were awarded Director's Discretionary Time with the \textit{Chandra X-ray Observatory}.  We used the back-illuminated ACIS-S3 detector with the thin filter, and the 1/8 subarray to maintain sub-second temporal resolution.   \askap\ was observed on 2021~February~17.61 for 25.1\,ks (observation ID \dataset[24966]{https://cxc.cfa.harvard.edu/cdaftp/byobsid/6/24966/}).  We filtered the data to 0.3--10\,keV.  There are 0 events within $1\arcsec$, and based on the observed background rate we set a 95\% upper limit of $1.0\times 10^{-4}\,{\rm s}^{-1}$.
Likewise, we estimate the upper limit of the X-ray luminosity (0.3--8\,keV) based on the \textit{Chandra} non-detection to be $\sim 5.0\times10^{31}(d/10\,{\rm kpc})^2$\,erg\,s$^{-1}$.

\subsection{Near-Infrared Data}
We searched for near-IR counterparts in the VISTA Variable in the Via Lactea Survey \citep[VVV,][]{2010NewA...15..433M}.
There is no counterpart visible in the VVV DR2 catalog. We find $3\sigma$ upper limits of $J>19.25$, $H>17.65$\,mag, and $K_{\rm s}>16.70$\,mag from VVV within a $2\farcs5$ radius (corresponding to a 5$\sigma$ positional error).

We observed the source using Gemini Flamingos-2 in $J$-band ($1.2\,\mu$m) for 40 mins on 2021~April~28 and 2021~April~29, and in $K_s$-band ($2.15\,\mu$m) for 18.5 mins on 2021~May~24 (project code GS-2021A-FT-210). We used Gemini {\sc Dragons} \citep{2019ASPC..523..321L} to reduce the data and {\sc Sextractor} \citep{1996A&AS..117..393B} to perform the photometry.

We used the VVV catalog as both astrometric and photometric references to correct the Gemini data. 
For astrometry, we used 340 sources that we identified as not blended or badly saturated for $J$-band and 96 sources for $K_s$-band. The  uncertainty is $\approx 0.15\arcsec$ in each coordinate. 
For photometry, we used fewer sources to avoid sources that showed signs of saturation or non-linearity. We used 270 sources in $J$-band and 90 sources in $K_s$-band. We estimated zero-point uncertainties of 0.02 mag for $J$-band and 0.04 mag for $K_s$-band.
The seeing of  both observations was $\sim0\farcs7$. 

\begin{figure}[hbt!]
\plotone{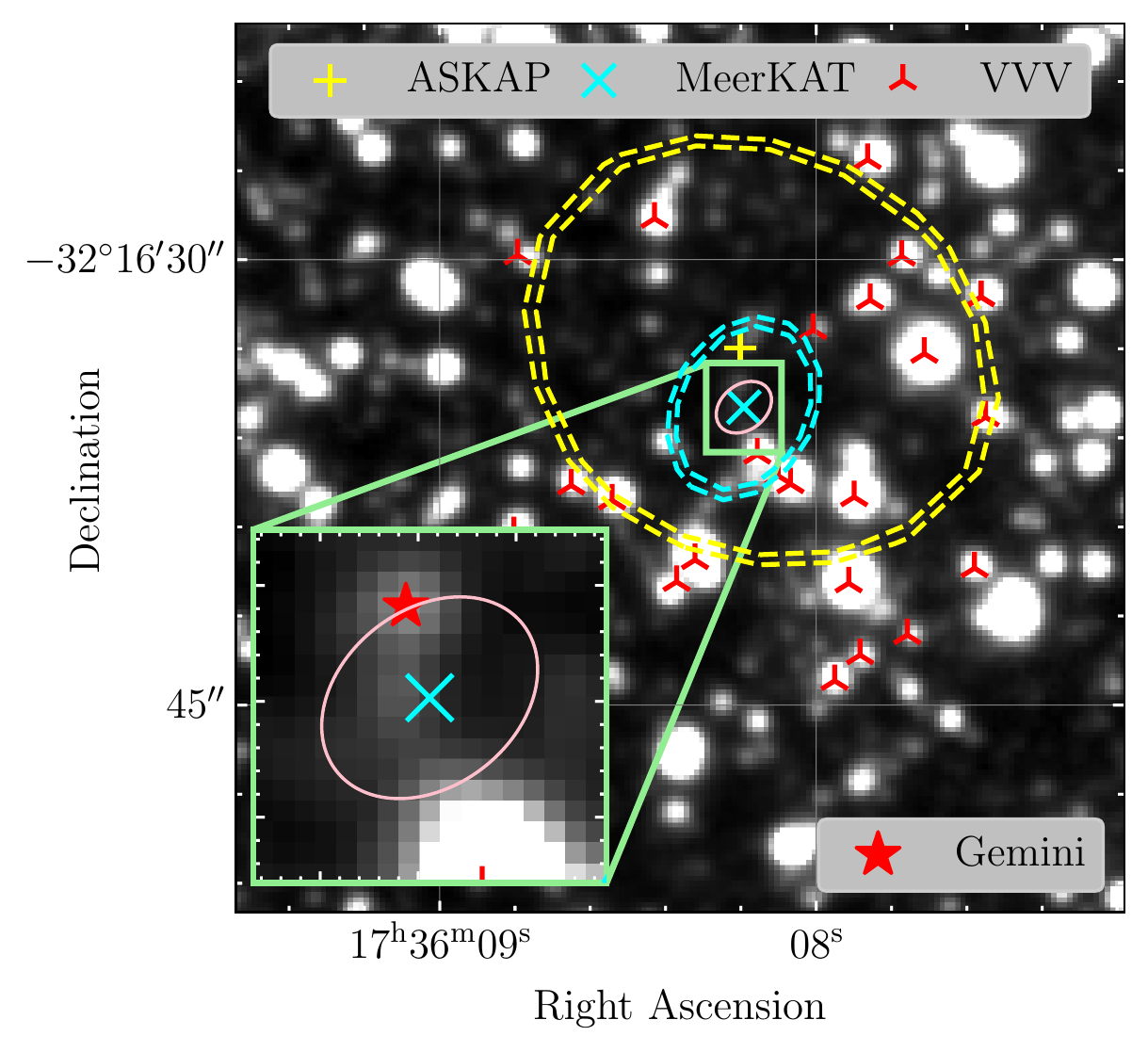}
    \caption{The Gemini $J$-band (1.2\,$\mu$m) image ($30\arcsec$ on a side, $\sim$2 times the ASKAP synthesized beam) of \askap. 
    The yellow contours show the ASKAP detection, while the cyan contours show the MeerKAT detection. 
    The best-fit positions from ASKAP and MeerKAT  are shown as yellow $+$ and cyan $\times$, respectively. 
    Inverted Y's show the sources from the VVV catalog \citep{2010NewA...15..433M}. 
    The small pink contour is the best astrometry constraint from MeerKAT (at 5$\sigma$ confidence level).
    We show one well-detected source from Gemini observations that is within $2\farcs5$ of the radio position as the red star in the inset; there is a fainter source just to the south of that, but it is consistent with our upper limits.
    }
    \label{fig:J1736-3216_gemini}
\end{figure}

There is a faint source within $2\farcs5$ of the radio position with $J = 20.8\pm0.2\,{\rm mag}$ and ${K_s} = 17.6\pm0.1\,{\rm mag}$.
This infrared source is just within the $5\sigma$ error circle of the radio position (Figure~\ref{fig:J1736-3216_gemini}), therefore we consider it unlikely to be associated with the radio source, but we examine this in more detail in Section~\ref{sec:stars}.  Finally, just to the south of that source is a fainter source visible in both $J$ and $K_s$ bands but with magnitudes at or fainter than our 3$\sigma$ limit.  We are unable to measure its properties reliably, but given the density of such sources in the image we do not believe the association to be significant.

\subsection{Archival Radio Data}
This source was not detected in previous radio surveys including the  quick look images from the Karl G.\ Jansky Very Large Array Sky Survey \citep[VLASS;][]{2020PASP..132c5001L}, the TIFR GMRT Sky Survey \citep[TGSS;][]{2017A&A...598A..78I},  the GaLactic and Extragalactic All-sky MWA \citep[GLEAM;][]{2015PASA...32...25W, 2017MNRAS.464.1146H}, the NRAO VLA Sky Survey \citep[NVSS;][]{1998AJ....115.1693C} and the second epoch Molonglo Galactic Plane Survey \citep[MGPS-2;][]{2007MNRAS.382..382M}. These limits are included in Table~\ref{tab:radio_sum}. 
We have also searched for any archival VLA and ATCA data but did not find any other observation that covers our source.

\begin{deluxetable*}{cccccccc}
\tablecaption{Radio observation summary for \askap. Non-detections are denoted by 3$\sigma$ upper-limits based on the local noise. }
\label{tab:radio_sum}
\tablecolumns{8}
\tablehead{
\colhead{Telescope} & 
\colhead{Mode} & 
\colhead{Start} & 
\colhead{Duration} & 
\colhead{Frequency Range} & 
\colhead{$S_{\rm Stokes I}$} & 
\colhead{$S_{\rm Stokes V}$} & 
\colhead{Survey} \\ 
\colhead{} & 
\colhead{} & 
\colhead{(UT)} & 
\colhead{(h)} & 
\colhead{(MHz)} & 
\colhead{(mJy/beam)} & 
\colhead{(mJy/beam)} & 
\colhead{} 
} 
\startdata
VLA & Imaging & 1996~May--1996~Jun & \nodata & 1363--1447 & $< 1.71$ & \nodata & NVSS\\ 
Molonglo & Imaging & 1997~Jul--2007~May & $\sim 11$ & 841.5--844.5 & $< 15.6$ & \nodata & MGPS-2\\ 
MWA & Imaging & 2014~Jun & \nodata & 170--231 & $< 1800$ & \nodata & GLEAM\\ 
GMRT & Imaging & 2016~Mar & \nodata & 140--156 & $< 27.0$ & \nodata & TGSS\\ 
EVLA & Imaging & 2018~Feb~11 15:47 & \nodata & 2000--4000 & $< 0.36$ & \nodata & VLASS\\ 
ASKAP & Imaging & 2019~Apr~28 20:47 & 0.25 & 744--1032 & $< 1.47$ & \nodata & RACS\\ 
ASKAP & Imaging & 2019~Apr~28 21:03 & 0.25 & 744--1032 & $< 1.47$ & \nodata & RACS\\ 
ASKAP & Imaging & 2019~Aug~28 10:01 & 0.2 & 744--1032 & $< 1.44$ & \nodata & VAST-P1\\ 
ASKAP & Imaging & 2019~Aug~28 11:04 & 0.2 & 744--1032 & $< 1.44$ & \nodata & VAST-P1\\ 
ASKAP & Imaging & 2019~Oct~30 04:02 & 0.2 & 744--1032 & $< 1.32$ & \nodata & VAST-P1\\ 
ASKAP & Imaging & 2019~Oct~30 04:14 & 0.2 & 744--1032 & $< 1.39$ & \nodata & VAST-P1\\ 
ASKAP & Imaging & 2020~Jan~11 02:14 & 0.2 & 744--1032 & $12.24 \pm 0.89$ & $3.80 \pm 0.77$ & VAST-P1\\ 
ASKAP & Imaging & 2020~Jan~11 02:56 & 0.2 & 744--1032 & $15.51 \pm 0.48$ & $3.48 \pm 0.32$ & VAST-P1\\ 
ASKAP & Imaging & 2020~Jan~18 02:56 & 0.2 & 744--1032 & $4.15 \pm 0.86$ & $< 2.19$ & VAST-P1\\ 
ASKAP & Imaging & 2020~Jan~19 02:27 & 0.2 & 744--1032 & $10.82 \pm 0.85$ & $3.17 \pm 0.66$ & VAST-P1\\ 
ASKAP & Imaging & 2020~Feb~01 01:31 & 0.2 & 744--1032 & $6.28 \pm 0.47$ & $< 1.14$ & VAST-P1\\ 
Parkes & Pulsar & 2020~Apr~12 20:50 & 0.5 & 704--4032 & $< 0.05$\tablenotemark{a} & \nodata & \\ 
ASKAP & Imaging & 2020~Jun~20 14:17 & 0.25 & 744--1032 & $6.90 \pm 0.35$ & $< 0.99$ & VAST-P1\\
Parkes & Pulsar & 2020~Jul~29 12:29 & 0.5 & 704--4032 & $< 0.05$\tablenotemark{a} & \nodata & \\ 
ASKAP & Imaging & 2020~Aug~29 09:46 & 0.2 & 744--1032 & $2.51 \pm 0.49$ & $< 1.11$ & VAST-P1\\ 
ASKAP & Imaging & 2020~Nov~01 02:16 & 10 & 799--1087 & $\sim 50 \pm 10$\tablenotemark{b} & \nodata & \\ 
EVLA & Imaging & 2020~Nov~06 20:42 & 0.5 & 2000--4000 & $< 0.57$ & \nodata & VLASS\\ 
ASKAP & Imaging & 2020~Nov~15 06:10 & 0.25 & 1295--1439 & $< 0.18$ & \nodata &  RACS-mid\\ 
MeerKAT & Imaging\&Pulsar & 2020~Nov~19 14:45 & 0.2 & 856--1712 & $< 0.18$ & \nodata & \\ 
MeerKAT & Imaging\&Pulsar & 2020~Nov~21 16:04 & 0.2 & 856--1712 & $< 0.16$ & \nodata & \\ 
MeerKAT & Imaging\&Pulsar & 2020~Dec~11 12:15 & 0.2 & 856--1712 & $< 0.14$ & \nodata & \\ 
MeerKAT & Imaging\&Pulsar & 2020~Dec~23 09:09 & 0.2 & 856--1712 & $< 0.14$ & \nodata & \\ 
ASKAP & Imaging & 2020~Dec~28 04:09 & 0.25 & 1295--1439 & $< 0.90$ & \nodata &  RACS-mid\\ 
MeerKAT & Imaging\&Pulsar & 2021~Jan~06 10:05 & 0.2 & 856--1712 & $< 0.12$ & \nodata &  \\ 
MeerKAT & Imaging\&Pulsar & 2021~Jan~19 09:13 & 0.2 & 856--1712 & $< 0.12$ & \nodata &  \\ 
MeerKAT & Imaging\&Pulsar & 2021~Feb~07 06:09 & 0.2 & 856--1712 & $5.67 \pm 0.04$\tablenotemark{c} & $0.46 \pm 0.03$ & \\ 
ASKAP & Imaging & 2021~Feb~08 01:23 & 0.25 & 1295-1439 & $2.40 \pm 0.33$ & \nodata &  RACS-mid\\ 
MeerKAT & Imaging\&Pulsar & 2021~Feb~09 01:55 & 0.2 & 856--1712 & $0.92 \pm 0.06$\tablenotemark{d} & $-0.15 \pm 0.03$ &  \\ 
MeerKAT & Imaging\&Pulsar & 2021~Feb~09 09:59 & 0.2 & 856--1712 & $0.82 \pm 0.07$\tablenotemark{d} & $-0.18 \pm 0.03$ &  \\ 
MeerKAT & Imaging\&Pulsar & 2021~Feb~11 01:45 & 0.2 & 544--1088 & $< 1.56$\tablenotemark{e} & \nodata &  \\ 
MeerKAT & Imaging\&Pulsar & 2021~Feb~11 09:50 & 0.2 & 544--1088 & $< 0.73$\tablenotemark{e} & \nodata &  \\ 
MeerKAT & Imaging\&Pulsar & 2021~Feb~12 06:19 & 0.2 & 544--1088 & $< 0.89$\tablenotemark{e} & \nodata &  \\ 
MeerKAT & Imaging\&Pulsar & 2021~Feb~14 06:09 & 0.2 & 856--1712 & $< 0.13$ & \nodata &  \\ 
MeerKAT & Imaging\&Pulsar & 2021~Feb~14 10:49 & 0.2 & 856--1712 & $< 0.18$ & \nodata &  \\ 
ATCA & Imaging & 2021~Apr~25 13:37 & 1.3 & 4500--6500 & $< 0.078$ & \nodata & \\ 
ATCA & Imaging & 2021~Apr~25 13:37 & 1.3 & 8000--10000 & $< 0.060$ & \nodata & \\ 
ATCA & Imaging & 2021~Apr~25 14:03 & 1.3 & 1100--3100 & $4.41 \pm 0.14$\tablenotemark{f} & $0.29 \pm 0.05$ & \\ 
\enddata 
\tablenotetext{a}{The upper limit is derivated from Equation~\ref{eq:pulsar_snr}. We assumed the duty cycle of the pulsar (W/P) to be 10\%.}

\tablenotetext{b}{The location of the source is close to the edge of the primary beam. The systematic error can be as high as $\sim 10\,{\rm mJy}$.}

\tablenotetext{c}{The spectral index across the bandpass is $\alpha = -2.7\pm 0.1$. RM is $-11.8 \pm 0.8$\,rad\,m$^{-2}$ after ionospheric RM correction.}

\tablenotemark{d}{The spectral index across the bandpass is $\alpha = -3.4\pm 0.3$. RM is $-64.0 \pm 1.5$\,rad\,m$^{-2}$ after ionospheric RM correction.}

\tablenotemark{e}{We combined these three UHF observations and got a detection with flux density of $0.73 \pm 0.17$\,mJy\,beam$^{-1}$}

\tablenotetext{f}{The spectral index across the bandpass is $\alpha = -5.6\pm 0.3$. }
\end{deluxetable*}

\section{Discussion} \label{sec:discuss}
We can summarize the most important characteristics of \askap\ before we discuss interpretations:
\begin{itemize}
    \item Factor of $>100$ variability over a timescale of a week at 900\,MHz with a peak flux density of $\sim 10\,$mJy.
    \item Persistent emission for a few weeks, but can decline as fast as 1\,day.
    \item High degree of circular polarization and steep radio spectrum.  
    \item High degree of linear polarization with a small RM, and depolarization toward the lower frequencies. RM changes significantly across the observations within three days.
    \item No radio pulsations (searching the DM from 0--3000\,{\rm pc}\,{\rm cm}$^{-3}$ and exploring the acceleration up to 150\,{\rm m}\,{\rm s}$^{-1}$).
    \item No counterpart at near-infrared (down to $J = 20.8\,{\rm mag}$ and $K_s = 17.6\,{\rm mag}$) or X-ray wavelengths (with upper limits of $\sim 5.0\times10^{31}\,{\rm erg}\,{\rm s}^{-1}$).
\end{itemize}

\begin{figure}[hbt!]
\plotone{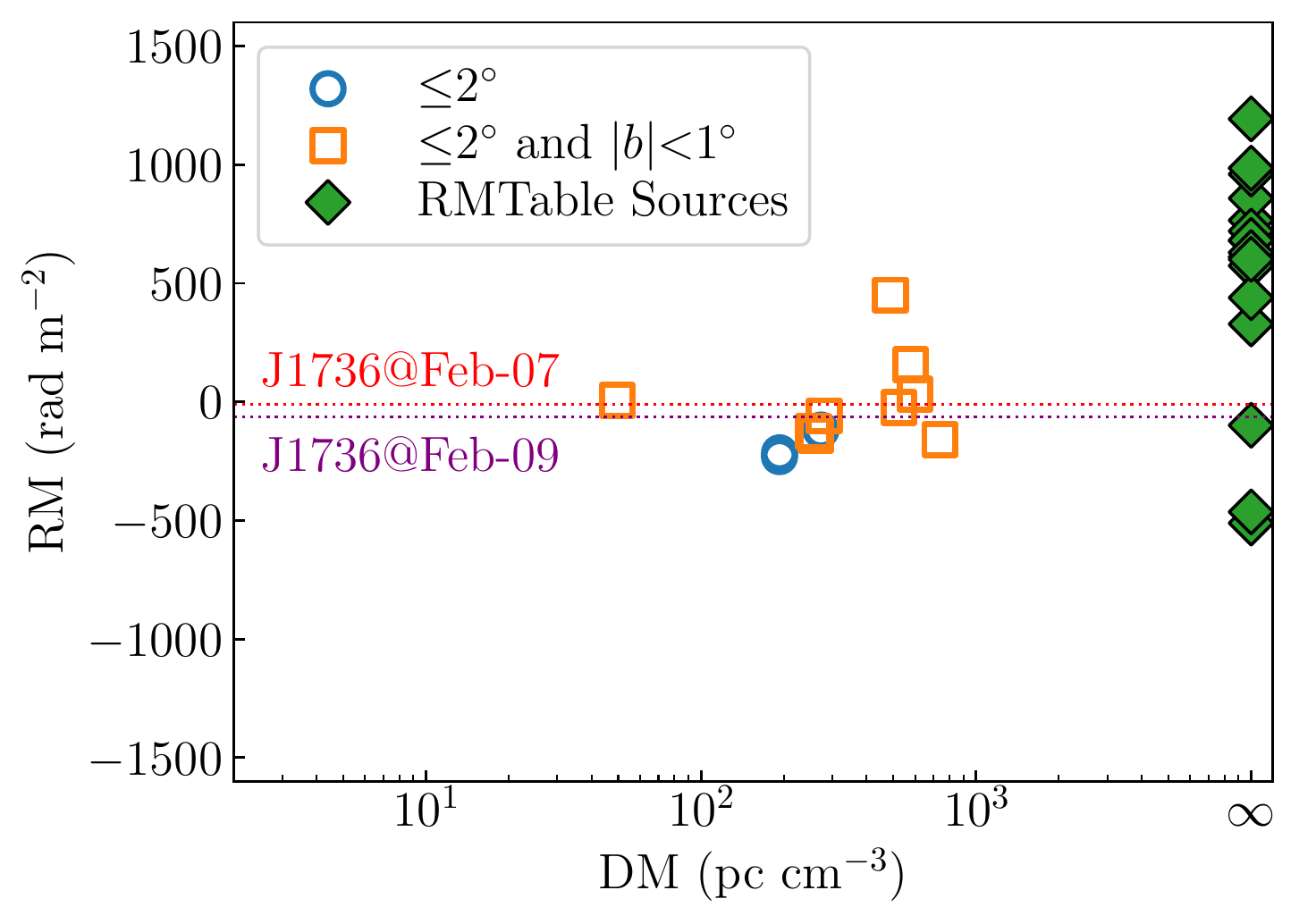}
    \caption{Dispersion measure (DM) versus rotation measure (RM)  for pulsars within 2$\degr$ of the \askap\ from the ATNF pulsar catalog \citep{2005AJ....129.1993M}. 
    Orange squares show pulsars near the Galactic Plane ($\vert b\vert<$ 1$\degr$) and blue circles show pulsars at higher latitudes.
    We also show sources within 2$\degr$ of the source from the RMTable catalog (Van Eck et al., \textit{in prep.}) as green diamonds (we plot them with DM $= \infty$ as most of them are extragalactic).
    The red and purple dashed lines show the RMs from our observations on 2021~February~07 and 2021~February~09 respectively.
    }
    \label{fig:J1736-3216_rmdm}
\end{figure}

Based on its low RM, \askap\ may be a Galactic source. We show the pulsars with known RM and DM within $2\degr$ of the source from the ATNF pulsar catalog \citep{2005AJ....129.1993M}\footnote{\url{http://www.atnf.csiro.au/research/pulsar/psrcat}} and extragalactic sources with known RM within $2\degr$ from RMTable\footnote{\url{https://github.com/CIRADA-Tools/RMTable}} (v0.1.8, Van Eck et al., \textit{in prep.}) in Figure~\ref{fig:J1736-3216_rmdm}. The absolute values of the RMs for almost all nearby sources are much higher than that for our source. 
Furthermore, according to \citet{2021arXiv210201709H}, the RM towards the direction of the source is $\sim$ +450\,rad\,m$^{-2}$, mainly contributed by the Milky Way. If we assume the source is extragalactic, a low RM for our source would require a large $\sim -450$\,rad\,m$^{-2}$ additional contribution to cancel the Galactic RM. 

The shortest rise and decay timescales we can constrain for our source are $\tau\sim 1$\,day\change{, based on the factor of $\sim 2$ rise between 2020~January~18 and 2020~January~19, and the $\sim$day-long decay following the MeerKAT detection on 2021~February~7, although the rise in particular is only weakly constrained} . If we assume that the emitting region is less than $c\tau$ in size, then the brightness temperature of our source is $T_{\rm B}\sim 10^{12}\,{\rm K}(d/1\,{\rm Mpc})^2$. 
The low RM for our source suggests that it is nearby, with $d\lesssim10$\,kpc. If there is not any shorter timescale variability, we can constrain that $T_{\rm B} \lesssim 10^{8}$\,K, which is far lower than the limit for coherent emission, $\sim10^{12}$\,K \citep{1994ApJ...426...51R}. However, this limit can not help us discriminate between coherent and incoherent source, as some coherent emission can have brightness temperature well below $10^{12}\,$K \citep[e.g., type II, III solar bursts,][]{2014RAA....14..773R}.  Even so, the high degree of circular polarization suggests some coherent process such as electron cyclotron maser emission may be operating \citep[e.g.,][ and see below]{1985ARA&A..23..169D,2021MNRAS.502.5438P}.

Significant changes in rotation measure as seen for \askap\ are rare. 
Sources with short timescale RM variations are usually extragalactic, such as AGNs with extreme environments \citep[e.g.,][]{2003ApJ...589..126Z, 2017MNRAS.469.1612L, 2019MNRAS.485.3600A}, and some FRBs \citep[FRB~121102,][]{2021ApJ...908L..10H}. 
RM variations for Galactic sources are usually slow and small \citep[e.g.,][]{2011Ap&SS.335..485Y, 2021arXiv210405723W} except the Galactic Center magnetar \object[PSR J1745-2900]{PSR~J1745--2900}: \citet{2018ApJ...852L..12D} found large changes in observed RM for PSR~J1745--2900 by up to $3500\,{\rm rad}\,{\rm m}^{-2}$ over four years. Even more interestingly, they found that the RM for PSR~J1745--2900 changed by about 7.4\,rad\,m$^{-2}$ per day in 2017. The RM variations is thought to come from a minimum scale of magneto-ionic fluctuations in the scattering screen. 

As we see no change in the intrinsic polarization angle for our source (Appendix~\ref{sec:pol}), we infer that the RM variation for \askap\ is not intrinsic to the source but is probably external, related to a change in the intervening interstellar medium (ISM). With only two RM measurements, it is hard to put a strong constraint on the property of the ISM along the line of sight. However, given the observational features of \askap, we can still describe the medium as well as the source more broadly.

Based on the typical magnetic field values for interstellar medium \citep{2001RvMP...73.1031F,2017ARA&A..55..111H}, the length scale of the Faraday region to give the change in RM is $l_{\rm RM}\sim 250\,{\rm pc}\,(B/2\,\mu{\rm G})^{-1}\,(n_e/10^{-1}\,{\rm cm}^{-3})^{-1}$, where $B$ is the magnetic field and $n_e$ is the electron density of the interstellar medium. 
Since we see no turnover in our radio spectrum, this suggests that the turnover frequency should be lower than $\sim 1\,{\rm GHz}$ if the source is a synchrotron emitter, which means the magnetic field of the source is $\lesssim 3\times10^4\,{\rm G}$ \citep[e.g.,][]{1981ARA&A..19..373K}: which is consistent with the argument above but not very constraining; moreover, the high degree of circular polarization suggests that this is not typical synchrotron emission. 
The optical depth of free-free absorption at the frequencies we observed should be much smaller than one, which implies $n_e \ll 10^2\,{\rm cm}^{-3}\,(T/10^4\,K)^{0.675}(l_{\rm abs}/250\,{\rm pc})^{-0.5}$, where $T$ is the temperature and $l_{\rm abs}$ is the length scale of the absorber \citep[e.g.,][]{1989agna.book.....O}: again, consistent but not necessarily constraining. 
If we assume there is no change in magnetic field, the RM variation implies a DM variation to be $\sim 30\,{\rm pc}\,{\rm cm}^{-3} (B/2\,\mu{\rm G})^{-1}$ in three days, much higher than those measured in pulsar timing \citep[e.g.,][]{2007MNRAS.378..493Y,2013ApJ...762...94D,2018ApJ...861..132L,2020A&A...644A.153D}. It is still high ($\sim 1\,{\rm pc}\,{\rm cm}^{-3}\,{\rm yr}^{-1}$) even if we assume the magnetic field can be as high as that near the Galactic Center \citep[$\sim0.8\,$mG,][]{2013Natur.501..391E}.

Only a few types of radio sources are known to emit circular polarization at more than a few percent of their total intensity emission at low frequencies ($<5$\,GHz). These include  stars \citep[e.g.,][]{2017ApJ...836L..30L} and pulsars \citep[e.g.,][]{2018MNRAS.474.4629J}. Circular polarization has also been seen from jets in binaries but the fractional polarization is low, $\sim 0.5\%$ (e.g., \citealt{2003Ap&SS.288...79F,2003NewAR..47..609M}), with similar values seen in extragalactic sources \citep[e.g.,][]{2003Ap&SS.288..105M}.
And indeed recent circular polarization searches have identified both new pulsars \citep{2019ApJ...884...96K} and the first brown dwarf discovered at radio wavelength \citep{2020ApJ...903L..33V}.
In this section we discuss these possibilities.

\subsection{Stellar interpretation}
\label{sec:stars}

\begin{figure}[hbt!]
\plotone{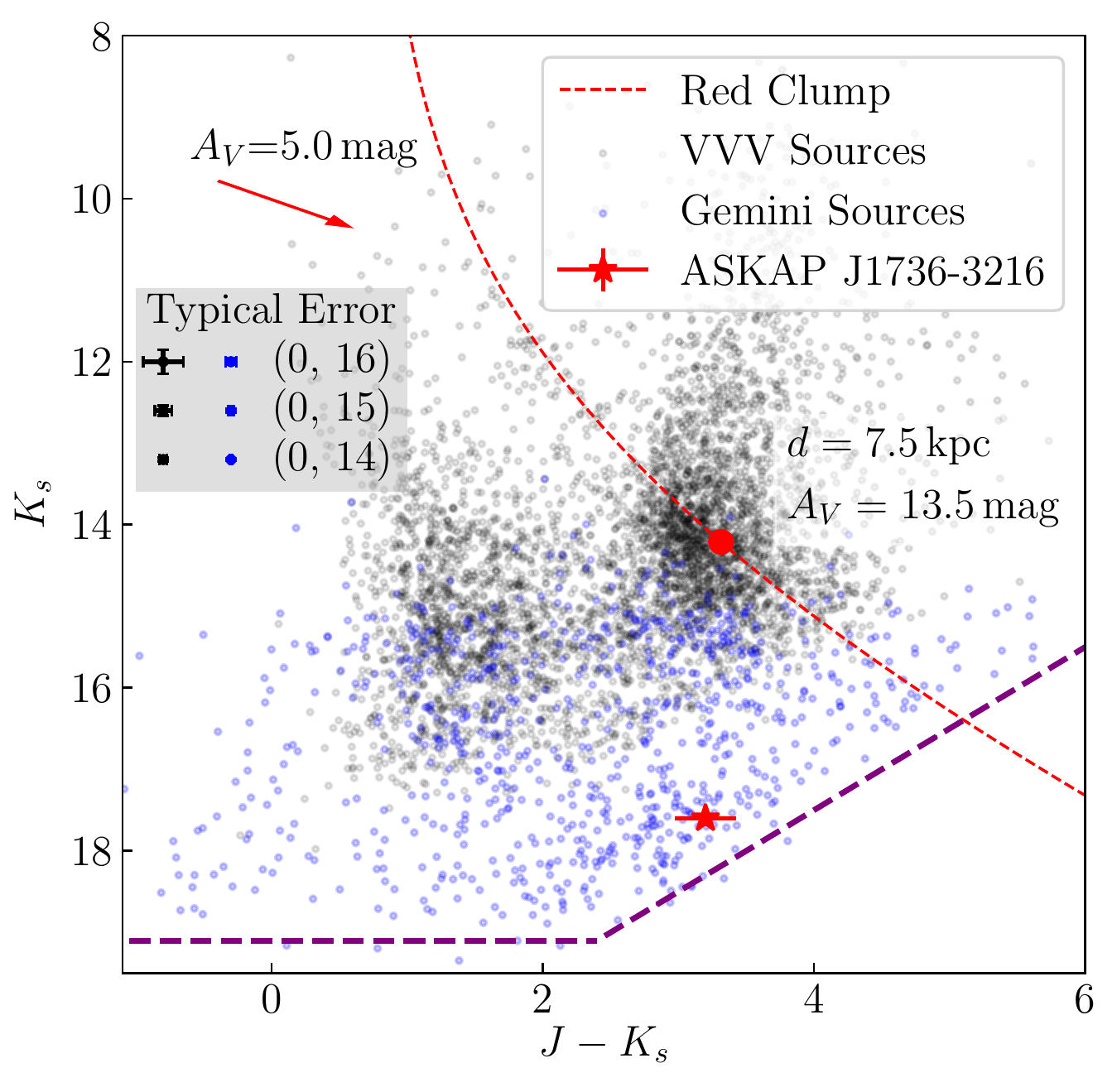}
    \caption{Color-magnitude diagram for the field of \askap. We plot $J-K_s$ color versus $K_s$ magnitude. We show sources from VVV and our deeper Gemini observations (both $3\arcmin$ in radius) as black dots and blue dots respectively. The red star shows the possible infra-red counterpart candidate of \askap. 
    We also plot the error for certain pairs of ($J-K_s$, $K_s$) values in the center left as a reference. 
    Purple dashed line shows the detection thresholds of our Gemini observation, with $J < 21.5\,{\rm mag}$ and $K_s < 19.1\,{\rm mag}$. 
    The red dashed line shows the location of the red clump for different distances. We assume the intrinsic color for the red clump to be $J-K_s = 0.75$ and the intrinsic luminosity $M_K = -1.65$ \citep{1992ApJS...83..111W, 2000MNRAS.317L..45H}. We adopt the extinction coefficients in \citet{2013MNRAS.430.2188Y} and assume an average extinction in the visual band of $A_V/d \approx 1.8\,{\rm mag}\,{\rm kpc}^{-1}$ \citep{1992dge..book.....W}.
    A reddening vector for $A_V = 5\,{\rm mag}$ is also plotted.
    }
    \label{fig:J1736-3216_CMD}
\end{figure}

Low-mass flare stars and chromospherically-active \change{binaries} such as RS CVns often show polarized flares \citep[e.g.,][]{2019MNRAS.488..559Z, 1987AJ.....93.1220M}. We show the color-magnitude diagram for sources within the field of our Gemini observation in Figure~\ref{fig:J1736-3216_CMD}, with additional sources from VVV. 
We investigate the possibility that the Gemini source in Figure~\ref{fig:J1736-3216_gemini} is a nearby cool dwarf associated with \askap\ (RS~CVns would be far brighter, e.g., \citealt{2020MNRAS.491..560D}). According to \citet{2013ApJS..208....9P}\footnote{\url{http://www.pas.rochester.edu/~emamajek/EEM_dwarf_UBVIJHK_colors_Teff.txt}.}  cool dwarfs (spectral type of M/L/T/Y) have typical colors of $J-K_s$ from $-1.0$ to $1.0$ and  absolute magnitudes in $K_s$ band of $M_{K_s} \gtrsim 6$. 
For a cool dwarf with observed color of $J-K_s \approx 3.0$, it would need at least an extinction in V-band of $A_V \approx 12\,{\rm mag}$ (we use the extinction coefficients in \citealt{2013MNRAS.430.2188Y}). With an average extinction of $A_V/d \approx 1.8\,{\rm mag}\,{\rm kpc}^{-1}$ \citep{1992dge..book.....W}, it requires a source at a distance of $\sim 7\,{\rm kpc}$, which implies the magnitude in $K_s$-band would be $\gtrsim 21\,{\rm mag}$ (including the effect of extinction). As our source is about 3 magnitudes brighter than this limit, it is hard for this source to be a cool dwarf: more likely is a more distant red giant branch/red clump star.  In general it does not stand out at all compared to the surrounding population, suggesting that it is not a unique object.  We come to a similar but less robust conclusion about the fainter object in Figure~\ref{fig:J1736-3216_gemini}, which we cannot measure reliably \citep[also see][]{2008ApJ...687..262K}.

\begin{figure}[hbt!]
\plotone{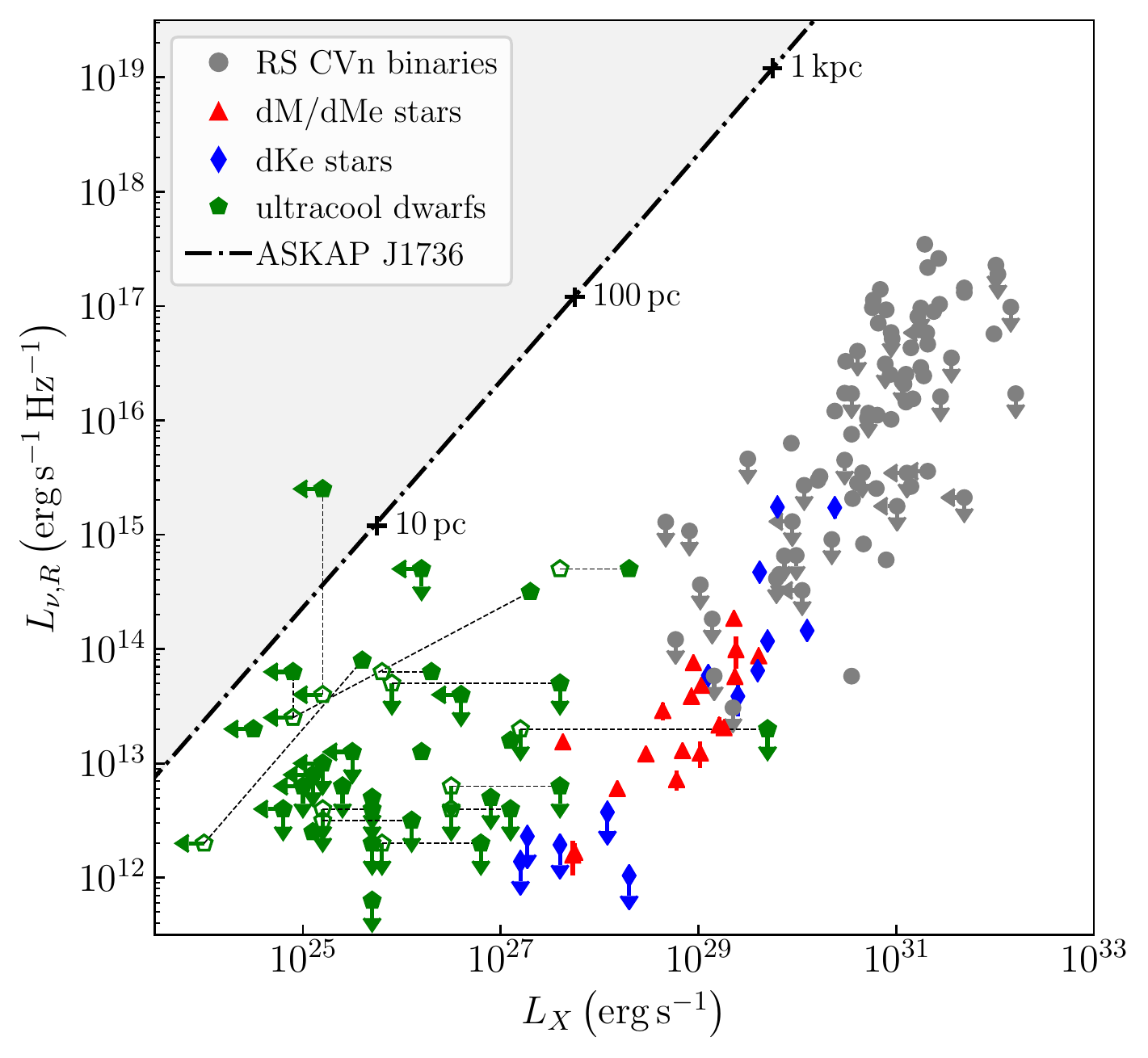}
    \caption{Soft X-ray versus radio luminosity plot for active stars from \citet{1993ApJ...405L..63G, 1994A&A...285..621B, 2014ApJ...785....9W} and references therein, adapted from Figure~12 of \citet{2020MNRAS.491..560D}. 
    Gray circles are RS~CVn binaries, red triangles are dM/dMe stars, blue diamonds are dKe stars, and green pentagons are ultracool dwarfs .  Black dashed lines connect the same source at different states, quiescent state as hollow markers and flaring state as solid markers.  We plot the X-ray luminosity upper limit (0.04--2\,keV, based on the model we assumed earlier) for \askap\ at different distances (as labeled) as the black dot-dashed line, limiting the source to the shaded region to the upper left.
    }
    \label{fig:J1736-3216_GB}
\end{figure}

The high radio flux density of \askap, together with non-detections at X-ray and near-IR wavelengths, also makes a stellar interpretation unlikely.  X-ray and radio luminosities for various types of active stars are typically correlated (the G\"{u}del-Benz relation; \citealt{1993ApJ...405L..63G,2020MNRAS.491..560D}).  In contrast, \askap\ has an X-ray upper limit too low by at least 2 orders of magnitude.  Even for ultracool dwarfs \citep[known to be radio over-luminous relative to their X-ray luminosity; e.g.,][]{ 2014ApJ...785....9W}, the X-ray limit of our source is lower than \change{most of the} ultracool dwarfs (Figure~\ref{fig:J1736-3216_GB})\footnote{Also see \url{https://github.com/AstroLaura/GuedelPlot}.}. 

\begin{figure}
\plotone{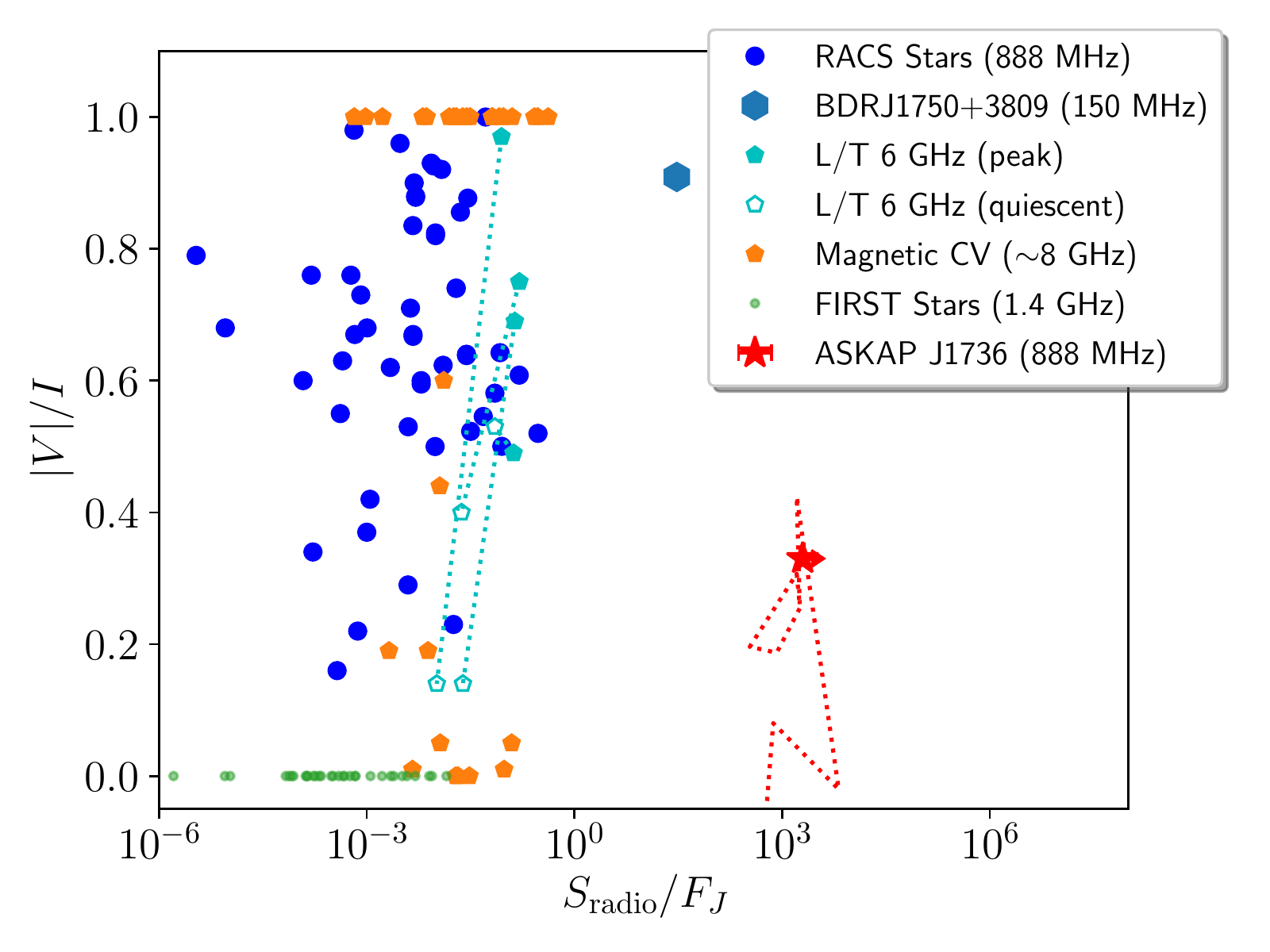}
    \caption{Fractional circular polarization versus radio-to-near-IR flux ratio for stellar sources.  We show  stars measured in the Faint Images of the Radio Sky at Twenty Centimeters (FIRST) survey at 1.4\,GHz as small green circles \citep[][no polarization information was available]{1999AJ....117.1568H}, magnetic CVs measured at 8\,GHz as orange pentagons \citep{2020AdSpR..66.1226B}, auroral emission from L/T dwarfs measured at 6\,GHz as the cyan pentagons for quiescence (open symbols) and peak (filled symbols;  \citealt{2016ApJ...818...24K}),  the T dwarf BDR~J1750+3809 measured at 150\,MHz as the blue hexagon \citep{2020ApJ...903L..33V}, and stars identified in RACS as blue circles \citep{2021MNRAS.502.5438P}. \askap\ is the large red star.  When available, dashed lines connect different radio states for the same source.  The near-infrared data were taken from VVV \citep{2010NewA...15..433M} and the Two Micron All Sky Survey \citep[2MASS;][]{2006AJ....131.1163S}.}
    \label{fig:star_flux_ratio}
\end{figure}

Similarly, based on the brightest possible object that we cannot rule out in infrared (excluding the object in Figure~\ref{fig:J1736-3216_gemini}), we measure $J > 20.8$\,mag from our Gemini observation.  Empirically, we can examine different types of active stars with circularly polarized emission (Figure~\ref{fig:star_flux_ratio}).  The vast majority of stars across different types (L/T dwarfs, magnetic CVs, and radio flux-limited samples) have radio to near-IR flux ratios of $\lesssim 1$.  For the radio-discovered T dwarf BDR~J1750+3809, the ratio is near 10.  Except for the youngest, most energetic pulsars, this ratio is typically $\gtrsim 10^3$ \citep[e.g.,][]{2016MNRAS.455.1746Z}.  
\askap\ itself has a ratio $>10^3$, depending on the radio state.

We can do the same analysis a different way, based on the ratio of radio to bolometric flux from ultra-cool dwarfs.  We determined a lower limit on the distance of a stellar/substellar counterpart (spectral type from late L to mid-M) to be $\sim$150--1400\,pc based on the observed population of ultra-cool dwarfs \citep{2008AJ....136.1290R}. 
At this distance, we would expect low extinction, about 0.5\,mag in $J$-band.
Based on this lower limit on the distance (applying the extinction correction), we calculated upper limits on the radio flux density at 888\,MHz to be $< 0.3 - 0.6$\,$\mu$Jy, assuming $L_{\rm radio}/L_{\rm bol}=10^{-7}$ \citep[which is the typical value for M dwarfs,][]{2010ApJ...709..332B}. The ratio of radio luminosity to  bolometric luminosity for later L dwarfs can be as high as $10^{-5}$: for BDR~J1750+3809, it can reach $2\times10^{-5}$.
The limit would give an expected radio flux density of $<120$\,$\mu$Jy. 
Even for a slightly beamed emission \citep[such as for Jupiter,][]{2016MNRAS.463.2202B}, the expected flux density would be $\lesssim$ 0.9\,mJy.
This is considerably lower than our measured values of $\sim$10\,mJy, suggesting that \askap\ is either a star with an extreme near-IR to radio ratio or another kind of source entirely.

To summarize, we excluded \askap\ as a star based on
\begin{itemize}
    \item Compared to its color ($J-K_s$), the IR source that we detect is too bright in $K_s$-band (Figure~\ref{fig:J1736-3216_CMD}).
    \item The ratio of X-ray luminosity to radio luminosity is too low for stars (Figure~\ref{fig:J1736-3216_GB}).
    \item The source is too bright in radio compared to $J$-band (Figure~\ref{fig:star_flux_ratio}).
\end{itemize}

\subsection{Pulsar interpretation}
\label{sec:pulsar}
Though we found no pulsations in our data, the high degree of polarization and steep spectrum suggest the source may be a pulsar.
We can use our MeerKAT observations to constrain the pulsar-like properties of \askap. 
The expected signal-to-noise ratio of a pulsar at the beam center can be estimated as \citep{2012hpa..book.....L}:
\begin{equation}\label{eq:pulsar_snr}
    {\rm S/N}_{\rm exp} = \frac{SG\sqrt{N_{\rm pol}\tau_{\rm obs}\Delta\nu}}{T_{\rm sys}\beta}
    \sqrt{\frac{P-W}{W}}
\end{equation}
where $S$ is the flux density of the pulsar, $G=2.8$\,K/Jy is the gain of the MeerKAT telescope, $N_{\rm pol}=2$ is the number of polarizations recorded, $\tau_{\rm obs}=700$\,s is the length of the observation, $\Delta\nu\sim 856$\,MHz is the bandwidth, $T_{\rm sys}\sim40$\,K is the system temperature (which includes the sky temperature in this direction), $\beta$ is a correction factor due to downsampling, $W$ is the pulse width of the pulsar and $P$ is the period of pulsar.
The effective pulse width is a combination of its intrinsic pulse width, pulse broadening due to dispersion, and scattering: 
\begin{equation}
    W_{\rm e} = \sqrt{W^2 + \delta t_{\rm disp}^2 + \delta t_{\rm scat}^2}
\end{equation}
where $W_{\rm e}$ is the effective pulse width, $\delta t_{\rm disp} = 8.3\times10^6{\rm DM}\,\nu^{-3}_{\rm MHz}\,{\rm \delta \nu}\,{\rm ms}$ ($\delta \nu$ is the channel bandwidth in  units of MHz) is the smearing time due to dispersion across a channel observed at frequency $\nu_{\rm MHz}$, and $\delta t_{\rm scat}$ is the smearing time due to scattering.
We considered scattering as a function of DM based on \citet[][note that this is consistent with the very high degree of scattering from \citealt{2021arXiv210600386C}]{2004ApJ...605..759B}.

\begin{figure}[hbt!]
    \plotone{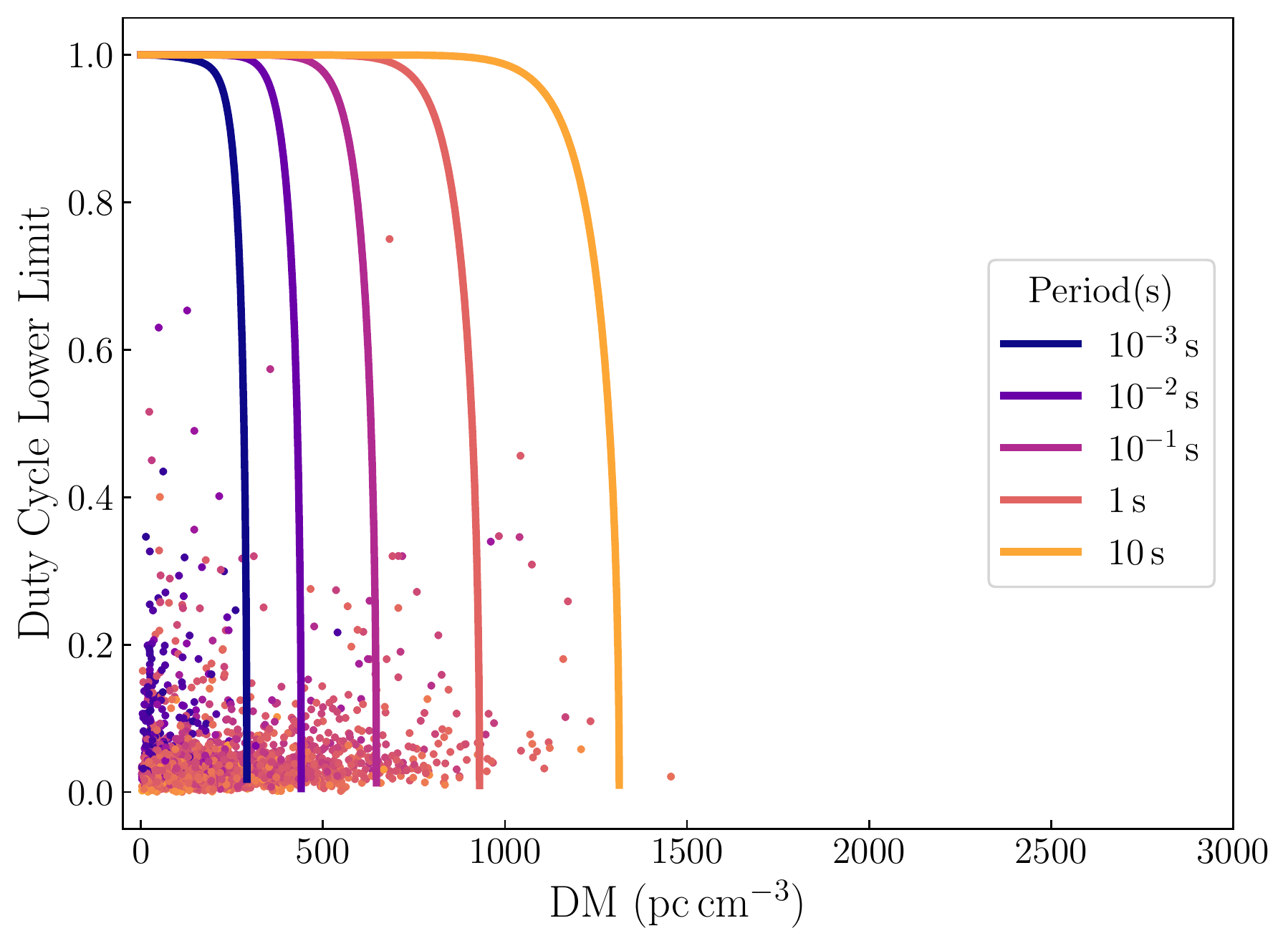}
    \caption{Duty cycle lower limit for a non-detection in \change{the} pulsar search for the MeerKAT \change{data from}  2021~February~07 \change{(at 856--1712\,MHz)}. We considered pulse broadening effects from dispersion and scattering \citep{2004ApJ...605..759B} as a function of DM. We also plot DM versus duty cycle (width of pulse at 50\% peak/period) of the pulsars in the ATNF pulsar catalog \citep{2005AJ....129.1993M} with colors to indicate their period.}
    \label{fig:J1736-3216_pulsar}
\end{figure}

A wide effective pulse width can reduce the pulsation SNR. For example, \citet{hyman2021extreme} argues that C1709--3918 and C1748--2827,  with steep spectra, 10\%--20\% circular polarization, but no pulsations detected, may be pulsars with scatter-broadened pulses.
At the most conservative, if \askap\ is a pulsar with 1\,ms pulsation period, considering the effect of dispersion and scattering, the non-detection in our MeerKAT pulsar search with S/N = 9 threshold suggests the duty cycle ($W/P$) of the pulsar would be $>99\%$ for a source with a DM below $200\,{\rm pc\,cm}^{-3}$ ($\sim 3\,{\rm kpc}$ based on YMW16). For longer pulse periods we would have a duty cycle limit of $>99$\% at DMs up to $\lesssim 1000\,{\rm pc\,cm}^{-3}$ ($\sim 6\,{\rm kpc}$ based on YMW16). 
Compared to pulsars in ATNF pulsar catalog, the highest duty cycle is $\sim 80\%$ (see Figure~\ref{fig:J1736-3216_pulsar}). However, at the highest DMs considered in our search (up to $3000\,{\rm pc\,cm}^{-3}$), we would not be sensitive to even the longest period pulsars with typical scattering behavior.
Observing at higher frequencies can help us minimize the effect of scattering.   We will also employ fast folding algorithms \citep{1969IEEEP..57..724S} to search for longer periods when they become available for MeerKAT data.

An alternative way to smear pulsations would be through orbital acceleration in a tight binary \citep[e.g.,][]{2018ApJ...864...16M,2018MNRAS.474.5008D}.  Our MeerKAT searches were shorter than the Parkes observations, so most binary orbits would not be too smeared out.  Based on the range of accelerations searched, we exclude pulsars in a binary system with orbital period $P_B \gtrsim 5\,$hrs (assuming circular edge-on orbit, pulsar mass of  $1.4\,M_\sun$, and companion mass of  $0.1\,M_\sun$).

The decline in flux seen in our MeerKAT detections (lower right panel of Figure~\ref{fig:J1736-3216_lc}) is a factor of $>$10 faster than the initial detections seen with ASKAP (upper right panel in Figure~\ref{fig:J1736-3216_lc}), with several intermediate values between the ``high" state and non-detections.  This  suggests that what we see  is not ``on versus off” behavior, like might be expected for a standard intermittent pulsar \citep{2006Sci...312..549K,2009ASSL..357...67L}.  

These intermediate flux levels may also rule out effects such as random sampling of eclipses from a ``black widow" \citep[e.g.,][]{1988Natur.333..237F} or ``redback" \citep[e.g.,][]{2013IAUS..291..127R} system, where radio pulses can be periodically eclipsed when the companion wind's obscures the line of sight, and this can both smear out pulsations \citep[e.g.,][]{1996ApJ...465L.119S} and block the continuum flux \citep{2016MNRAS.459.2681B, 2020MNRAS.494.2948P}. Typical orbital periods for those  are $<10\,$hr, so samples days/weeks apart would be very unlikely to end up during the short ($<1\,$hr) ingress/egress periods.  Some systems have been observed to have more complex flux density/eclipse variations \citep[e.g.,][]{ 2020MNRAS.494.2948P}, but still generally not the large degree of continuum flux variability seen here.

Similarly, the precession of a pulsar will result in emission that comes and goes with a timescale of hours \citep[e.g.,][]{2006MNRAS.365L..16Z}. 
The multiple detections with fading behavior over 50 hours in 2021 February, and multiple non-detections over three months make eclipsing and precession unlikely interpretations. Hence we conclude the observed emission is unlike to be due to common pulsar-related origins.  

\begin{figure}[hbt!]
\plotone{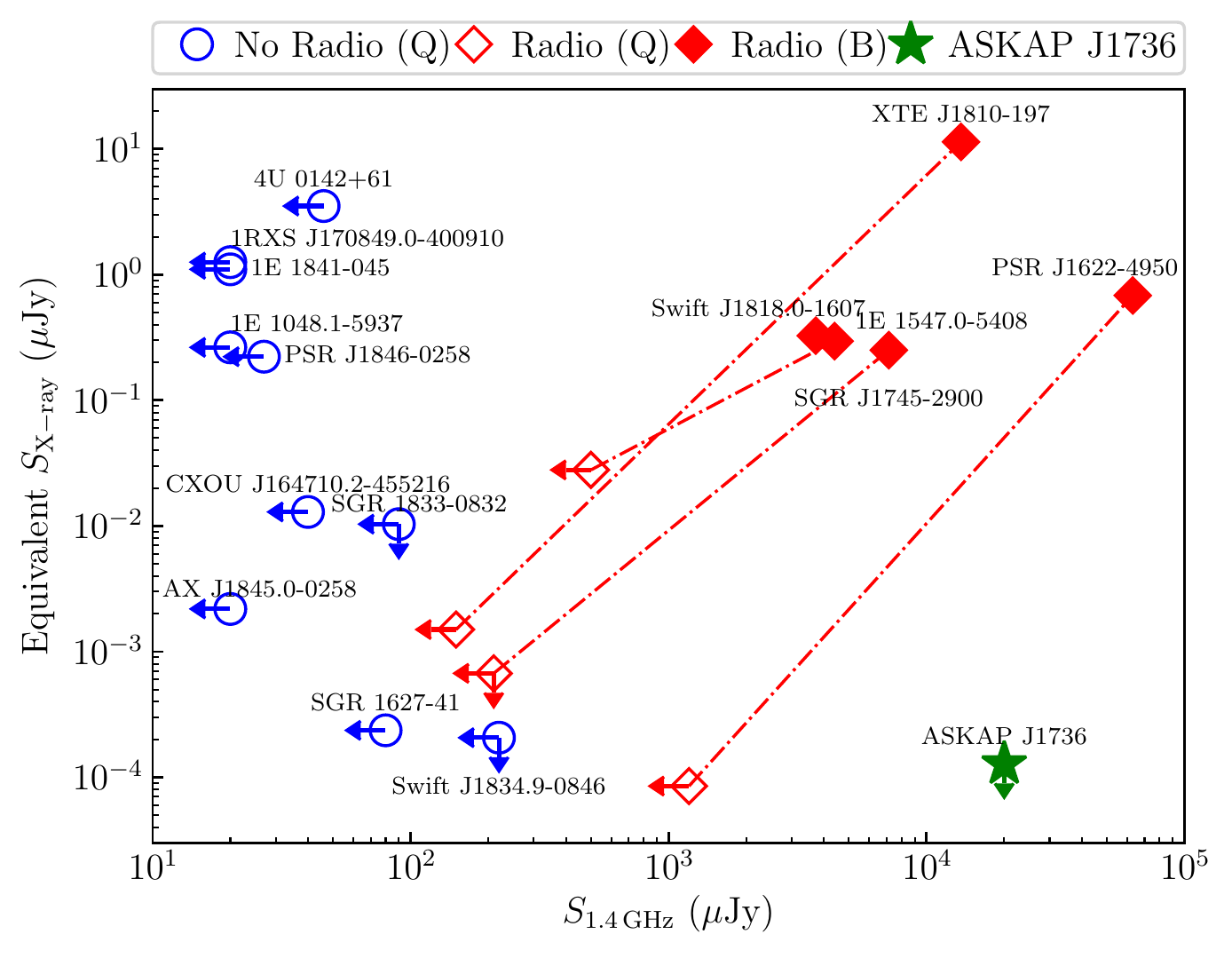}
    \caption{Equivalent X-ray spectral flux density versus radio  flux density for magnetars from the McGill Online Magnetar Catalog \citep{2014ApJS..212....6O}, along with \askap\ (X-ray upper limit from the \textit{Chandra} observation). We show magnetars with no radio pulsation as blue circles, magnetars with radio pulsations as red diamonds and our source as a green star. Hollow markers and solid markers represent the source in quiescent (Q) and outburst (B) state respectively. The red lines connect the same source but with a different state.
    There are some missing or outdated values in the original catalog. We used new  1.4\,GHz flux densities for \object[Swift J1818.0-1607]{Swift~J1818.0--1607} \citep{2020ApJ...896L..37L}, \object[SGR J1745-2900]{SGR~J1745--2900} \citep[][scaling the flux to 1.4\,GHz with a spectral index of $-1$]{2013MNRAS.435L..29S} and \object[PSR J1622-4950]{PSR~J1622--4950} \citep{2018ApJ...856..180C}. We adopt the unabsorbed X-ray flux for five magetars at outburst state from \citet{2008ApJ...676.1178H, 2018ApJ...856..180C, 2013ApJ...770L..23M, 2019ApJ...874L..25G, 2020ApJ...896L..30E}. 
    Note that the radio flux density at an outburst state may not match the X-ray flux corresponding to the same outburst event, as not all sources were measured in both bands for the same outbursts. Radio fluxes of magnetars can be very variable on very short timescales and we used the average flux density here.
    }
    \label{fig:J1736-3216_magnetar}
\end{figure}

Magnetars are neutron stars with extreme strong magnetic fields (up to $\sim 10^{15}\,{\rm G}$; \citealt{1992ApJ...392L...9D,2017ARA&A..55..261K}). 
There are 31 known magnetars and magnetar candidates to date\footnote{\url{http://www.physics.mcgill.ca/~pulsar/magnetar/main.html}} \citep{2014ApJS..212....6O}, but only five are detected in the radio as pulsars \citep{2006Natur.442..892C, 2007ApJ...666L..93C, 2010ApJ...721L..33L, 2013Natur.501..391E, 2013MNRAS.435L..29S, 2013ApJ...775L..34R, 2020ATel13553....1K, 2020ApJ...896L..37L}. 
All the radio detections of magnetars happened during  periods of X-ray outburst \citep{2017ARA&A..55..261K,2021ASSL..461...97E}, and faded eventually.
Magnetars with confirmed radio pulsations show large pulse-to-pulse variability, including pulse morphology \citep{2017ARA&A..55..261K} and polarization \citep[e.g.,][]{2019ApJ...874L..14D}. 
The persistent X-ray luminosity for these radio magnetars is typically $\sim 10^{33}$\,erg\,s$^{-1}$ \citep{2012ApJ...748L..12R}, and  can reach as high as $\sim 10^{36}\,{\rm erg}\,{\rm s}^{-1}$ during an outburst \citep[e.g.,][]{2011ASSP...21..247R}. Our upper limit  based on the \textit{Chandra} observation is comparable to the persistent luminosity of radio magnetars but much lower than those during outbursts (Figure~\ref{fig:J1736-3216_magnetar}).
All radio magnetars show very high degrees of polarization, but their flat radio spectra \citep{2013MNRAS.435L..29S}, in contrast to what we see for \askap, makes a magnetar an unlikely interpretation (although see \citealt{2018ApJ...866..160P}). 
Similarly, the rotation period of magnetars is typically $\sim 1-10\,$s \citep{2017ARA&A..55..261K}, and that range is excluded based on our MeerKAT searches for most sources (${\rm DM}\lesssim 1000\,{\rm pc}\,{\rm cm}^{-3}$, corresponding to $\lesssim 6$\,kpc based on YMW16;  Figure~\ref{fig:J1736-3216_pulsar}).
As we discussed earlier,  pulsations can be smeared out due to scattering, but the Galactic Center magnetar \object[PSR J1745–2900]{PSR~J1745$-$2900} has scattering of only $1.3\,$s at 1\,GHz, considerably lower than that expected from DM models \citep{2014ApJ...780L...3S,2018ApJ...866..160P}, so we may actually be sensitive to higher DMs than Figure~\ref{fig:J1736-3216_pulsar} implies. 
Regardless, higher radio frequency observations may help to rule out or confirm a magnetar origin.
\change{We noted that our search did not exclude sources with extreme long period, such as an ultra long period magnetar 1E 161348-5055.1 \citep[with a rotation period of 6.67\,hrs,][]{2006Sci...313..814D}. Further monitoring observations may help us find such periodic activity.}

\subsection{Other transient classes}

We now consider whether \askap\ could be an X-ray binary or extragalactic transient. 
The polarization and extremely steep spectrum are inconsistent with expectations for emission produced by a steady jet ($\alpha\sim 0$ at these radio frequencies) such as from low-mass X-ray binaries, or optically thin ejecta ($\alpha\sim -0.7$) such as gamma-ray bursts \citep[e.g.,][]{2006csxs.book..381F}.
The short timescale ($\sim$days) of our source also rules out
sources such as supernovae \citep[$\sim$years; e.g.,][]{2015A&ARv..23....3D} and tidal disruption events \citep[$\sim$months; e.g.,][]{2021arXiv210414580G}.

\subsection{Variability due to extrinsic effects}
The large variability ($\sim$100$\times$) is inconsistent with standard diffractive scintillation, which has a modulation index of order unity \citep[e.g.,][]{1991ApJ...376..123C,1992RSPTA.341..151N} for compact sources. Refractive scintillation will produce even less variability, particularly due to the proximity of \askap\ to the Galactic centre. While the total electron column density is unclear due to the unknown source distance, the expected variability due to refractive scintillation at 900\,MHz ranges from a few tens of percent if it is nearby, to as little as 2\% if it is more distance \citep{1998MNRAS.294..307W,2002astro.ph..7156C}.
 
Intraday variables (IDVs) similarly have typical modulation of up to a factor of two \citep[e.g.,][]{1992A&A...258..279Q}, although it can be a slightly higher \citep[e.g.,][]{2000ApJ...529L..65D}.  The linearly polarized flux can vary with higher amplitude and on faster timescales, but the polarization fraction is $<10$\% \citep{2003A&A...401..161K}, so inconsistent with \askap.

However, we consider whether the observed emission could be caused by other forms of extrinsic variability, or a combination of both extrinsic and intrinsic effects. For example, one could invoke a compact radio source undergoing an Extreme Scattering Event \citep[ESE; e.g.,][]{1987Natur.326..675F, 2016Sci...351..354B}, in which the emission is lensed by plasma in the intervening medium; however, this does not explain the high circular polarization we observe. In this case, the lightcurve variability would be caused by propagation effects, while the change in polarization between the two periods of detectability would be intrinsic to the source. 

Similar variability could be caused by gravitational lensing or plasma lensing. 
In general, gravitational lensing is achromatic while plasma lensing is highly chromatic \citep[e.g.,][]{2020arXiv200616263W}, but this is only true when the source is unresolved by the lens.  If the source has finite size with spectral variations across it, then even gravitational lensing can have a chromatic effect as different regions are magnified/demagnified. For instance, a star with an active region could have different parts of that region magnified, which could increase the radio flux relative to other bands (Sec.~\ref{sec:stars}) and  give rise to highly polarized emission.  However, it might still be difficult to explain multiple lensing events with similar magnifications.  Further multi-wavelength searches during bright states and better characterization of the light curve could help resolve this scenario.

\subsection{A GCRT-like interpretation}
As the source is located only 4 degrees from the Galactic Center, we consider whether it could be another Galactic Center Radio Transient.

The GCRT sources share some properties with \askap. GCRT~J1742--3001 has a spectral index of $\lesssim -2$ and GCRT~J1745--3009 has a spectral index varying from $-4$ to $-13$, while that for our source varies from $-2.7$ to $-5.6$. 
Both our source and GCRT~J1745--3009 are highly polarized. GCRT~J1745--3009 was $\sim$100\% circular polarized at 325\,MHz \citep{2010ApJ...712L...5R}. Our source was found to be 100\% linearly polarized at $\sim$1.6\,GHz and as high as $\sim$40\% circularly polarized at $\sim$0.9\,GHz.
The source would have a flux density of $\sim$0.25\,Jy extrapolated to 300\,MHz, which is comparable to that for GCRT~J1745--3009 ($\sim0.3$\,Jy). There is no X-ray detection for any of the GCRT sources when they are radio-bright. 

However, some properties of our source are different from those of the GCRTs.   GCRT~J1745--3009 is thought to be a coherent emitter based on the very rapid variability ($\sim 10\,$min), while our source shows no rapid variability and therefore may not emit coherently.  The variability timescale for GCRT~J1742--3001 is of order of one month, comparable with the initial ``flare'' detected in ASKAP but much longer than the timescale for the latest detections. GCRT~J1745--3009 varies on much faster timescales: it emits flare-like emission for about 10 mins out of 77 mins period at a relative constant flux density. 
\citet{2007ApJ...660L.121H} showed that GCRT~J1745--3009 has been detected in three different states. We have detected \askap\ in two significant observation states so far (bright for a week versus fast  fading). In general the sparse observations of \askap\ and the GCRTs limits conclusions based on their temporal properties, and it is not even clear that all of the GCRTs share a common origin. Further monitoring will help resolve this.

\section{Conclusions}
\label{sec:conc}
We have presented the discovery and characterization of \askap: a highly-polarized, variable radio source located near the Galactic Center and with no clear multi-wavelength counterpart.  
We have largely ruled out most possible origins of \askap\ including stars, normal neutron stars, and  X-ray binaries. An intriguing remaining possibility comes from similarities to steep-spectrum radio sources discovered in recent  imaging surveys \citep[e.g.,][]{2018MNRAS.474.5008D, 2018ApJ...864...16M}. Galactic sources with steep spectra are usually pulsars \citep[e.g.,][]{2013MNRAS.431.1352B}. However, pulsation searches for most of these sources have been unsuccessful \citep[e.g.,][]{2000AJ....119.2376C, 2018ApJ...864...16M, 2019ApJ...876...20H, 2021RNAAS...5...21C}.
As discussed by \citet{2018ApJ...864...16M} and \citet{2018MNRAS.474.5008D}, the explanations for unsuccessful pulsar searching include short period or eccentric binary systems \citep{2015MNRAS.450.2922N}, scattering in the interstellar medium, bias towards short period pulsars in the searching, or alignment of the magnetic and  rotation  \citep{1985MNRAS.212..489P}.  Our searches, especially the short MeerKAT observations, should have had sufficient sensitivity to detect binary systems, but the other two effects may be at play here as well.  Or, these sources along with \askap\ may belong to a new class of steep spectrum sources, possibly related to the GCRTs.  In order to constrain the origin of \askap, continued radio monitoring, pulsations searches at higher frequencies,  and  multi-wavelength observations are necessary.

\askap\ is one of the first sources identified from our searches for transient, polarized sources in the VAST-P1 Survey \citep{2021arXiv210806039M}, and while it is among the most extreme in terms of its variability and polarization properties, it is not the only transient polarized source.  However, most other such sources have straightforward identification with known stars (\citealt{2021arXiv210806039M}; Pritchard et al. \textit{in prep}).  Some do not, and these are the subject of further investigation (e.g., Y.~Wang et al. \textit{in prep}).  
\askap\ is further notable for its location toward the Galactic Center, although we do not yet know whether that is a coincidence or if that location is related to its nature: similar questions could be raised about the GCRT sources. Future comprehensive searches will quantify the exact number of such sources at different locations in the sky, including the Galactic plane, high-latitude regions, and the Magellanic Clouds (see \citealt{2021arXiv210806039M} for the VAST Pilot-1 sky coverage).
We found three variable sources above a modulation index of 0.9, from which \askap\ easily stood out as it is the most variable source, the only polarized source, and the only source with no clear infra-red counterpart.  Given that \askap\ is typically not detected and can turn off on timescales from several weeks to as quickly as a day, our sparse sampling (12 epochs over 16 months) suggests that there could be other similar sources in these fields.   
Increasing the survey cadence and comparing the results of this search to other regions will help us understand how truly unique \askap\ is and whether it is related to the Galactic plane, which should ultimately help us deduce its nature.


\section*{Acknowledgements}
We thank Elaine Sadler, Ron Ekers, Mark Walker, and Shami Chatterjee for useful discussions.
We thank the MeerKAT, \textit{Swift} and \textit{Chandra} directors for approving our DDT observations, and S.~Goedhart and S.~Buchner for assistance in scheduling and conducting MeerKAT observations.
TM acknowledges the support of the Australian Research Council through grant DP190100561.
DK and AO are supported by NSF grant AST-1816492. NR is supported by an ERC Consolidator Grant ``MAGNESIA" (No.\ 817661), Catalan grant SGR2017-1383, and Spanish grant PGC2018-095512-BI00. 
BWS acknowledges funding from the European Research Council (ERC) under the European Union’s Horizon 2020 research and innovation programme (grant agreement No. 694745) and thanks the entire MeerTRAP team for developing the MeerTRAP hardware and software.
GRS is supported by NSERC Discovery Grants RGPIN-2016-06569 and RGPIN-2021-04001.
SD is the recipient of an Australian Research Council Discovery Early Career Award (DE210101738) funded by the Australian Government.

The Dunlap Institute is funded through an endowment established by the David Dunlap family and the University of Toronto. B.M.G. acknowledges the support of the Natural Sciences and Engineering Research Council of Canada (NSERC) through grant RGPIN-2015-05948, and of the Canada Research Chairs program.
This research was supported by the Sydney Informatics Hub (SIH), a core research facility at the University of Sydney.
This work was also supported by software support resources awarded under the Astronomy Data and Computing Services (ADACS) Merit Allocation Program. ADACS is funded from the Astronomy National Collaborative Research Infrastructure Strategy (NCRIS) allocation provided by the Australian Government and managed by Astronomy Australia Limited (AAL).
Parts of this research were conducted by the Australian Research Council Centre of Excellence for Gravitational Wave Discovery (OzGrav), project number CE170100004. 
The Australian Square Kilometre Array Pathfinder is part of the Australia Telescope National Facility which is managed by CSIRO. 
Operation of ASKAP is funded by the Australian Government with support from the National Collaborative Research Infrastructure Strategy. ASKAP uses the resources of the Pawsey Supercomputing Centre. Establishment of ASKAP, the Murchison Radio-astronomy Observatory and the Pawsey Supercomputing Centre are initiatives of the Australian Government, with support from the Government of Western Australia and the Science and Industry Endowment Fund. 
We acknowledge the Wajarri Yamatji as the traditional owners of the Murchison Radio-astronomy Observatory site.
The MeerKAT telescope is operated by the South African Radio Astronomy Observatory, which is a facility
of the National Research Foundation, an agency of the Department of Science and Innovation.  
The scientific results reported in this article are based in part  on observations made by the \textit{Chandra X-ray Observatory}. This work made use of data supplied by the UK Swift Science Data Centre at the University of Leicester.
The Australia Telescope Compact Array is part of the Australia Telescope National Facility which is funded by the Australian Government for operation as a National Facility managed by CSIRO.
This research has made use of the VizieR catalogue access tool, CDS, Strasbourg, France. 
This research has made use of NASA's Astrophysics Data System Bibliographic Services.


\facility{ASKAP, Parkes, ATCA, MeerKAT, Swift, Chandra, Gemini}
\software{
ASKAPsoft \citep{2019ascl.soft12003G},
VAST Transient detection pipeline \citep{2021arXiv210105898P},
Presto \citep{2001PhDT.......123R},
Oxcat \citep[\url{https://github.com/IanHeywood/oxkat}]{2020ascl.soft09003H},
CASA \citep{2007ASPC..376..127M}, 
Wsclean \citep{2014MNRAS.444..606O}, 
IonFR \citep[\url{https://github.com/csobey/ionFR}]{2013A&A...552A..58S},
Peasoup (\url{https://github.com/ewanbarr/peasoup}),
PulsarX (\url{https://github.com/ypmen/PulsarX}),
Miriad \citep{1995ASPC...77..433S},
Dragons \citep{2019ASPC..523..321L},
Sextractor \citep{1996A&AS..117..393B},
matplotlib \citep{2007CSE.....9...90H},
scipy \citep{2020NatMe..17..261V},
astropy \citep{2013A&A...558A..33A, 2018AJ....156..123A}.
}



\appendix
\restartappendixnumbering
\section{Polarization Verification}
\label{sec:pol}
\begin{figure}[hbt!]
\plotone{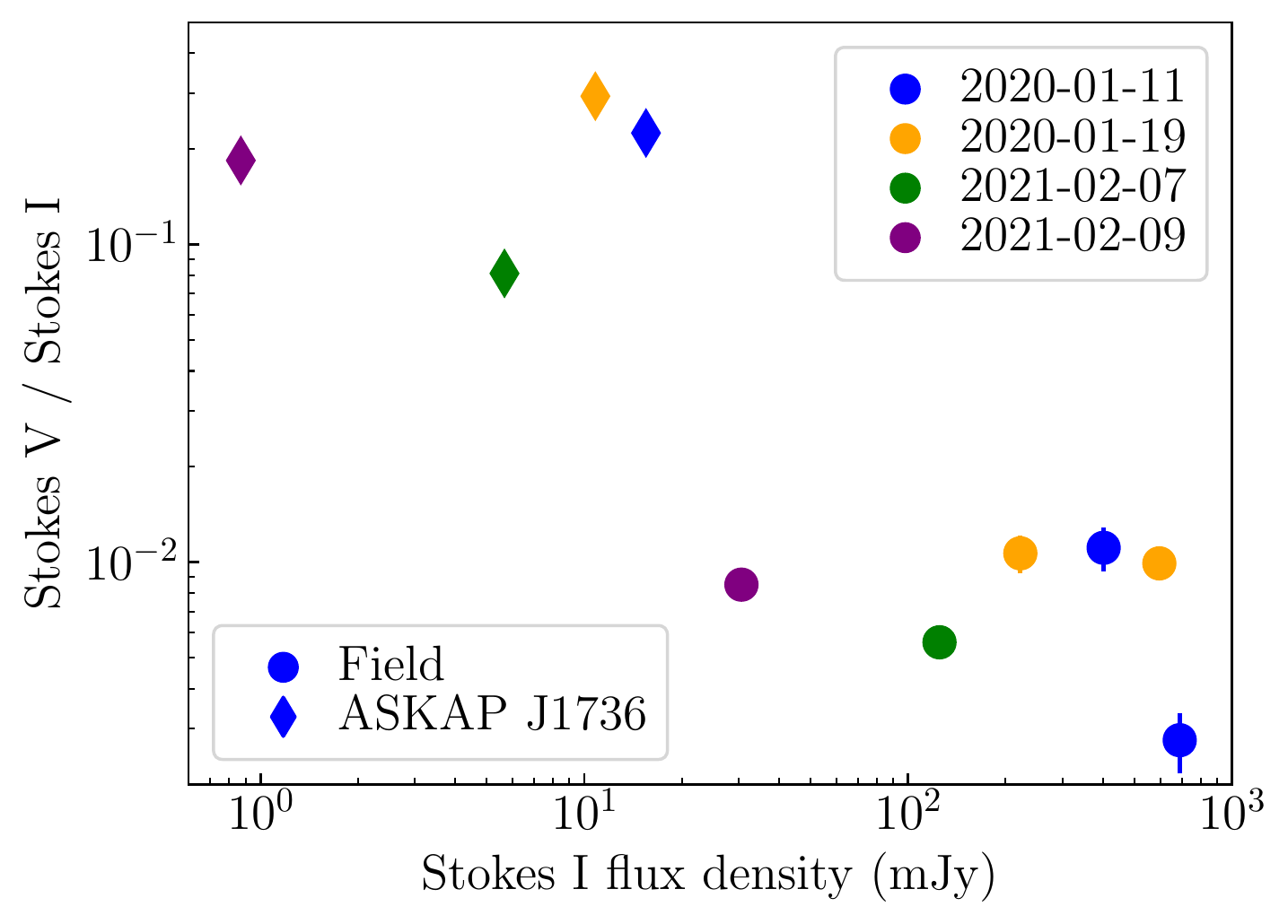}
    \caption{Fractional circular polarization in our images. We show the $V/I$ flux density fraction against Stokes I flux density in four ASKAP observations with $V/I$ detections. \askap\ is shown as a diamond, and field sources (dominated by leakage) are shown as circles.  All sources are detected at $> 5\sigma$ in the Stokes V images, but the field sources have $V/I<1$\%. }
    \label{fig:J1736-3216_leakage}
\end{figure}

\askap\ appears to be circularly polarized. We examined if the Stokes V detection is intrinsic or is the result of polarization leakage. We identified a few field sources with Stokes V detections at $>5\sigma$ significance in individual observations. As shown in Figure~\ref{fig:J1736-3216_leakage}, the field sources with Stokes V detections are usually bright sources (detection SNRs $> 100$), and Stokes V detections are due to a modest level of leakage ($< 1\%$). We can confirm that the circular polarization from our source is real, as the fractional circular polarization is much higher than 1\% (also see \citealt{2019ApJ...884...96K}).

\begin{figure}[hbt!]
\plotone{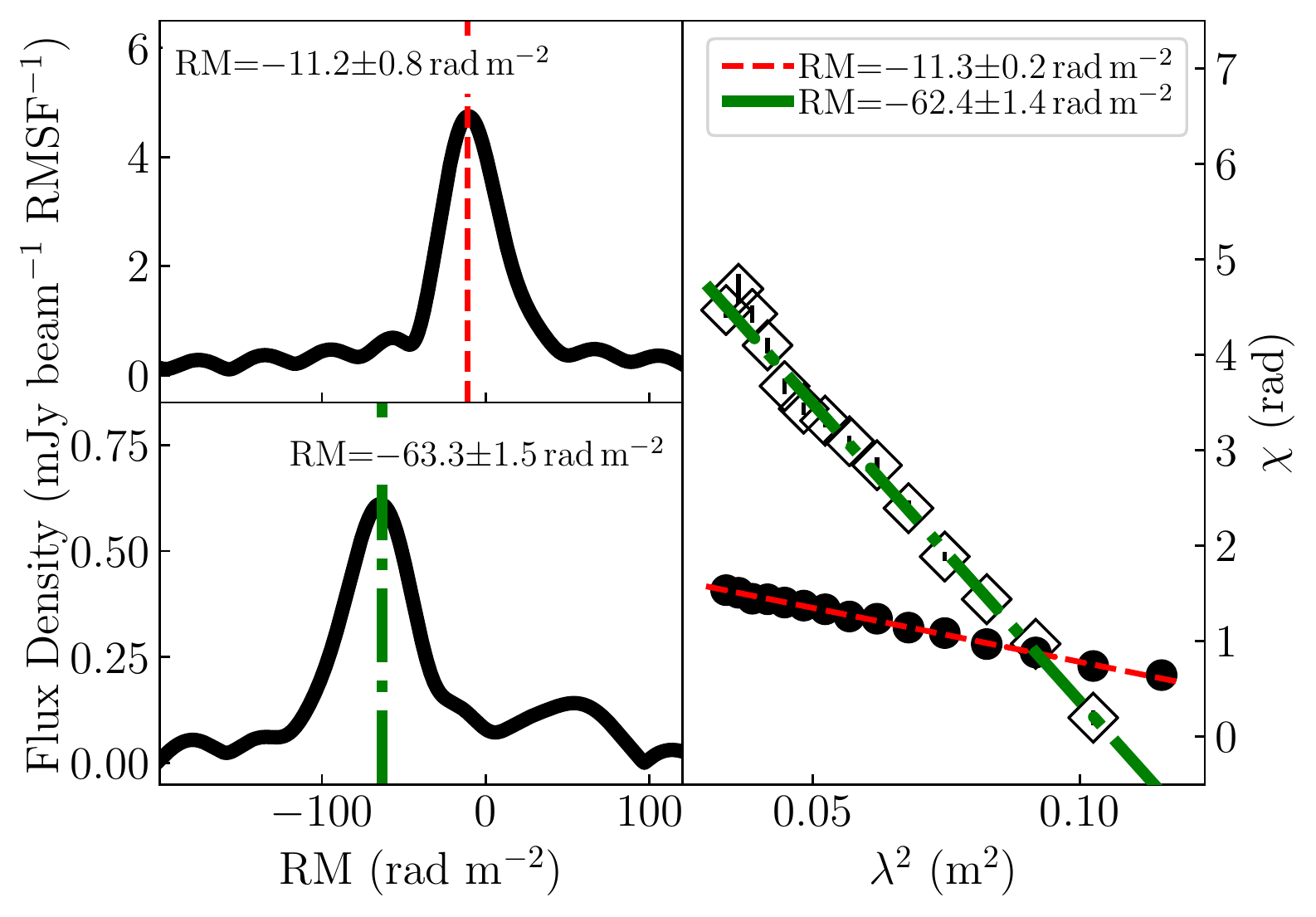}
    \caption{Faraday dispersion function (FDF) plot and $\lambda^2 - \chi$ (relative position angle) plot. 
    The left side shows the FDF plots (with RMclean run), where the upper left panel is for 2021~February~07 (red dashed) and lower left panel for 2021~February~09 (green dashdot). 
    The right side shows the $\lambda^2-\chi$ plot with best-fit lines to each observation, where circles and red dashed line correspond to  2021~February~07 and diamonds and green dashdot line correspond to  2021~February~09. 
    We also show the RMs measured by different methods (in the unit of rad\,m$^{-2}$) on each panel: Left - RM-synthesis; Right - direct $\lambda^2-\chi$  fitting.}
    \label{fig:J1736-3216_RM}
\end{figure}

\begin{figure}
\plotone{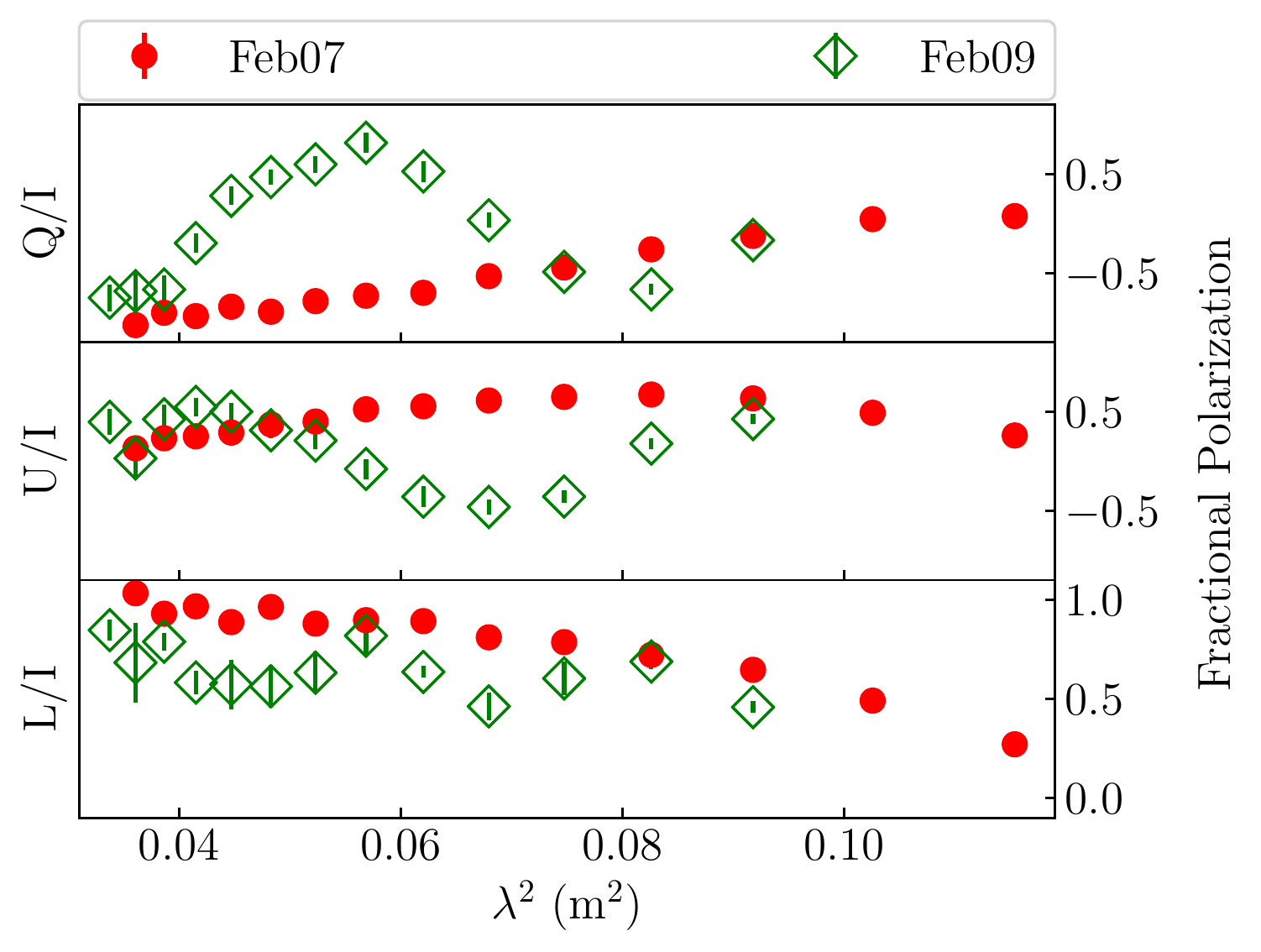}
    \caption{Fractional Stokes Q, U and linear polarization as a function of $\lambda^2$. Red circles show the data from 2021~February~07, while green diamonds show the data from 2021~February~09}
    \label{fig:J1736-3216_QUspec}
\end{figure}

We attempted to verify whether the change in rotation measure for \askap\ was due to instrumental effects or if it was intrinsic to \askap.
Besides measuring the RM value based on RM-synthesis and after RMClean, we also used direct $\lambda^2-\chi$ fitting method to measure the RMs. As is shown in Figure~\ref{fig:J1736-3216_RM} the RMs we measured from different methods are consistent and both of methods show clear changes between the epochs. 
Stokes Q and Stokes U spectra (Figure~\ref{fig:J1736-3216_QUspec}) clearly show that the RM is different in the two epochs.
We also found a linearly polarized field source (J173641.8$-$320029) with RM of $+259.9 \pm 1.9$\,rad m$^{-2}$ and $+256.2 \pm 2.2$\,rad m$^{-2}$ in the two epochs. This field source demonstrates that the RM stability between epochs is suitable to draw the conclusion about the temporal variability of \askap's RM. 

The absence of a dedicated polarization calibration means that we cannot trust the absolute intrinsic polarization angle of our data. However, the changes of intrinsic polarization angle between epochs for \askap\ and the field source (J173641.8$-$320029) are consistent. The intrinsic polarization angle for our source changed from $109.7\pm0.7\,{\rm deg}$ to $18.6\pm5.4\,{\rm deg}$, while that for the field source changed from $116.0\pm20.0\,{\rm deg}$ to $23.9\pm22.6\,{\rm deg}$. Therefore it is likely that the intrinsic polarization angle for \askap\ did not change between  epochs.

\bibliography{bibliography}{}

\begin{thebibliography}{}
\expandafter\ifx\csname natexlab\endcsname\relax\def\natexlab#1{#1}\fi
\providecommand{\url}[1]{\href{#1}{#1}}
\providecommand{\dodoi}[1]{doi:~\href{http://doi.org/#1}{\nolinkurl{#1}}}
\providecommand{\doeprint}[1]{\href{http://ascl.net/#1}{\nolinkurl{http://ascl.net/#1}}}
\providecommand{\doarXiv}[1]{\href{https://arxiv.org/abs/#1}{\nolinkurl{https://arxiv.org/abs/#1}}}

\bibitem[{{Anderson} {et~al.}(2019){Anderson}, {O'Sullivan}, {Heald},
  {Hodgson}, {Pasetto}, \& {Gaensler}}]{2019MNRAS.485.3600A}
{Anderson}, C.~S., {O'Sullivan}, S.~P., {Heald}, G.~H., {et~al.} 2019, \mnras,
  485, 3600, \dodoi{10.1093/mnras/stz377}

\bibitem[{{Astropy Collaboration} {et~al.}(2013){Astropy Collaboration},
  {Robitaille}, {Tollerud}, {Greenfield}, {Droettboom}, {Bray}, {Aldcroft},
  {Davis}, {Ginsburg}, {Price-Whelan}, {Kerzendorf}, {Conley}, {Crighton},
  {Barbary}, {Muna}, {Ferguson}, {Grollier}, {Parikh}, {Nair}, {Unther},
  {Deil}, {Woillez}, {Conseil}, {Kramer}, {Turner}, {Singer}, {Fox}, {Weaver},
  {Zabalza}, {Edwards}, {Azalee Bostroem}, {Burke}, {Casey}, {Crawford},
  {Dencheva}, {Ely}, {Jenness}, {Labrie}, {Lim}, {Pierfederici}, {Pontzen},
  {Ptak}, {Refsdal}, {Servillat}, \& {Streicher}}]{2013A&A...558A..33A}
{Astropy Collaboration}, {Robitaille}, T.~P., {Tollerud}, E.~J., {et~al.} 2013,
  \aap, 558, A33, \dodoi{10.1051/0004-6361/201322068}

\bibitem[{{Astropy Collaboration} {et~al.}(2018){Astropy Collaboration},
  {Price-Whelan}, {Sip{\H{o}}cz}, {G{\"u}nther}, {Lim}, {Crawford}, {Conseil},
  {Shupe}, {Craig}, {Dencheva}, {Ginsburg}, {VanderPlas}, {Bradley},
  {P{\'e}rez-Su{\'a}rez}, {de Val-Borro}, {Aldcroft}, {Cruz}, {Robitaille},
  {Tollerud}, {Ardelean}, {Babej}, {Bach}, {Bachetti}, {Bakanov}, {Bamford},
  {Barentsen}, {Barmby}, {Baumbach}, {Berry}, {Biscani}, {Boquien}, {Bostroem},
  {Bouma}, {Brammer}, {Bray}, {Breytenbach}, {Buddelmeijer}, {Burke},
  {Calderone}, {Cano Rodr{\'\i}guez}, {Cara}, {Cardoso}, {Cheedella}, {Copin},
  {Corrales}, {Crichton}, {D'Avella}, {Deil}, {Depagne}, {Dietrich}, {Donath},
  {Droettboom}, {Earl}, {Erben}, {Fabbro}, {Ferreira}, {Finethy}, {Fox},
  {Garrison}, {Gibbons}, {Goldstein}, {Gommers}, {Greco}, {Greenfield},
  {Groener}, {Grollier}, {Hagen}, {Hirst}, {Homeier}, {Horton}, {Hosseinzadeh},
  {Hu}, {Hunkeler}, {Ivezi{\'c}}, {Jain}, {Jenness}, {Kanarek}, {Kendrew},
  {Kern}, {Kerzendorf}, {Khvalko}, {King}, {Kirkby}, {Kulkarni}, {Kumar},
  {Lee}, {Lenz}, {Littlefair}, {Ma}, {Macleod}, {Mastropietro}, {McCully},
  {Montagnac}, {Morris}, {Mueller}, {Mumford}, {Muna}, {Murphy}, {Nelson},
  {Nguyen}, {Ninan}, {N{\"o}the}, {Ogaz}, {Oh}, {Parejko}, {Parley}, {Pascual},
  {Patil}, {Patil}, {Plunkett}, {Prochaska}, {Rastogi}, {Reddy Janga},
  {Sabater}, {Sakurikar}, {Seifert}, {Sherbert}, {Sherwood-Taylor}, {Shih},
  {Sick}, {Silbiger}, {Singanamalla}, {Singer}, {Sladen}, {Sooley},
  {Sornarajah}, {Streicher}, {Teuben}, {Thomas}, {Tremblay}, {Turner},
  {Terr{\'o}n}, {van Kerkwijk}, {de la Vega}, {Watkins}, {Weaver}, {Whitmore},
  {Woillez}, {Zabalza}, \& {Astropy Contributors}}]{2018AJ....156..123A}
{Astropy Collaboration}, {Price-Whelan}, A.~M., {Sip{\H{o}}cz}, B.~M., {et~al.}
  2018, \aj, 156, 123, \dodoi{10.3847/1538-3881/aabc4f}

\bibitem[{{Bannister} {et~al.}(2016){Bannister}, {Stevens}, {Tuntsov},
  {Walker}, {Johnston}, {Reynolds}, \& {Bignall}}]{2016Sci...351..354B}
{Bannister}, K.~W., {Stevens}, J., {Tuntsov}, A.~V., {et~al.} 2016, Science,
  351, 354, \dodoi{10.1126/science.aac7673}

\bibitem[{Barr(2017)}]{barr_2017}
Barr, E.~D. 2017, Proceedings of the International Astronomical Union, 13,
  175–178, \dodoi{10.1017/S1743921317009036}

\bibitem[{{Barrett} {et~al.}(2020){Barrett}, {Dieck}, {Beasley}, {Mason}, \&
  {Singh}}]{2020AdSpR..66.1226B}
{Barrett}, P., {Dieck}, C., {Beasley}, A.~J., {Mason}, P.~A., \& {Singh}, K.~P.
  2020, Advances in Space Research, 66, 1226, \dodoi{10.1016/j.asr.2020.04.007}

\bibitem[{{Bates} {et~al.}(2013){Bates}, {Lorimer}, \&
  {Verbiest}}]{2013MNRAS.431.1352B}
{Bates}, S.~D., {Lorimer}, D.~R., \& {Verbiest}, J.~P.~W. 2013, \mnras, 431,
  1352, \dodoi{10.1093/mnras/stt257}

\bibitem[{{Benz} \& {G\"{u}del}(1994)}]{1994A&A...285..621B}
{Benz}, A.~O., \& {G\"{u}del}, M. 1994, \aap, 285, 621

\bibitem[{{Berger} {et~al.}(2010){Berger}, {Basri}, {Fleming}, {Giampapa},
  {Gizis}, {Liebert}, {Mart{\'\i}n}, {Phan-Bao}, \&
  {Rutledge}}]{2010ApJ...709..332B}
{Berger}, E., {Basri}, G., {Fleming}, T.~A., {et~al.} 2010, \apj, 709, 332,
  \dodoi{10.1088/0004-637X/709/1/332}

\bibitem[{{Bertin} \& {Arnouts}(1996)}]{1996A&AS..117..393B}
{Bertin}, E., \& {Arnouts}, S. 1996, \aaps, 117, 393,
  \dodoi{10.1051/aas:1996164}

\bibitem[{{Bhat} {et~al.}(2004){Bhat}, {Cordes}, {Camilo}, {Nice}, \&
  {Lorimer}}]{2004ApJ...605..759B}
{Bhat}, N.~D.~R., {Cordes}, J.~M., {Camilo}, F., {Nice}, D.~J., \& {Lorimer},
  D.~R. 2004, \apj, 605, 759, \dodoi{10.1086/382680}

\bibitem[{{Bower} {et~al.}(2005){Bower}, {Roberts}, {Yusef-Zadeh}, {Backer},
  {Cotton}, {Goss}, {Lang}, \& {Lithwick}}]{2005ApJ...633..218B}
{Bower}, G.~C., {Roberts}, D.~A., {Yusef-Zadeh}, F., {et~al.} 2005, \apj, 633,
  218, \dodoi{10.1086/444587}

\bibitem[{{Broderick} {et~al.}(2016){Broderick}, {Fender}, {Breton}, {Stewart},
  {Rowlinson}, {Swinbank}, {Hessels}, {Staley}, {van der Horst}, {Bell},
  {Carbone}, {Cendes}, {Corbel}, {Eisl{\"o}ffel}, {Falcke}, {Grie{\ss}meier},
  {Hassall}, {Jonker}, {Kramer}, {Kuniyoshi}, {Law}, {Markoff}, {Molenaar},
  {Pietka}, {Scheers}, {Serylak}, {Stappers}, {ter Veen}, {van Leeuwen},
  {Wijers}, {Wijnands}, {Wise}, \& {Zarka}}]{2016MNRAS.459.2681B}
{Broderick}, J.~W., {Fender}, R.~P., {Breton}, R.~P., {et~al.} 2016, \mnras,
  459, 2681, \dodoi{10.1093/mnras/stw794}

\bibitem[{{Burningham} {et~al.}(2016){Burningham}, {Hardcastle}, {Nichols},
  {Casewell}, {Littlefair}, {Stark}, {Burleigh}, {Metchev}, {Tannock}, {van
  Weeren}, {Williams}, \& {Wynn}}]{2016MNRAS.463.2202B}
{Burningham}, B., {Hardcastle}, M., {Nichols}, J.~D., {et~al.} 2016, \mnras,
  463, 2202, \dodoi{10.1093/mnras/stw2065}

\bibitem[{{Burrows} {et~al.}(2005){Burrows}, {Hill}, {Nousek}, {Kennea},
  {Wells}, {Osborne}, {Abbey}, {Beardmore}, {Mukerjee}, {Short}, {Chincarini},
  {Campana}, {Citterio}, {Moretti}, {Pagani}, {Tagliaferri}, {Giommi},
  {Capalbi}, {Tamburelli}, {Angelini}, {Cusumano}, {Br{\"a}uninger}, {Burkert},
  \& {Hartner}}]{2005SSRv..120..165B}
{Burrows}, D.~N., {Hill}, J.~E., {Nousek}, J.~A., {et~al.} 2005, \ssr, 120,
  165, \dodoi{10.1007/s11214-005-5097-2}

\bibitem[{{Camilo} {et~al.}(2007){Camilo}, {Ransom}, {Halpern}, \&
  {Reynolds}}]{2007ApJ...666L..93C}
{Camilo}, F., {Ransom}, S.~M., {Halpern}, J.~P., \& {Reynolds}, J. 2007, \apjl,
  666, L93, \dodoi{10.1086/521826}

\bibitem[{{Camilo} {et~al.}(2006){Camilo}, {Ransom}, {Halpern}, {Reynolds},
  {Helfand}, {Zimmerman}, \& {Sarkissian}}]{2006Natur.442..892C}
{Camilo}, F., {Ransom}, S.~M., {Halpern}, J.~P., {et~al.} 2006, \nat, 442, 892,
  \dodoi{10.1038/nature04986}

\bibitem[{{Camilo} {et~al.}(2021){Camilo}, {Ransom}, {Halpern}, \&
  {Roshi}}]{2021arXiv210600386C}
{Camilo}, F., {Ransom}, S.~M., {Halpern}, J.~P., \& {Roshi}, D.~A. 2021, \apj,
  in press, arXiv:2106.00386.
\newblock \doarXiv{2106.00386}

\bibitem[{{Camilo} {et~al.}(2018){Camilo}, {Scholz}, {Serylak}, {Buchner},
  {Merryfield}, {Kaspi}, {Archibald}, {Bailes}, {Jameson}, {van Straten},
  {Sarkissian}, {Reynolds}, {Johnston}, {Hobbs}, {Abbott}, {Adam}, {Adams},
  {Alberts}, {Andreas}, {Asad}, {Baker}, {Baloyi}, {Bauermeister}, {Baxana},
  {Bennett}, {Bernardi}, {Booisen}, {Booth}, {Botha}, {Boyana}, {Brederode},
  {Burger}, {Cheetham}, {Conradie}, {Conradie}, {Davidson}, {De Bruin}, {de
  Swardt}, {de Villiers}, {de Villiers}, {de Villiers}, {de Villiers}, {De
  Waal}, {Dikgale}, {du Toit}, {du Toit}, {Esterhuyse}, {Fanaroff}, {Fataar},
  {Foley}, {Foster}, {Fourie}, {Gamatham}, {Gatsi}, {Geschke}, {Goedhart},
  {Grobler}, {Gumede}, {Hlakola}, {Hokwana}, {Hoorn}, {Horn}, {Horrell},
  {Hugo}, {Isaacson}, {Jacobs}, {Jansen van Rensburg}, {Jonas}, {Jordaan},
  {Joubert}, {Joubert}, {J{\'o}zsa}, {Julie}, {Julius}, {Kapp}, {Karastergiou},
  {Karels}, {Kariseb}, {Karuppusamy}, {Kasper}, {Knox-Davies}, {Koch},
  {Kotz{\'e}}, {Krebs}, {Kriek}, {Kriel}, {Kusel}, {Lamoor}, {Lehmensiek},
  {Liebenberg}, {Liebenberg}, {Lord}, {Lunsky}, {Mabombo}, {Macdonald},
  {Macfarlane}, {Madisa}, {Mafhungo}, {Magnus}, {Magozore}, {Mahgoub}, {Main},
  {Makhathini}, {Malan}, {Malgas}, {Manley}, {Manzini}, {Marais}, {Marais},
  {Marais}, {Maree}, {Martens}, {Matshawule}, {Matthysen}, {Mauch}, {McNally},
  {Merry}, {Millenaar}, {Mjikelo}, {Mkhabela}, {Mnyandu}, {Moeng}, {Mokone},
  {Monama}, {Montshiwa}, {Moss}, {Mphego}, {New}, {Ngcebetsha}, {Ngoasheng},
  {Niehaus}, {Ntuli}, {Nzama}, {Obies}, {Obrocka}, {Ockards}, {Olyn}, {Oozeer},
  {Otto}, {Padayachee}, {Passmoor}, {Patel}, {Paula}, {Peens-Hough},
  {Pholoholo}, {Prozesky}, {Rakoma}, {Ramaila}, {Rammala}, {Ramudzuli},
  {Rasivhaga}, {Ratcliffe}, {Reader}, {Renil}, {Richter}, {Robyntjies},
  {Rosekrans}, {Rust}, {Salie}, {Sambu}, {Schollar}, {Schwardt}, {Seranyane},
  {Sethosa}, {Sharpe}, {Siebrits}, {Sirothia}, {Slabber}, {Smirnov}, {Smith},
  {Sofeya}, {Songqumase}, {Spann}, {Stappers}, {Steyn}, {Steyn}, {Strong},
  {Struthers}, {Stuart}, {Sunnylall}, {Swart}, {Taljaard}, {Tasse}, {Taylor},
  {Theron}, {Thondikulam}, {Thorat}, {Tiplady}, {Toruvanda}, {van Aardt}, {van
  Balla}, {van den Heever}, {van der Byl}, {van der Merwe}, {van der Merwe},
  {van Niekerk}, {van Rooyen}, {van Staden}, {van Tonder}, {van Wyk}, {Wait},
  {Walker}, {Wallace}, {Welz}, {Williams}, {Xaia}, {Young}, \&
  {Zitha}}]{2018ApJ...856..180C}
{Camilo}, F., {Scholz}, P., {Serylak}, M., {et~al.} 2018, \apj, 856, 180,
  \dodoi{10.3847/1538-4357/aab35a}

\bibitem[{{Chiti} {et~al.}(2016){Chiti}, {Chatterjee}, {Wharton}, {Cordes},
  {Lazio}, {Kaplan}, {Bower}, \& {Croft}}]{2016ApJ...833...11C}
{Chiti}, A., {Chatterjee}, S., {Wharton}, R., {et~al.} 2016, \apj, 833, 11,
  \dodoi{10.3847/0004-637X/833/1/11}

\bibitem[{{Condon} {et~al.}(1998){Condon}, {Cotton}, {Greisen}, {Yin},
  {Perley}, {Taylor}, \& {Broderick}}]{1998AJ....115.1693C}
{Condon}, J.~J., {Cotton}, W.~D., {Greisen}, E.~W., {et~al.} 1998, \aj, 115,
  1693, \dodoi{10.1086/300337}

\bibitem[{{Cordes} \& {Lazio}(1991)}]{1991ApJ...376..123C}
{Cordes}, J.~M., \& {Lazio}, T.~J. 1991, \apj, 376, 123, \dodoi{10.1086/170261}

\bibitem[{{Cordes} \& {Lazio}(2002)}]{2002astro.ph..7156C}
{Cordes}, J.~M., \& {Lazio}, T.~J.~W. 2002, arXiv e-prints, astro.
\newblock \doarXiv{astro-ph/0207156}

\bibitem[{{Crawford} {et~al.}(2000){Crawford}, {Kaspi}, \&
  {Bell}}]{2000AJ....119.2376C}
{Crawford}, F., {Kaspi}, V.~M., \& {Bell}, J.~F. 2000, \aj, 119, 2376,
  \dodoi{10.1086/301329}

\bibitem[{{Crawford} {et~al.}(2021){Crawford}, {Margeson}, {Nguyen}, {Saigal},
  {Young}, {Agarwal}, \& {Aggarwal}}]{2021RNAAS...5...21C}
{Crawford}, F., {Margeson}, J., {Nguyen}, B., {et~al.} 2021, Research Notes of
  the American Astronomical Society, 5, 21, \dodoi{10.3847/2515-5172/abe0b9}

\bibitem[{{Dai} {et~al.}(2019){Dai}, {Lower}, {Bailes}, {Camilo}, {Halpern},
  {Johnston}, {Kerr}, {Reynolds}, {Sarkissian}, \&
  {Scholz}}]{2019ApJ...874L..14D}
{Dai}, S., {Lower}, M.~E., {Bailes}, M., {et~al.} 2019, \apjl, 874, L14,
  \dodoi{10.3847/2041-8213/ab0e7a}

\bibitem[{{Davies} {et~al.}(1976){Davies}, {Walsh}, {Browne}, {Edwards}, \&
  {Noble}}]{1976Natur.261..476D}
{Davies}, R.~D., {Walsh}, D., {Browne}, I.~W.~A., {Edwards}, M.~R., \& {Noble},
  R.~G. 1976, \nat, 261, 476, \dodoi{10.1038/261476a0}

\bibitem[{{de Gasperin} {et~al.}(2018){de Gasperin}, {Intema}, \&
  {Frail}}]{2018MNRAS.474.5008D}
{de Gasperin}, F., {Intema}, H.~T., \& {Frail}, D.~A. 2018, \mnras, 474, 5008,
  \dodoi{10.1093/mnras/stx3125}

\bibitem[{{De Luca} {et~al.}(2006){De Luca}, {Caraveo}, {Mereghetti}, {Tiengo},
  \& {Bignami}}]{2006Sci...313..814D}
{De Luca}, A., {Caraveo}, P.~A., {Mereghetti}, S., {Tiengo}, A., \& {Bignami},
  G.~F. 2006, Science, 313, 814, \dodoi{10.1126/science.1129185}

\bibitem[{{Demorest} {et~al.}(2013){Demorest}, {Ferdman}, {Gonzalez}, {Nice},
  {Ransom}, {Stairs}, {Arzoumanian}, {Brazier}, {Burke-Spolaor}, {Chamberlin},
  {Cordes}, {Ellis}, {Finn}, {Freire}, {Giampanis}, {Jenet}, {Kaspi}, {Lazio},
  {Lommen}, {McLaughlin}, {Palliyaguru}, {Perrodin}, {Shannon}, {Siemens},
  {Stinebring}, {Swiggum}, \& {Zhu}}]{2013ApJ...762...94D}
{Demorest}, P.~B., {Ferdman}, R.~D., {Gonzalez}, M.~E., {et~al.} 2013, \apj,
  762, 94, \dodoi{10.1088/0004-637X/762/2/94}

\bibitem[{{Dennett-Thorpe} \& {de Bruyn}(2000)}]{2000ApJ...529L..65D}
{Dennett-Thorpe}, J., \& {de Bruyn}, A.~G. 2000, \apjl, 529, L65,
  \dodoi{10.1086/312459}

\bibitem[{{Desvignes} {et~al.}(2018){Desvignes}, {Eatough}, {Pen}, {Lee},
  {Mao}, {Karuppusamy}, {Schnitzeler}, {Falcke}, {Kramer}, {Wucknitz},
  {Spitler}, {Torne}, {Liu}, {Bower}, {Cognard}, {Lyne}, \&
  {Stappers}}]{2018ApJ...852L..12D}
{Desvignes}, G., {Eatough}, R.~P., {Pen}, U.~L., {et~al.} 2018, \apjl, 852,
  L12, \dodoi{10.3847/2041-8213/aaa2f8}

\bibitem[{{Donner} {et~al.}(2020){Donner}, {Verbiest}, {Tiburzi},
  {Os{\l}owski}, {K{\"u}nsem{\"o}ller}, {Bak Nielsen}, {Grie{\ss}meier},
  {Serylak}, {Kramer}, {Anderson}, {Wucknitz}, {Keane}, {Kondratiev}, {Sobey},
  {McKee}, {Bilous}, {Breton}, {Br{\"u}ggen}, {Ciardi}, {Hoeft}, {van Leeuwen},
  \& {Vocks}}]{2020A&A...644A.153D}
{Donner}, J.~Y., {Verbiest}, J.~P.~W., {Tiburzi}, C., {et~al.} 2020, \aap, 644,
  A153, \dodoi{10.1051/0004-6361/202039517}

\bibitem[{{Driessen} {et~al.}(2020){Driessen}, {McDonald}, {Buckley}, {Caleb},
  {Kotze}, {Potter}, {Rajwade}, {Rowlinson}, {Stappers}, {Tremou}, {Woudt},
  {Fender}, {Armstrong}, {Groot}, {Heywood}, {Horesh}, {van der Horst},
  {Koerding}, {McBride}, {Miller-Jones}, {Mooley}, \&
  {Wijers}}]{2020MNRAS.491..560D}
{Driessen}, L.~N., {McDonald}, I., {Buckley}, D.~A.~H., {et~al.} 2020, \mnras,
  491, 560, \dodoi{10.1093/mnras/stz3027}

\bibitem[{{Dubner} \& {Giacani}(2015)}]{2015A&ARv..23....3D}
{Dubner}, G., \& {Giacani}, E. 2015, \aapr, 23, 3,
  \dodoi{10.1007/s00159-015-0083-5}

\bibitem[{{Dulk}(1985)}]{1985ARA&A..23..169D}
{Dulk}, G.~A. 1985, \araa, 23, 169, \dodoi{10.1146/annurev.aa.23.090185.001125}

\bibitem[{{Duncan} \& {Thompson}(1992)}]{1992ApJ...392L...9D}
{Duncan}, R.~C., \& {Thompson}, C. 1992, \apjl, 392, L9, \dodoi{10.1086/186413}

\bibitem[{{Eatough} {et~al.}(2013){Eatough}, {Falcke}, {Karuppusamy}, {Lee},
  {Champion}, {Keane}, {Desvignes}, {Schnitzeler}, {Spitler}, {Kramer},
  {Klein}, {Bassa}, {Bower}, {Brunthaler}, {Cognard}, {Deller}, {Demorest},
  {Freire}, {Kraus}, {Lyne}, {Noutsos}, {Stappers}, \&
  {Wex}}]{2013Natur.501..391E}
{Eatough}, R.~P., {Falcke}, H., {Karuppusamy}, R., {et~al.} 2013, \nat, 501,
  391, \dodoi{10.1038/nature12499}

\bibitem[{{Esposito} {et~al.}(2021){Esposito}, {Rea}, \&
  {Israel}}]{2021ASSL..461...97E}
{Esposito}, P., {Rea}, N., \& {Israel}, G.~L. 2021, {Magnetars: A Short Review
  and Some Sparse Considerations}, ed. T.~M. {Belloni}, M.~{M{\'e}ndez}, \&
  C.~{Zhang}, Vol. 461, 97--142, \dodoi{10.1007/978-3-662-62110-3\_3}

\bibitem[{{Esposito} {et~al.}(2020){Esposito}, {Rea}, {Borghese}, {Coti
  Zelati}, {Vigan{\`o}}, {Israel}, {Tiengo}, {Ridolfi}, {Possenti}, {Burgay},
  {G{\"o}tz}, {Pintore}, {Stella}, {Dehman}, {Ronchi}, {Campana},
  {Garcia-Garcia}, {Graber}, {Mereghetti}, {Perna}, {Rodr{\'\i}guez Castillo},
  {Turolla}, \& {Zane}}]{2020ApJ...896L..30E}
{Esposito}, P., {Rea}, N., {Borghese}, A., {et~al.} 2020, \apjl, 896, L30,
  \dodoi{10.3847/2041-8213/ab9742}

\bibitem[{{Farnes} {et~al.}(2014){Farnes}, {Gaensler}, \&
  {Carretti}}]{2014ApJS..212...15F}
{Farnes}, J.~S., {Gaensler}, B.~M., \& {Carretti}, E. 2014, \apjs, 212, 15,
  \dodoi{10.1088/0067-0049/212/1/15}

\bibitem[{{Fender}(2003)}]{2003Ap&SS.288...79F}
{Fender}, R. 2003, \apss, 288, 79, \dodoi{10.1023/B:ASTR.0000004996.95929.b7}

\bibitem[{{Fender}(2006)}]{2006csxs.book..381F}
---. 2006, {Jets from X-ray binaries}, Vol.~39, 381--419

\bibitem[{{Fender} {et~al.}(2015){Fender}, {Stewart}, {Macquart}, {Donnarumma},
  {Murphy}, {Deller}, {Paragi}, \& {Chatterjee}}]{2015aska.confE..51F}
{Fender}, R., {Stewart}, A., {Macquart}, J.~P., {et~al.} 2015, in Advancing
  Astrophysics with the Square Kilometre Array (AASKA14), 51.
\newblock \doarXiv{1507.00729}

\bibitem[{{Ferri{\`e}re}(2001)}]{2001RvMP...73.1031F}
{Ferri{\`e}re}, K.~M. 2001, Reviews of Modern Physics, 73, 1031,
  \dodoi{10.1103/RevModPhys.73.1031}

\bibitem[{{Fiedler} {et~al.}(1987){Fiedler}, {Dennison}, {Johnston}, \&
  {Hewish}}]{1987Natur.326..675F}
{Fiedler}, R.~L., {Dennison}, B., {Johnston}, K.~J., \& {Hewish}, A. 1987,
  \nat, 326, 675, \dodoi{10.1038/326675a0}

\bibitem[{{Fruchter} {et~al.}(1988){Fruchter}, {Stinebring}, \&
  {Taylor}}]{1988Natur.333..237F}
{Fruchter}, A.~S., {Stinebring}, D.~R., \& {Taylor}, J.~H. 1988, \nat, 333,
  237, \dodoi{10.1038/333237a0}

\bibitem[{{Gehrels} {et~al.}(2004){Gehrels}, {Chincarini}, {Giommi}, {Mason},
  {Nousek}, {Wells}, {White}, {Barthelmy}, {Burrows}, {Cominsky}, {Hurley},
  {Marshall}, {M{\'e}sz{\'a}ros}, {Roming}, {Angelini}, {Barbier}, {Belloni},
  {Campana}, {Caraveo}, {Chester}, {Citterio}, {Cline}, {Cropper}, {Cummings},
  {Dean}, {Feigelson}, {Fenimore}, {Frail}, {Fruchter}, {Garmire}, {Gendreau},
  {Ghisellini}, {Greiner}, {Hill}, {Hunsberger}, {Krimm}, {Kulkarni}, {Kumar},
  {Lebrun}, {Lloyd-Ronning}, {Markwardt}, {Mattson}, {Mushotzky}, {Norris},
  {Osborne}, {Paczynski}, {Palmer}, {Park}, {Parsons}, {Paul}, {Rees},
  {Reynolds}, {Rhoads}, {Sasseen}, {Schaefer}, {Short}, {Smale}, {Smith},
  {Stella}, {Tagliaferri}, {Takahashi}, {Tashiro}, {Townsley}, {Tueller},
  {Turner}, {Vietri}, {Voges}, {Ward}, {Willingale}, {Zerbi}, \&
  {Zhang}}]{2004ApJ...611.1005G}
{Gehrels}, N., {Chincarini}, G., {Giommi}, P., {et~al.} 2004, \apj, 611, 1005,
  \dodoi{10.1086/422091}

\bibitem[{{Gezari}(2021)}]{2021arXiv210414580G}
{Gezari}, S. 2021, \araa, in press, arXiv:2104.14580.
\newblock \doarXiv{2104.14580}

\bibitem[{{Gotthelf} {et~al.}(2019){Gotthelf}, {Halpern}, {Alford}, {Mihara},
  {Negoro}, {Kawai}, {Dai}, {Lower}, {Johnston}, {Bailes}, {Os{\l}owski},
  {Camilo}, {Miyasaka}, \& {Madsen}}]{2019ApJ...874L..25G}
{Gotthelf}, E.~V., {Halpern}, J.~P., {Alford}, J.~A.~J., {et~al.} 2019, \apjl,
  874, L25, \dodoi{10.3847/2041-8213/ab101a}

\bibitem[{{G\"{u}del} \& {Benz}(1993)}]{1993ApJ...405L..63G}
{G\"{u}del}, M., \& {Benz}, A.~O. 1993, \apjl, 405, L63, \dodoi{10.1086/186766}

\bibitem[{{Guzman} {et~al.}(2019){Guzman}, {Whiting}, {Voronkov}, {Mitchell},
  {Ord}, {Collins}, {Marquarding}, {Lahur}, {Maher}, {Van Diepen}, {Bannister},
  {Wu}, {Lenc}, {Khoo}, \& {Bastholm}}]{2019ascl.soft12003G}
{Guzman}, J., {Whiting}, M., {Voronkov}, M., {et~al.} 2019, {ASKAPsoft: ASKAP
  science data processor software}.
\newblock \doeprint{1912.003}

\bibitem[{{Hallinan} {et~al.}(2007){Hallinan}, {Bourke}, {Lane}, {Antonova},
  {Zavala}, {Brisken}, {Boyle}, {Vrba}, {Doyle}, \&
  {Golden}}]{2007ApJ...663L..25H}
{Hallinan}, G., {Bourke}, S., {Lane}, C., {et~al.} 2007, \apjl, 663, L25,
  \dodoi{10.1086/519790}

\bibitem[{{Halpern} {et~al.}(2008){Halpern}, {Gotthelf}, {Reynolds}, {Ransom},
  \& {Camilo}}]{2008ApJ...676.1178H}
{Halpern}, J.~P., {Gotthelf}, E.~V., {Reynolds}, J., {Ransom}, S.~M., \&
  {Camilo}, F. 2008, \apj, 676, 1178, \dodoi{10.1086/527293}

\bibitem[{{Hammersley} {et~al.}(2000){Hammersley}, {Garz{\'o}n}, {Mahoney},
  {L{\'o}pez-Corredoira}, \& {Torres}}]{2000MNRAS.317L..45H}
{Hammersley}, P.~L., {Garz{\'o}n}, F., {Mahoney}, T.~J.,
  {L{\'o}pez-Corredoira}, M., \& {Torres}, M.~A.~P. 2000, \mnras, 317, L45,
  \dodoi{10.1046/j.1365-8711.2000.03858.x}

\bibitem[{{Han}(2017)}]{2017ARA&A..55..111H}
{Han}, J.~L. 2017, \araa, 55, 111, \dodoi{10.1146/annurev-astro-091916-055221}

\bibitem[{{Helfand} {et~al.}(1999){Helfand}, {Schnee}, {Becker}, {White}, \&
  {McMahon}}]{1999AJ....117.1568H}
{Helfand}, D.~J., {Schnee}, S., {Becker}, R.~H., {White}, R.~L., \& {McMahon},
  R.~G. 1999, \aj, 117, 1568, \dodoi{10.1086/300789}

\bibitem[{{Heywood}(2020)}]{2020ascl.soft09003H}
{Heywood}, I. 2020, {oxkat: Semi-automated imaging of MeerKAT observations}.
\newblock \doeprint{2009.003}

\bibitem[{{HI4PI Collaboration} {et~al.}(2016){HI4PI Collaboration}, {Ben
  Bekhti}, {Fl{\"o}er}, {Keller}, {Kerp}, {Lenz}, {Winkel}, {Bailin},
  {Calabretta}, {Dedes}, {Ford}, {Gibson}, {Haud}, {Janowiecki}, {Kalberla},
  {Lockman}, {McClure-Griffiths}, {Murphy}, {Nakanishi}, {Pisano}, \&
  {Staveley-Smith}}]{2016A&A...594A.116H}
{HI4PI Collaboration}, {Ben Bekhti}, N., {Fl{\"o}er}, L., {et~al.} 2016, \aap,
  594, A116, \dodoi{10.1051/0004-6361/201629178}

\bibitem[{{Hilmarsson} {et~al.}(2021){Hilmarsson}, {Michilli}, {Spitler},
  {Wharton}, {Demorest}, {Desvignes}, {Gourdji}, {Hackstein}, {Hessels},
  {Nimmo}, {Seymour}, {Kramer}, \& {Mckinven}}]{2021ApJ...908L..10H}
{Hilmarsson}, G.~H., {Michilli}, D., {Spitler}, L.~G., {et~al.} 2021, \apjl,
  908, L10, \dodoi{10.3847/2041-8213/abdec0}

\bibitem[{{Hobbs} {et~al.}(2020){Hobbs}, {Manchester}, {Dunning}, {Jameson},
  {Roberts}, {George}, {Green}, {Tuthill}, {Toomey}, {Kaczmarek}, {Mader},
  {Marquarding}, {Ahmed}, {Amy}, {Bailes}, {Beresford}, {Bhat}, {Bock},
  {Bourne}, {Bowen}, {Brothers}, {Cameron}, {Carretti}, {Carter}, {Castillo},
  {Chekkala}, {Cheng}, {Chung}, {Craig}, {Dai}, {Dawson}, {Dempsey}, {Doherty},
  {Dong}, {Edwards}, {Ergesh}, {Gao}, {Han}, {Hayman}, {Indermuehle},
  {Jeganathan}, {Johnston}, {Kanoniuk}, {Kesteven}, {Kramer}, {Leach},
  {Mcintyre}, {Moss}, {Os{\l}owski}, {Phillips}, {Pope}, {Preisig}, {Price},
  {Reeves}, {Reilly}, {Reynolds}, {Robishaw}, {Roush}, {Ruckley}, {Sadler},
  {Sarkissian}, {Severs}, {Shannon}, {Smart}, {Smith}, {Smith}, {Sobey},
  {Staveley-Smith}, {Tzioumis}, {van Straten}, {Wang}, {Wen}, \&
  {Whiting}}]{2020PASA...37...12H}
{Hobbs}, G., {Manchester}, R.~N., {Dunning}, A., {et~al.} 2020, \pasa, 37,
  e012, \dodoi{10.1017/pasa.2020.2}

\bibitem[{{Hotan} {et~al.}(2021){Hotan}, {Bunton}, {Chippendale}, {Whiting},
  {Tuthill}, {Moss}, {McConnell}, {Amy}, {Huynh}, {Allison}, {Anderson},
  {Bannister}, {Bastholm}, {Beresford}, {Bock}, {Bolton}, {Chapman}, {Chow},
  {Collier}, {Cooray}, {Cornwell}, {Diamond}, {Edwards}, {Feain}, {Franzen},
  {George}, {Gupta}, {Hampson}, {Harvey-Smith}, {Hayman}, {Heywood}, {Jacka},
  {Jackson}, {Jackson}, {Jeganathan}, {Johnston}, {Kesteven}, {Kleiner},
  {Koribalski}, {Lee-Waddell}, {Lenc}, {Lensson}, {Mackay}, {Mahony},
  {McClure-Griffiths}, {McConigley}, {Mirtschin}, {Ng}, {Norris}, {Pearce},
  {Phillips}, {Pilawa}, {Raja}, {Reynolds}, {Roberts}, {Roxby}, {Sadler},
  {Shields}, {Schinckel}, {Serra}, {Shaw}, {Sweetnam}, {Troup}, {Tzioumis},
  {Voronkov}, \& {Westmeier}}]{2021PASA...38....9H}
{Hotan}, A.~W., {Bunton}, J.~D., {Chippendale}, A.~P., {et~al.} 2021, \pasa,
  38, e009, \dodoi{10.1017/pasa.2021.1}

\bibitem[{{Hunter}(2007)}]{2007CSE.....9...90H}
{Hunter}, J.~D. 2007, Computing in Science and Engineering, 9, 90,
  \dodoi{10.1109/MCSE.2007.55}

\bibitem[{{Hurley-Walker} {et~al.}(2017){Hurley-Walker}, {Callingham},
  {Hancock}, {Franzen}, {Hindson}, {Kapi{\'n}ska}, {Morgan}, {Offringa},
  {Wayth}, {Wu}, {Zheng}, {Murphy}, {Bell}, {Dwarakanath}, {For}, {Gaensler},
  {Johnston-Hollitt}, {Lenc}, {Procopio}, {Staveley-Smith}, {Ekers}, {Bowman},
  {Briggs}, {Cappallo}, {Deshpande}, {Greenhill}, {Hazelton}, {Kaplan},
  {Lonsdale}, {McWhirter}, {Mitchell}, {Morales}, {Morgan}, {Oberoi}, {Ord},
  {Prabu}, {Shankar}, {Srivani}, {Subrahmanyan}, {Tingay}, {Webster},
  {Williams}, \& {Williams}}]{2017MNRAS.464.1146H}
{Hurley-Walker}, N., {Callingham}, J.~R., {Hancock}, P.~J., {et~al.} 2017,
  \mnras, 464, 1146, \dodoi{10.1093/mnras/stw2337}

\bibitem[{{Hutschenreuter} {et~al.}(2021){Hutschenreuter}, {Anderson}, {Betti},
  {Bower}, {Brown}, {Br{\"u}ggen}, {Carretti}, {Clarke}, {Clegg}, {Costa},
  {Croft}, {Van Eck}, {Gaensler}, {de Gasperin}, {Haverkorn}, {Heald}, {Hull},
  {Inoue}, {Johnston-Hollitt}, {Kaczmarek}, {Law}, {Ma}, {MacMahon}, {Mao},
  {Riseley}, {Roy}, {Shanahan}, {Shimwell}, {Stil}, {Sobey}, {O'Sullivan},
  {Tasse}, {Vacca}, {Vernstrom}, {Williams}, {Wright}, \&
  {En{\ss}lin}}]{2021arXiv210201709H}
{Hutschenreuter}, S., {Anderson}, C.~S., {Betti}, S., {et~al.} 2021, \aap,
  submitted, arXiv:2102.01709.
\newblock \doarXiv{2102.01709}

\bibitem[{{Hyman} {et~al.}(2019){Hyman}, {Frail}, {Deneva}, {Kassim},
  {McLaughlin}, {Kooi}, {Ray}, \& {Polisensky}}]{2019ApJ...876...20H}
{Hyman}, S.~D., {Frail}, D.~A., {Deneva}, J.~S., {et~al.} 2019, \apj, 876, 20,
  \dodoi{10.3847/1538-4357/ab11c8}

\bibitem[{{Hyman} {et~al.}(2002){Hyman}, {Lazio}, {Kassim}, \&
  {Bartleson}}]{2002AJ....123.1497H}
{Hyman}, S.~D., {Lazio}, T. J.~W., {Kassim}, N.~E., \& {Bartleson}, A.~L. 2002,
  \aj, 123, 1497, \dodoi{10.1086/338905}

\bibitem[{{Hyman} {et~al.}(2005){Hyman}, {Lazio}, {Kassim}, {Ray}, {Markwardt},
  \& {Yusef-Zadeh}}]{2005Natur.434...50H}
{Hyman}, S.~D., {Lazio}, T. J.~W., {Kassim}, N.~E., {et~al.} 2005, \nat, 434,
  50, \dodoi{10.1038/nature03400}

\bibitem[{{Hyman} {et~al.}(2007){Hyman}, {Roy}, {Pal}, {Lazio}, {Ray},
  {Kassim}, \& {Bhatnagar}}]{2007ApJ...660L.121H}
{Hyman}, S.~D., {Roy}, S., {Pal}, S., {et~al.} 2007, \apjl, 660, L121,
  \dodoi{10.1086/518245}

\bibitem[{{Hyman} {et~al.}(2009){Hyman}, {Wijnands}, {Lazio}, {Pal},
  {Starling}, {Kassim}, \& {Ray}}]{2009ApJ...696..280H}
{Hyman}, S.~D., {Wijnands}, R., {Lazio}, T. J.~W., {et~al.} 2009, \apj, 696,
  280, \dodoi{10.1088/0004-637X/696/1/280}

\bibitem[{{Hyman} {et~al.}(2021){Hyman}, {Frail}, {Deneva}, {Kassim},
  {Giacintucci}, {Kooi}, {Lazio}, {Joyner}, {Peters}, {Gajjar}, \&
  {Siemion}}]{hyman2021extreme}
{Hyman}, S.~D., {Frail}, D.~A., {Deneva}, J.~S., {et~al.} 2021, \mnras,
  submitted, arXiv:2105.03282.
\newblock \doarXiv{2105.03282}

\bibitem[{{Intema} {et~al.}(2017){Intema}, {Jagannathan}, {Mooley}, \&
  {Frail}}]{2017A&A...598A..78I}
{Intema}, H.~T., {Jagannathan}, P., {Mooley}, K.~P., \& {Frail}, D.~A. 2017,
  \aap, 598, A78, \dodoi{10.1051/0004-6361/201628536}

\bibitem[{{Johnston} \& {Kulkarni}(1991)}]{1991ApJ...368..504J}
{Johnston}, H.~M., \& {Kulkarni}, S.~R. 1991, \apj, 368, 504,
  \dodoi{10.1086/169715}

\bibitem[{{Johnston} \& {Kerr}(2018)}]{2018MNRAS.474.4629J}
{Johnston}, S., \& {Kerr}, M. 2018, \mnras, 474, 4629,
  \dodoi{10.1093/mnras/stx3095}

\bibitem[{{Jonas} \& {MeerKAT Team}(2016)}]{2016mks..confE...1J}
{Jonas}, J., \& {MeerKAT Team}. 2016, in MeerKAT Science: On the Pathway to the
  SKA, 1

\bibitem[{{Kao} {et~al.}(2016){Kao}, {Hallinan}, {Pineda}, {Escala},
  {Burgasser}, {Bourke}, \& {Stevenson}}]{2016ApJ...818...24K}
{Kao}, M.~M., {Hallinan}, G., {Pineda}, J.~S., {et~al.} 2016, \apj, 818, 24,
  \dodoi{10.3847/0004-637X/818/1/24}

\bibitem[{{Kaplan} {et~al.}(2008){Kaplan}, {Hyman}, {Roy}, {Bandyopadhyay},
  {Chakrabarty}, {Kassim}, {Lazio}, \& {Ray}}]{2008ApJ...687..262K}
{Kaplan}, D.~L., {Hyman}, S.~D., {Roy}, S., {et~al.} 2008, \apj, 687, 262,
  \dodoi{10.1086/591436}

\bibitem[{{Kaplan} {et~al.}(2019){Kaplan}, {Dai}, {Lenc}, {Zic}, {Swiggum},
  {Murphy}, {Anderson}, {Cameron}, {Dobie}, {Hobbs}, {Kaczmarek}, {Lynch}, \&
  {Toomey}}]{2019ApJ...884...96K}
{Kaplan}, D.~L., {Dai}, S., {Lenc}, E., {et~al.} 2019, \apj, 884, 96,
  \dodoi{10.3847/1538-4357/ab397f}

\bibitem[{{Karuppusamy} {et~al.}(2020){Karuppusamy}, {Desvignes}, {Kramer},
  {Porayko}, {Champion}, {Torne}, {Stappers}, {van der Horst}, {Kouveliotou},
  \& {O'Connor}}]{2020ATel13553....1K}
{Karuppusamy}, R., {Desvignes}, G., {Kramer}, M., {et~al.} 2020, The
  Astronomer's Telegram, 13553, 1

\bibitem[{{Kaspi} \& {Beloborodov}(2017)}]{2017ARA&A..55..261K}
{Kaspi}, V.~M., \& {Beloborodov}, A.~M. 2017, \araa, 55, 261,
  \dodoi{10.1146/annurev-astro-081915-023329}

\bibitem[{{Kellermann} \& {Pauliny-Toth}(1981)}]{1981ARA&A..19..373K}
{Kellermann}, K.~I., \& {Pauliny-Toth}, I.~I.~K. 1981, \araa, 19, 373,
  \dodoi{10.1146/annurev.aa.19.090181.002105}

\bibitem[{{Kramer} {et~al.}(2006){Kramer}, {Lyne}, {O'Brien}, {Jordan}, \&
  {Lorimer}}]{2006Sci...312..549K}
{Kramer}, M., {Lyne}, A.~G., {O'Brien}, J.~T., {Jordan}, C.~A., \& {Lorimer},
  D.~R. 2006, Science, 312, 549, \dodoi{10.1126/science.1124060}

\bibitem[{{Kraus} {et~al.}(2003){Kraus}, {Krichbaum}, {Wegner}, {Witzel},
  {Cim{\`o}}, {Quirrenbach}, {Britzen}, {Fuhrmann}, {Lobanov}, {Naundorf},
  {Otterbein}, {Peng}, {Risse}, {Ros}, \& {Zensus}}]{2003A&A...401..161K}
{Kraus}, A., {Krichbaum}, T.~P., {Wegner}, R., {et~al.} 2003, \aap, 401, 161,
  \dodoi{10.1051/0004-6361:20030118}

\bibitem[{{Labrie} {et~al.}(2019){Labrie}, {Anderson}, {C{\'a}rdenes},
  {Simpson}, \& {Turner}}]{2019ASPC..523..321L}
{Labrie}, K., {Anderson}, K., {C{\'a}rdenes}, R., {Simpson}, C., \& {Turner},
  J. E.~H. 2019, in Astronomical Society of the Pacific Conference Series, Vol.
  523, Astronomical Data Analysis Software and Systems XXVII, ed. P.~J.
  {Teuben}, M.~W. {Pound}, B.~A. {Thomas}, \& E.~M. {Warner}, 321

\bibitem[{{Lacy} {et~al.}(2020){Lacy}, {Baum}, {Chandler}, {Chatterjee},
  {Clarke}, {Deustua}, {English}, {Farnes}, {Gaensler}, {Gugliucci},
  {Hallinan}, {Kent}, {Kimball}, {Law}, {Lazio}, {Marvil}, {Mao}, {Medlin},
  {Mooley}, {Murphy}, {Myers}, {Osten}, {Richards}, {Rosolowsky}, {Rudnick},
  {Schinzel}, {Sivakoff}, {Sjouwerman}, {Taylor}, {White}, {Wrobel},
  {Andernach}, {Beasley}, {Berger}, {Bhatnager}, {Birkinshaw}, {Bower},
  {Brandt}, {Brown}, {Burke-Spolaor}, {Butler}, {Comerford}, {Demorest}, {Fu},
  {Giacintucci}, {Golap}, {G{\"u}th}, {Hales}, {Hiriart}, {Hodge}, {Horesh},
  {Ivezi{\'c}}, {Jarvis}, {Kamble}, {Kassim}, {Liu}, {Loinard}, {Lyons},
  {Masters}, {Mezcua}, {Moellenbrock}, {Mroczkowski}, {Nyland}, {O'Dea},
  {O'Sullivan}, {Peters}, {Radford}, {Rao}, {Robnett}, {Salcido}, {Shen},
  {Sobotka}, {Witz}, {Vaccari}, {van Weeren}, {Vargas}, {Williams}, \&
  {Yoon}}]{2020PASP..132c5001L}
{Lacy}, M., {Baum}, S.~A., {Chandler}, C.~J., {et~al.} 2020, \pasp, 132,
  035001, \dodoi{10.1088/1538-3873/ab63eb}

\bibitem[{{Lam} {et~al.}(2018){Lam}, {Ellis}, {Grillo}, {Jones}, {Hazboun},
  {Brook}, {Turner}, {Chatterjee}, {Cordes}, {Lazio}, {DeCesar}, {Arzoumanian},
  {Blumer}, {Cromartie}, {Demorest}, {Dolch}, {Ferdman}, {Ferrara}, {Fonseca},
  {Garver-Daniels}, {Gentile}, {Gupta}, {Lorimer}, {Lynch}, {Madison},
  {McLaughlin}, {Ng}, {Nice}, {Pennucci}, {Ransom}, {Spiewak}, {Stairs},
  {Stinebring}, {Stovall}, {Swiggum}, {Vigeland}, \&
  {Zhu}}]{2018ApJ...861..132L}
{Lam}, M.~T., {Ellis}, J.~A., {Grillo}, G., {et~al.} 2018, \apj, 861, 132,
  \dodoi{10.3847/1538-4357/aac770}

\bibitem[{{Lazio} {et~al.}(2006){Lazio}, {Deneva}, {Bower}, {Cordes}, {Hyman},
  {Backer}, {Bhat}, {Chatterjee}, {Demorest}, {Ransom}, \&
  {Vlemmings}}]{2006JPhCS..54..110L}
{Lazio}, J., {Deneva}, J.~S., {Bower}, G.~C., {et~al.} 2006, in Journal of
  Physics Conference Series, Vol.~54, Journal of Physics Conference Series,
  110--114, \dodoi{10.1088/1742-6596/54/1/019}

\bibitem[{{Levin} {et~al.}(2010){Levin}, {Bailes}, {Bates}, {Bhat}, {Burgay},
  {Burke-Spolaor}, {D'Amico}, {Johnston}, {Keith}, {Kramer}, {Milia},
  {Possenti}, {Rea}, {Stappers}, \& {van Straten}}]{2010ApJ...721L..33L}
{Levin}, L., {Bailes}, M., {Bates}, S., {et~al.} 2010, \apjl, 721, L33,
  \dodoi{10.1088/2041-8205/721/1/L33}

\bibitem[{{Lico} {et~al.}(2017){Lico}, {G{\'o}mez}, {Asada}, \&
  {Fuentes}}]{2017MNRAS.469.1612L}
{Lico}, R., {G{\'o}mez}, J.~L., {Asada}, K., \& {Fuentes}, A. 2017, \mnras,
  469, 1612, \dodoi{10.1093/mnras/stx960}

\bibitem[{{Lorimer} \& {Kramer}(2012)}]{2012hpa..book.....L}
{Lorimer}, D.~R., \& {Kramer}, M. 2012, {Handbook of Pulsar Astronomy}

\bibitem[{{Lower} {et~al.}(2020){Lower}, {Shannon}, {Johnston}, \&
  {Bailes}}]{2020ApJ...896L..37L}
{Lower}, M.~E., {Shannon}, R.~M., {Johnston}, S., \& {Bailes}, M. 2020, \apjl,
  896, L37, \dodoi{10.3847/2041-8213/ab9898}

\bibitem[{{Lynch} {et~al.}(2017){Lynch}, {Lenc}, {Kaplan}, {Murphy}, \&
  {Anderson}}]{2017ApJ...836L..30L}
{Lynch}, C.~R., {Lenc}, E., {Kaplan}, D.~L., {Murphy}, T., \& {Anderson}, G.~E.
  2017, \apjl, 836, L30, \dodoi{10.3847/2041-8213/aa5ffd}

\bibitem[{{Lyne}(2009)}]{2009ASSL..357...67L}
{Lyne}, A.~G. 2009, {Intermittent Pulsars}, ed. W.~{Becker}, Vol. 357, 67,
  \dodoi{10.1007/978-3-540-76965-1_4}

\bibitem[{{Maan} {et~al.}(2018){Maan}, {Bassa}, {van Leeuwen}, {Krishnakumar},
  \& {Joshi}}]{2018ApJ...864...16M}
{Maan}, Y., {Bassa}, C., {van Leeuwen}, J., {Krishnakumar}, M.~A., \& {Joshi},
  B.~C. 2018, \apj, 864, 16, \dodoi{10.3847/1538-4357/aad4ad}

\bibitem[{{Macquart}(2003)}]{2003NewAR..47..609M}
{Macquart}, J.-P. 2003, \nar, 47, 609, \dodoi{10.1016/S1387-6473(03)00104-0}

\bibitem[{{Macquart} {et~al.}(2003){Macquart}, {Wu}, {Hannikainen}, {Sault}, \&
  {Jauncey}}]{2003Ap&SS.288..105M}
{Macquart}, J.-P., {Wu}, K., {Hannikainen}, D.~C., {Sault}, R.~J., \&
  {Jauncey}, D.~L. 2003, \apss, 288, 105,
  \dodoi{10.1023/B:ASTR.0000004998.46724.eb}

\bibitem[{{Manchester} {et~al.}(2005){Manchester}, {Hobbs}, {Teoh}, \&
  {Hobbs}}]{2005AJ....129.1993M}
{Manchester}, R.~N., {Hobbs}, G.~B., {Teoh}, A., \& {Hobbs}, M. 2005, \aj, 129,
  1993, \dodoi{10.1086/428488}

\bibitem[{{McConnell} {et~al.}(2020){McConnell}, {Hale}, {Lenc}, {Banfield},
  {Heald}, {Hotan}, {Leung}, {Moss}, {Murphy}, {O'Brien}, {Pritchard}, {Raja},
  {Sadler}, {Stewart}, {Thomson}, {Whiting}, {Allison}, {Amy}, {Anderson},
  {Ball}, {Bannister}, {Bell}, {Bock}, {Bolton}, {Bunton}, {Chippendale},
  {Collier}, {Cooray}, {Cornwell}, {Diamond}, {Edwards}, {Gupta}, {Hayman},
  {Heywood}, {Jackson}, {Koribalski}, {Lee-Waddell}, {McClure-Griffiths}, {Ng},
  {Norris}, {Phillips}, {Reynolds}, {Roxby}, {Schinckel}, {Shields},
  {Tremblay}, {Tzioumis}, {Voronkov}, \& {Westmeier}}]{2020PASA...37...48M}
{McConnell}, D., {Hale}, C.~L., {Lenc}, E., {et~al.} 2020, \pasa, 37, e048,
  \dodoi{10.1017/pasa.2020.41}

\bibitem[{{McMullin} {et~al.}(2007){McMullin}, {Waters}, {Schiebel}, {Young},
  \& {Golap}}]{2007ASPC..376..127M}
{McMullin}, J.~P., {Waters}, B., {Schiebel}, D., {Young}, W., \& {Golap}, K.
  2007, in Astronomical Society of the Pacific Conference Series, Vol. 376,
  Astronomical Data Analysis Software and Systems XVI, ed. R.~A. {Shaw},
  F.~{Hill}, \& D.~J. {Bell}, 127

\bibitem[{{Minniti} {et~al.}(2010){Minniti}, {Lucas}, {Emerson}, {Saito},
  {Hempel}, {Pietrukowicz}, {Ahumada}, {Alonso}, {Alonso-Garcia}, {Arias},
  {Bandyopadhyay}, {Barb{\'a}}, {Barbuy}, {Bedin}, {Bica}, {Borissova},
  {Bronfman}, {Carraro}, {Catelan}, {Clari{\'a}}, {Cross}, {de Grijs},
  {D{\'e}k{\'a}ny}, {Drew}, {Fari{\~n}a}, {Feinstein}, {Fern{\'a}ndez
  Laj{\'u}s}, {Gamen}, {Geisler}, {Gieren}, {Goldman}, {Gonzalez}, {Gunthardt},
  {Gurovich}, {Hambly}, {Irwin}, {Ivanov}, {Jord{\'a}n}, {Kerins}, {Kinemuchi},
  {Kurtev}, {L{\'o}pez-Corredoira}, {Maccarone}, {Masetti}, {Merlo},
  {Messineo}, {Mirabel}, {Monaco}, {Morelli}, {Padilla}, {Palma}, {Parisi},
  {Pignata}, {Rejkuba}, {Roman-Lopes}, {Sale}, {Schreiber}, {Schr{\"o}der},
  {Smith}, {}, {Soto}, {Tamura}, {Tappert}, {Thompson}, {Toledo}, {Zoccali}, \&
  {Pietrzynski}}]{2010NewA...15..433M}
{Minniti}, D., {Lucas}, P.~W., {Emerson}, J.~P., {et~al.} 2010, \na, 15, 433,
  \dodoi{10.1016/j.newast.2009.12.002}

\bibitem[{{Mori} {et~al.}(2013){Mori}, {Gotthelf}, {Zhang}, {An}, {Baganoff},
  {Barri{\`e}re}, {Beloborodov}, {Boggs}, {Christensen}, {Craig}, {Dufour},
  {Grefenstette}, {Hailey}, {Harrison}, {Hong}, {Kaspi}, {Kennea}, {Madsen},
  {Markwardt}, {Nynka}, {Stern}, {Tomsick}, \& {Zhang}}]{2013ApJ...770L..23M}
{Mori}, K., {Gotthelf}, E.~V., {Zhang}, S., {et~al.} 2013, \apjl, 770, L23,
  \dodoi{10.1088/2041-8205/770/2/L23}

\bibitem[{{Murphy} {et~al.}(2007){Murphy}, {Mauch}, {Green}, {Hunstead},
  {Piestrzynska}, {Kels}, \& {Sztajer}}]{2007MNRAS.382..382M}
{Murphy}, T., {Mauch}, T., {Green}, A., {et~al.} 2007, \mnras, 382, 382,
  \dodoi{10.1111/j.1365-2966.2007.12379.x}

\bibitem[{{Murphy} {et~al.}(2013){Murphy}, {Chatterjee}, {Kaplan}, {Banyer},
  {Bell}, {Bignall}, {Bower}, {Cameron}, {Coward}, {Cordes}, {Croft}, {Curran},
  {Djorgovski}, {Farrell}, {Frail}, {Gaensler}, {Galloway}, {Gendre}, {Green},
  {Hancock}, {Johnston}, {Kamble}, {Law}, {Lazio}, {Lo}, {Macquart}, {Rea},
  {Rebbapragada}, {Reynolds}, {Ryder}, {Schmidt}, {Soria}, {Stairs}, {Tingay},
  {Torkelsson}, {Wagstaff}, {Walker}, {Wayth}, \&
  {Williams}}]{2013PASA...30....6M}
{Murphy}, T., {Chatterjee}, S., {Kaplan}, D.~L., {et~al.} 2013, \pasa, 30,
  e006, \dodoi{10.1017/pasa.2012.006}

\bibitem[{{Murphy} {et~al.}(2021){Murphy}, {Kaplan}, {Stewart}, {O'Brien},
  {Lenc}, {Pintaldi}, {Pritchard}, {Dobie}, {Fox}, {Leung}, {An}, {Bell},
  {Broderick}, {Chatterjee}, {Dai}, {d'Antonio}, {Doyle}, {Gaensler}, {Heald},
  {Horesh}, {Jones}, {McConnell}, {Moss}, {Raja}, {Ramsay}, {Ryder}, {Sadler},
  {Sivakoff}, {Wang}, {Wang}, {Wheatland}, {Whiting}, {Allison}, {Anderson},
  {Ball}, {Bannister}, {Bock}, {Bolton}, {Bunton}, {Chekkala}, {Chippendale},
  {Cooray}, {Gupta}, {Hayman}, {Jeganathan}, {Koribalski}, {Lee-Waddell},
  {Mahony}, {Marvil}, {McClure-Griffiths}, {Mirtschin}, {Ng}, {Pearce},
  {Phillips}, \& {Voronkov}}]{2021arXiv210806039M}
{Murphy}, T., {Kaplan}, D.~L., {Stewart}, A.~J., {et~al.} 2021, arXiv e-prints,
  arXiv:2108.06039.
\newblock \doarXiv{2108.06039}

\bibitem[{{Mutel} {et~al.}(1987){Mutel}, {Morris}, {Doiron}, \&
  {Lestrade}}]{1987AJ.....93.1220M}
{Mutel}, R.~L., {Morris}, D.~H., {Doiron}, D.~J., \& {Lestrade}, J.~F. 1987,
  \aj, 93, 1220, \dodoi{10.1086/114402}

\bibitem[{{Narayan}(1992)}]{1992RSPTA.341..151N}
{Narayan}, R. 1992, Philosophical Transactions of the Royal Society of London
  Series A, 341, 151, \dodoi{10.1098/rsta.1992.0090}

\bibitem[{{Ng} {et~al.}(2015){Ng}, {Champion}, {Bailes}, {Barr}, {Bates},
  {Bhat}, {Burgay}, {Burke-Spolaor}, {Flynn}, {Jameson}, {Johnston}, {Keith},
  {Kramer}, {Levin}, {Petroff}, {Possenti}, {Stappers}, {van Straten},
  {Tiburzi}, {Eatough}, \& {Lyne}}]{2015MNRAS.450.2922N}
{Ng}, C., {Champion}, D.~J., {Bailes}, M., {et~al.} 2015, \mnras, 450, 2922,
  \dodoi{10.1093/mnras/stv753}

\bibitem[{{Offringa} {et~al.}(2014){Offringa}, {McKinley}, {Hurley-Walker},
  {Briggs}, {Wayth}, {Kaplan}, {Bell}, {Feng}, {Neben}, {Hughes}, {Rhee},
  {Murphy}, {Bhat}, {Bernardi}, {Bowman}, {Cappallo}, {Corey}, {Deshpande},
  {Emrich}, {Ewall-Wice}, {Gaensler}, {Goeke}, {Greenhill}, {Hazelton},
  {Hindson}, {Johnston-Hollitt}, {Jacobs}, {Kasper}, {Kratzenberg}, {Lenc},
  {Lonsdale}, {Lynch}, {McWhirter}, {Mitchell}, {Morales}, {Morgan},
  {Kudryavtseva}, {Oberoi}, {Ord}, {Pindor}, {Procopio}, {Prabu}, {Riding},
  {Roshi}, {Shankar}, {Srivani}, {Subrahmanyan}, {Tingay}, {Waterson},
  {Webster}, {Whitney}, {Williams}, \& {Williams}}]{2014MNRAS.444..606O}
{Offringa}, A.~R., {McKinley}, B., {Hurley-Walker}, N., {et~al.} 2014, \mnras,
  444, 606, \dodoi{10.1093/mnras/stu1368}

\bibitem[{{Olausen} \& {Kaspi}(2014)}]{2014ApJS..212....6O}
{Olausen}, S.~A., \& {Kaspi}, V.~M. 2014, \apjs, 212, 6,
  \dodoi{10.1088/0067-0049/212/1/6}

\bibitem[{{Osterbrock}(1989)}]{1989agna.book.....O}
{Osterbrock}, D.~E. 1989, {Astrophysics of gaseous nebulae and active galactic
  nuclei} (Sausalito, CA: University Science Books)

\bibitem[{{Pearlman} {et~al.}(2018){Pearlman}, {Majid}, {Prince}, {Kocz}, \&
  {Horiuchi}}]{2018ApJ...866..160P}
{Pearlman}, A.~B., {Majid}, W.~A., {Prince}, T.~A., {Kocz}, J., \& {Horiuchi},
  S. 2018, \apj, 866, 160, \dodoi{10.3847/1538-4357/aade4d}

\bibitem[{{Pecaut} \& {Mamajek}(2013)}]{2013ApJS..208....9P}
{Pecaut}, M.~J., \& {Mamajek}, E.~E. 2013, \apjs, 208, 9,
  \dodoi{10.1088/0067-0049/208/1/9}

\bibitem[{{Perry} \& {Lyne}(1985)}]{1985MNRAS.212..489P}
{Perry}, T.~E., \& {Lyne}, A.~G. 1985, \mnras, 212, 489,
  \dodoi{10.1093/mnras/212.2.489}

\bibitem[{{Pintaldi} {et~al.}(2021){Pintaldi}, {Stewart}, {O'Brien}, {Kaplan},
  \& {Murphy}}]{2021arXiv210105898P}
{Pintaldi}, S., {Stewart}, A., {O'Brien}, A., {Kaplan}, D., \& {Murphy}, T.
  2021, arXiv e-prints, arXiv:2101.05898.
\newblock \doarXiv{2101.05898}

\bibitem[{{Polzin} {et~al.}(2020){Polzin}, {Breton}, {Bhattacharyya},
  {Scholte}, {Sobey}, \& {Stappers}}]{2020MNRAS.494.2948P}
{Polzin}, E.~J., {Breton}, R.~P., {Bhattacharyya}, B., {et~al.} 2020, \mnras,
  494, 2948, \dodoi{10.1093/mnras/staa596}

\bibitem[{{Pritchard} {et~al.}(2021){Pritchard}, {Murphy}, {Zic}, {Lynch},
  {Heald}, {Kaplan}, {Anderson}, {Banfield}, {Hale}, {Hotan}, {Lenc}, {Leung},
  {McConnell}, {Moss}, {Raja}, {Stewart}, \& {Whiting}}]{2021MNRAS.502.5438P}
{Pritchard}, J., {Murphy}, T., {Zic}, A., {et~al.} 2021, \mnras, 502, 5438,
  \dodoi{10.1093/mnras/stab299}

\bibitem[{{Quirrenbach} {et~al.}(1992){Quirrenbach}, {Witzel}, {Kirchbaum},
  {Hummel}, {Wegner}, {Schalinski}, {Ott}, {Alberdi}, \&
  {Rioja}}]{1992A&A...258..279Q}
{Quirrenbach}, A., {Witzel}, A., {Kirchbaum}, T.~P., {et~al.} 1992, \aap, 258,
  279

\bibitem[{{Ransom}(2001)}]{2001PhDT.......123R}
{Ransom}, S.~M. 2001, PhD thesis, Harvard University

\bibitem[{{Rea} \& {Esposito}(2011)}]{2011ASSP...21..247R}
{Rea}, N., \& {Esposito}, P. 2011, Astrophysics and Space Science Proceedings,
  21, 247, \dodoi{10.1007/978-3-642-17251-9\_21}

\bibitem[{{Rea} {et~al.}(2012){Rea}, {Pons}, {Torres}, \&
  {Turolla}}]{2012ApJ...748L..12R}
{Rea}, N., {Pons}, J.~A., {Torres}, D.~F., \& {Turolla}, R. 2012, \apjl, 748,
  L12, \dodoi{10.1088/2041-8205/748/1/L12}

\bibitem[{{Rea} {et~al.}(2013){Rea}, {Esposito}, {Pons}, {Turolla}, {Torres},
  {Israel}, {Possenti}, {Burgay}, {Vigan{\`o}}, {Papitto}, {Perna}, {Stella},
  {Ponti}, {Baganoff}, {Haggard}, {Camero-Arranz}, {Zane}, {Minter},
  {Mereghetti}, {Tiengo}, {Sch{\"o}del}, {Feroci}, {Mignani}, \&
  {G{\"o}tz}}]{2013ApJ...775L..34R}
{Rea}, N., {Esposito}, P., {Pons}, J.~A., {et~al.} 2013, \apjl, 775, L34,
  \dodoi{10.1088/2041-8205/775/2/L34}

\bibitem[{{Readhead}(1994)}]{1994ApJ...426...51R}
{Readhead}, A. C.~S. 1994, \apj, 426, 51, \dodoi{10.1086/174038}

\bibitem[{{Reid} \& {Ratcliffe}(2014)}]{2014RAA....14..773R}
{Reid}, H. A.~S., \& {Ratcliffe}, H. 2014, Research in Astronomy and
  Astrophysics, 14, 773, \dodoi{10.1088/1674-4527/14/7/003}

\bibitem[{{Reid} {et~al.}(2008){Reid}, {Cruz}, {Kirkpatrick}, {Allen},
  {Mungall}, {Liebert}, {Lowrance}, \& {Sweet}}]{2008AJ....136.1290R}
{Reid}, I.~N., {Cruz}, K.~L., {Kirkpatrick}, J.~D., {et~al.} 2008, \aj, 136,
  1290, \dodoi{10.1088/0004-6256/136/3/1290}

\bibitem[{{Roberts}(2013)}]{2013IAUS..291..127R}
{Roberts}, M. S.~E. 2013, in Neutron Stars and Pulsars: Challenges and
  Opportunities after 80 years, ed. J.~{van Leeuwen}, Vol. 291 (Cambridge, UK:
  Cambridge University Press), 127--132, \dodoi{10.1017/S174392131202337X}

\bibitem[{{Roy} {et~al.}(2010){Roy}, {Hyman}, {Pal}, {Lazio}, {Ray}, \&
  {Kassim}}]{2010ApJ...712L...5R}
{Roy}, S., {Hyman}, S.~D., {Pal}, S., {et~al.} 2010, \apjl, 712, L5,
  \dodoi{10.1088/2041-8205/712/1/L5}

\bibitem[{{Sault} {et~al.}(1995){Sault}, {Teuben}, \&
  {Wright}}]{1995ASPC...77..433S}
{Sault}, R.~J., {Teuben}, P.~J., \& {Wright}, M.~C.~H. 1995, in Astronomical
  Society of the Pacific Conference Series, Vol.~77, Astronomical Data Analysis
  Software and Systems IV, ed. R.~A. {Shaw}, H.~E. {Payne}, \& J.~J.~E.
  {Hayes}, 433.
\newblock \doarXiv{astro-ph/0612759}

\bibitem[{{Shannon} \& {Johnston}(2013)}]{2013MNRAS.435L..29S}
{Shannon}, R.~M., \& {Johnston}, S. 2013, \mnras, 435, L29,
  \dodoi{10.1093/mnrasl/slt088}

\bibitem[{{Skrutskie} {et~al.}(2006){Skrutskie}, {Cutri}, {Stiening},
  {Weinberg}, {Schneider}, {Carpenter}, {Beichman}, {Capps}, {Chester},
  {Elias}, {Huchra}, {Liebert}, {Lonsdale}, {Monet}, {Price}, {Seitzer},
  {Jarrett}, {Kirkpatrick}, {Gizis}, {Howard}, {Evans}, {Fowler}, {Fullmer},
  {Hurt}, {Light}, {Kopan}, {Marsh}, {McCallon}, {Tam}, {Van Dyk}, \&
  {Wheelock}}]{2006AJ....131.1163S}
{Skrutskie}, M.~F., {Cutri}, R.~M., {Stiening}, R., {et~al.} 2006, \aj, 131,
  1163, \dodoi{10.1086/498708}

\bibitem[{{Sotomayor-Beltran} {et~al.}(2013){Sotomayor-Beltran}, {Sobey},
  {Hessels}, {de Bruyn}, {Noutsos}, {Alexov}, {Anderson}, {Asgekar}, {Avruch},
  {Beck}, {Bell}, {Bell}, {Bentum}, {Bernardi}, {Best}, {Birzan}, {Bonafede},
  {Breitling}, {Broderick}, {Brouw}, {Br{\"u}ggen}, {Ciardi}, {de Gasperin},
  {Dettmar}, {van Duin}, {Duscha}, {Eisl{\"o}ffel}, {Falcke}, {Fallows},
  {Fender}, {Ferrari}, {Frieswijk}, {Garrett}, {Grie{\ss}meier}, {Grit},
  {Gunst}, {Hassall}, {Heald}, {Hoeft}, {Horneffer}, {Iacobelli}, {Juette},
  {Karastergiou}, {Keane}, {Kohler}, {Kramer}, {Kondratiev}, {Koopmans},
  {Kuniyoshi}, {Kuper}, {van Leeuwen}, {Maat}, {Macario}, {Markoff}, {McKean},
  {Mulcahy}, {Munk}, {Orru}, {Paas}, {Pandey-Pommier}, {Pilia}, {Pizzo},
  {Polatidis}, {Reich}, {R{\"o}ttgering}, {Serylak}, {Sluman}, {Stappers},
  {Tagger}, {Tang}, {Tasse}, {ter Veen}, {Vermeulen}, {van Weeren}, {Wijers},
  {Wijnholds}, {Wise}, {Wucknitz}, {Yatawatta}, \&
  {Zarka}}]{2013A&A...552A..58S}
{Sotomayor-Beltran}, C., {Sobey}, C., {Hessels}, J.~W.~T., {et~al.} 2013, \aap,
  552, A58, \dodoi{10.1051/0004-6361/201220728}

\bibitem[{{Spitler} {et~al.}(2014){Spitler}, {Lee}, {Eatough}, {Kramer},
  {Karuppusamy}, {Bassa}, {Cognard}, {Desvignes}, {Lyne}, {Stappers}, {Bower},
  {Cordes}, {Champion}, \& {Falcke}}]{2014ApJ...780L...3S}
{Spitler}, L.~G., {Lee}, K.~J., {Eatough}, R.~P., {et~al.} 2014, \apjl, 780,
  L3, \dodoi{10.1088/2041-8205/780/1/L3}

\bibitem[{{Staelin}(1969)}]{1969IEEEP..57..724S}
{Staelin}, D.~H. 1969, IEEE Proceedings, 57, 724,
  \dodoi{10.1109/PROC.1969.7051}

\bibitem[{{Staelin} \& {Reifenstein}(1968)}]{1968Sci...162.1481S}
{Staelin}, D.~H., \& {Reifenstein}, Edward~C., I. 1968, Science, 162, 1481,
  \dodoi{10.1126/science.162.3861.1481}

\bibitem[{{Stappers} {et~al.}(1996){Stappers}, {Bailes}, {Lyne}, {Manchester},
  {D'Amico}, {Tauris}, {Lorimer}, {Johnston}, \&
  {Sandhu}}]{1996ApJ...465L.119S}
{Stappers}, B.~W., {Bailes}, M., {Lyne}, A.~G., {et~al.} 1996, \apjl, 465,
  L119, \dodoi{10.1086/310148}

\bibitem[{{Swinbank} {et~al.}(2015){Swinbank}, {Staley}, {Molenaar}, {Rol},
  {Rowlinson}, {Scheers}, {Spreeuw}, {Bell}, {Broderick}, {Carbone}, {Garsden},
  {van der Horst}, {Law}, {Wise}, {Breton}, {Cendes}, {Corbel},
  {Eisl{\"o}ffel}, {Falcke}, {Fender}, {Grie{\ss}meier}, {Hessels}, {Stappers},
  {Stewart}, {Wijers}, {Wijnands}, \& {Zarka}}]{2015A&C....11...25S}
{Swinbank}, J.~D., {Staley}, T.~D., {Molenaar}, G.~J., {et~al.} 2015, Astronomy
  and Computing, 11, 25, \dodoi{10.1016/j.ascom.2015.03.002}

\bibitem[{{Vedantham} {et~al.}(2020){Vedantham}, {Callingham}, {Shimwell},
  {Dupuy}, {Best}, {Liu}, {Zhang}, {De}, {Lamy}, {Zarka}, {R{\"o}ttgering}, \&
  {Shulevski}}]{2020ApJ...903L..33V}
{Vedantham}, H.~K., {Callingham}, J.~R., {Shimwell}, T.~W., {et~al.} 2020,
  \apjl, 903, L33, \dodoi{10.3847/2041-8213/abc256}

\bibitem[{{Virtanen} {et~al.}(2020){Virtanen}, {Gommers}, {Oliphant},
  {Haberland}, {Reddy}, {Cournapeau}, {Burovski}, {Peterson}, {Weckesser},
  {Bright}, {van der Walt}, {Brett}, {Wilson}, {Millman}, {Mayorov}, {Nelson},
  {Jones}, {Kern}, {Larson}, {Carey}, {Polat}, {Feng}, {Moore}, {VanderPlas},
  {Laxalde}, {Perktold}, {Cimrman}, {Henriksen}, {Quintero}, {Harris},
  {Archibald}, {Ribeiro}, {Pedregosa}, {van Mulbregt}, \& {SciPy 1. 0
  Contributors}}]{2020NatMe..17..261V}
{Virtanen}, P., {Gommers}, R., {Oliphant}, T.~E., {et~al.} 2020, Nature
  Methods, 17, 261, \dodoi{10.1038/s41592-019-0686-2}

\bibitem[{{Wagner} \& {Er}(2020)}]{2020arXiv200616263W}
{Wagner}, J., \& {Er}, X. 2020, \aap, submitted, arXiv:2006.16263.
\newblock \doarXiv{2006.16263}

\bibitem[{{Wahl} {et~al.}(2021){Wahl}, {McLaughlin}, {Gentile}, {Jones},
  {Spiewak}, {Arzoumanian}, {Crowter}, {Demorest}, {DeCesar}, {Dolch}, {Ellis},
  {Ferdman}, {Ferrara}, {Fonseca}, {Garver-Daniels}, {Jones}, {Lam}, {Levin},
  {Lewandowska}, {Lorimer}, {Lynch}, {Madison}, {Ng}, {Nice}, {Pennucci},
  {Ransom}, {Ray}, {Stairs}, {Stovall}, {Swiggum}, \&
  {Zhu}}]{2021arXiv210405723W}
{Wahl}, H.~M., {McLaughlin}, M., {Gentile}, P.~A., {et~al.} 2021, \apj,
  submitted, arXiv:2104.05723.
\newblock \doarXiv{2104.05723}

\bibitem[{{Wainscoat} {et~al.}(1992){Wainscoat}, {Cohen}, {Volk}, {Walker}, \&
  {Schwartz}}]{1992ApJS...83..111W}
{Wainscoat}, R.~J., {Cohen}, M., {Volk}, K., {Walker}, H.~J., \& {Schwartz},
  D.~E. 1992, \apjs, 83, 111, \dodoi{10.1086/191733}

\bibitem[{{Walker}(1998)}]{1998MNRAS.294..307W}
{Walker}, M.~A. 1998, \mnras, 294, 307,
  \dodoi{10.1046/j.1365-8711.1998.01238.x}

\bibitem[{{Wayth} {et~al.}(2015){Wayth}, {Lenc}, {Bell}, {Callingham},
  {Dwarakanath}, {Franzen}, {For}, {Gaensler}, {Hancock}, {Hindson},
  {Hurley-Walker}, {Jackson}, {Johnston-Hollitt}, {Kapi{\'n}ska}, {McKinley},
  {Morgan}, {Offringa}, {Procopio}, {Staveley-Smith}, {Wu}, {Zheng}, {Trott},
  {Bernardi}, {Bowman}, {Briggs}, {Cappallo}, {Corey}, {Deshpande}, {Emrich},
  {Goeke}, {Greenhill}, {Hazelton}, {Kaplan}, {Kasper}, {Kratzenberg},
  {Lonsdale}, {Lynch}, {McWhirter}, {Mitchell}, {Morales}, {Morgan}, {Oberoi},
  {Ord}, {Prabu}, {Rogers}, {Roshi}, {Shankar}, {Srivani}, {Subrahmanyan},
  {Tingay}, {Waterson}, {Webster}, {Whitney}, {Williams}, \&
  {Williams}}]{2015PASA...32...25W}
{Wayth}, R.~B., {Lenc}, E., {Bell}, M.~E., {et~al.} 2015, \pasa, 32, e025,
  \dodoi{10.1017/pasa.2015.26}

\bibitem[{{Whittet}(1992)}]{1992dge..book.....W}
{Whittet}, D.~C.~B. 1992, {Dust in the galactic environment}

\bibitem[{{Williams} {et~al.}(2014){Williams}, {Cook}, \&
  {Berger}}]{2014ApJ...785....9W}
{Williams}, P.~K.~G., {Cook}, B.~A., \& {Berger}, E. 2014, \apj, 785, 9,
  \dodoi{10.1088/0004-637X/785/1/9}

\bibitem[{{Yan} {et~al.}(2011){Yan}, {Manchester}, {Hobbs}, {van Straten},
  {Reynolds}, {Wang}, {Bailes}, {Bhat}, {Burke-Spolaor}, {Champion},
  {Chaudhary}, {Coles}, {Hotan}, {Khoo}, {Oslowski}, {Sarkissian}, \&
  {Yardley}}]{2011Ap&SS.335..485Y}
{Yan}, W.~M., {Manchester}, R.~N., {Hobbs}, G., {et~al.} 2011, \apss, 335, 485,
  \dodoi{10.1007/s10509-011-0756-0}

\bibitem[{{Yao} {et~al.}(2017){Yao}, {Manchester}, \&
  {Wang}}]{2017ApJ...835...29Y}
{Yao}, J.~M., {Manchester}, R.~N., \& {Wang}, N. 2017, \apj, 835, 29,
  \dodoi{10.3847/1538-4357/835/1/29}

\bibitem[{{You} {et~al.}(2007){You}, {Hobbs}, {Coles}, {Manchester}, {Edwards},
  {Bailes}, {Sarkissian}, {Verbiest}, {van Straten}, {Hotan}, {Ord}, {Jenet},
  {Bhat}, \& {Teoh}}]{2007MNRAS.378..493Y}
{You}, X.~P., {Hobbs}, G., {Coles}, W.~A., {et~al.} 2007, \mnras, 378, 493,
  \dodoi{10.1111/j.1365-2966.2007.11617.x}

\bibitem[{{Yuan} {et~al.}(2013){Yuan}, {Liu}, \& {Xiang}}]{2013MNRAS.430.2188Y}
{Yuan}, H.~B., {Liu}, X.~W., \& {Xiang}, M.~S. 2013, \mnras, 430, 2188,
  \dodoi{10.1093/mnras/stt039}

\bibitem[{{Zavala} \& {Taylor}(2003)}]{2003ApJ...589..126Z}
{Zavala}, R.~T., \& {Taylor}, G.~B. 2003, \apj, 589, 126,
  \dodoi{10.1086/374619}

\bibitem[{{Zhao} {et~al.}(2020){Zhao}, {Morris}, \&
  {Goss}}]{2020ApJ...905..173Z}
{Zhao}, J.-H., {Morris}, M.~R., \& {Goss}, W.~M. 2020, \apj, 905, 173,
  \dodoi{10.3847/1538-4357/abc75e}

\bibitem[{{Zhao} {et~al.}(1992){Zhao}, {Roberts}, {Goss}, {Frail}, {Lo},
  {Subrahmanyan}, {Kesteven}, {Ekers}, {Allen}, {Burton}, \&
  {Spyromilio}}]{1992Sci...255.1538Z}
{Zhao}, J.-H., {Roberts}, D.~A., {Goss}, W.~M., {et~al.} 1992, Science, 255,
  1538, \dodoi{10.1126/science.255.5051.1538}

\bibitem[{{Zhu} \& {Xu}(2006)}]{2006MNRAS.365L..16Z}
{Zhu}, W.~W., \& {Xu}, R.~X. 2006, \mnras, 365, L16,
  \dodoi{10.1111/j.1745-3933.2005.00117.x}

\bibitem[{{Zic} {et~al.}(2019){Zic}, {Stewart}, {Lenc}, {Murphy}, {Lynch},
  {Kaplan}, {Hotan}, {Anderson}, {Bunton}, {Chippendale}, {Mader}, \&
  {Phillips}}]{2019MNRAS.488..559Z}
{Zic}, A., {Stewart}, A., {Lenc}, E., {et~al.} 2019, \mnras, 488, 559,
  \dodoi{10.1093/mnras/stz1684}

\bibitem[{{Zyuzin} {et~al.}(2016){Zyuzin}, {Zharikov}, {Shibanov}, {Danilenko},
  {Mennickent}, \& {Kirichenko}}]{2016MNRAS.455.1746Z}
{Zyuzin}, D., {Zharikov}, S., {Shibanov}, Y., {et~al.} 2016, \mnras, 455, 1746,
  \dodoi{10.1093/mnras/stv2401}

\end{thebibliography}
\bibliographystyle{aasjournal}

\end{document}